\documentclass{llncs}
\usepackage{multirow}
\usepackage{makeidx}  

\usepackage{amssymb,amsfonts,amsmath}
\usepackage{graphicx}
\usepackage{tabularx}
\usepackage{algorithm}
\usepackage{algorithmicx}

\usepackage{algpseudocode}
\usepackage{epsfig}
\usepackage{verbatim} 
\usepackage{epstopdf}

\usepackage{ntheorem}
\theoremseparator{.}   

\newtheorem{theo}{Theorem}

\newtheorem{hyp}{Working hypothesis}

\begin{document}

\pagestyle{headings}  
\title{Bose-Einstein Condensation  in Satisfiability Problems}
\titlerunning{}  
\author{Claudio Angione$^1$, Annalisa Occhipinti$^2$, Giovanni Stracquadanio $^3$ Giuseppe Nicosia$^4$}
\authorrunning{}   
%
\tocauthor{}

%
\institute{					$^1$Computer Laboratory - University of Cambridge, UK\\ 
               $^2$Dept. of Mathematics and Computer Science -  University of Catania, Italy\\
               $^3$Department of Biomedical Engineering - Johns Hopkins University, USA\\
               $^4$Dept. of Mathematics and Computer Science - University of Catania and Institute for Advanced Studies, Italy\\
	\texttt{\scriptsize claudio.angione@cl.cam.ac.uk; occhipinti@dmi.unict.it; stracquadanio@jhu.edu; nicosia@dmi.unict.it}\\
	\hfill \\
	\begin{flushleft}
	\end{flushleft}
}
\maketitle              

\begin{abstract} 
This paper is concerned with the complex behavior arising in satisfiability problems. We present a new statistical physics-based characterization of the satisfiability problem.
Specifically, we design an algorithm that is able to produce graphs starting from a $k$--SAT instance, in order to  analyze them and show whether a Bose-Einstein condensation occurs. We observe that, analogously to complex networks, the networks of $k$--SAT instances follow Bose statistics and can undergo Bose-Einstein condensation. In particular, $k$--SAT instances move from a fit-get-rich network to a winner-takes-all network as the ratio of clauses to variables decreases, and the phase transition of $k$--SAT approximates the critical temperature for the Bose-Einstein condensation. Finally, we employ the fitness-based
classification to enhance SAT solvers (e.g., ChainSAT) and obtain the
consistently highest performing SAT solver for CNF formulas, and
therefore a new class of efficient hardware and software verification
tools.
\end{abstract}

\section{Introduction}
\label{sec:intro}

Satisfiability (SAT) is a famous logical reasoning problem defined in terms of Boolean variables
and logical constraints ({\it clauses}) describing the relation among these variables.
Each such variable can be negated or not, that is, each variable (a {\it literal}) can be either {\it True} or {\it False}; the constraint is built as the OR function of the $k$ variables ($k$--SAT) \cite{GareyJohnson:complexity}.
In general,  propositional formulas are represented in Conjunctive Normal Form (CNF).
A CNF formula  consists of a conjunction of $m$ clauses, each
of which consists of a disjunction of $k$ literals.
SAT has received a great deal of theoretical and experimental study as
the paradigmatic ${\cal NP}$-complete problem as decision problem \cite{nature:99,science:02} and ${\cal NP}$-hard as solution when there are more than two literals for each clause \cite{zimmermann1997linear}.
The SAT problem is also crucial for solving large-scale computational problems, such as
AI planning, scheduling \cite{coelho2011multi}, protein structure prediction, haplotype inference, circuit-level prediction of crosstalk noise, computer chip verification, termination analysis in term-rewrite systems, model checking, and hardware and software verification \cite{clarke2008model,jain2009efficient}. Indeed, most verification tools consist of decision procedures to check the satisfiability of a given formula generated by the verification process. As a result, the subject of practical SAT solvers has received considerable research attention, and numerous solver algorithms have been proposed and implemented \cite{hamadi2009manysat,moskewicz2001chaff,ganai2002combining}. In particular, several SAT solvers rely on linear programming \cite{zimmermann1997linear} or tabu search \cite{nonobe1998tabu} and have been thoroughly analyzed in their worst cases \cite{mastrolilli2005maximum}. When we consider randomly generated instances, SAT is called random satisfiability problem. The original aim for inspecting random instances of $k$--SAT has been the desire to decipher the hardness and complexity of typical (standard) instances. For this reason, research works on $k$--SAT have been focused on developing algorithms for counting the number of solutions \cite{Birnbaum:99,Dubois:91,zhang}, and analyzing their computational complexity \cite{littman}.\\
\indent The cooperative dynamics of the interacting clauses can exhibit new
rich behavior that is not evident in the properties of the
individual clauses and literals (the elementary units) that make up a
SAT formula (the many-body system) of a very large numbers of these
units.
Standard experimental methods for studying ${\cal NP}$-complete
problems use a random generator of the problem instances and an
exact (possibly optimized by means of heuristics) algorithm to
solve them. By analyzing the results with proper measures (e.g., the number of recursive calls), one can obtain
important information about the problem, such as phase transitions, topological characterization of
the search space, and clusters of solutions \cite{montanari2008clusters}.
During the last twenty years, studies in theoretical computer science
have exploited new methodologies, based on statistical physics and experimental computer science,
for investigating the nature and properties of ${\cal NP}$-complete problems \cite{mitchell1992hard,hogg1996phase,martin2001statistical}.\\
\indent There is a deep connection between ${\cal NP}$-complete problems
and models studied in statistical physics. This connection leads to determining computational complexity from
characteristic \emph{phase transitions} in the $k$--SAT problem \cite{nature:99}. In Sherrington's work \cite{sherrington2010physics}, $k$--SAT is thought of as an extension of the Sherrington and Kirkpatrick's spin glass model \cite{sherrington1975solvable}. Moreover, its graph structure is an extension of the Erd\"{o}s-R\'{e}nyi random graphs; in particular, $k$--SAT models on Erd\"{o}s-R\'{e}nyi graphs showed the existence of free energy limits \cite{bayati2010combinatorial}. Although computer programs based on local dynamical algorithms are unable to reach the HARD-SAT phase in the neighborhood of the $k$--SAT phase transition, spin glasses techniques \cite{mezard2009information} allow to quantify the HARD-SAT region between the SAT and UNSAT ones. 
M{\'e}zard et al.\ \cite{science:02} showed the existence of an intermediate phase in $k$--SAT problems below
the phase transition threshold, and a powerful class of optimization algorithms was designed and tested successfully on the largest existing benchmark of $k$--SAT. Krzaka{\l}a et al.\ \cite{krzakala2007gibbs} discovered and analyzed four phase transitions in the solution space of random $k$--SAT. As the constraint density increases, clusters of solutions appear in the solution space; then, solutions condense over a few large clusters.
These results strengthen the link between computational models and properties of physical systems, and offer the possibility of new developments and discoveries in this research field.\\
%
\indent The goal of our research is to characterize the condensation phenomenon for $k$--SAT
problems by translating a formula into a graph $G=(V,E)$, and then to employ this characterization to improve the well-known ChainSAT algorithm \cite{alava2008circumspect}. Inspired by
Bianconi and  Barab{\'a}si's research work on Bose-Einstein condensation (BEC) in complex networks \cite{bianconi2001}, we design an algorithm that  produces graphs starting from a $k$--SAT instance
and associates each clause to a fitness value.
The phase diagram of the graph provided by the algorithm shows evidence of BEC for low values of the clauses-to-variables ratio. The BEC, from the very beginning, was associated to superfluidity: as London stated in 1938, ``the peculiar phase transition ($\lambda$ point) that liquid helium undergoes at \mbox{$2.19$ K}, most probably has to be regarded as the condensation phenomenon of the Bose-Einstein statistic'' \cite{london1938bose}. Hence, superfluidity in a $k$-SAT formula could be thought of as a consequence of the low constraint density that we find in the SAT phase.
Our results give new hints in understanding the complexity and the structure of a $k$--SAT instance in phase transition. The graph of a given instance allows us to satisfy it by finding a truth assignment only for the fittest clauses. Our approach makes use of complex networks in order to operate on the instance, without requiring a priori investigation of its solutions.

The rest of the paper is organized as follows. First, we give an overview of the Bose-Einstein distribution and tailor it to the satisfiability problem by translating a SAT formula into a graph. Then, we investigate two variants of our algorithm. We present experimental evidence supporting the hypothesis that the phase transition between solvable and unsatisfiable instances of $3$-SAT approximates the locus of the Bose-Einstein condensation in the phase diagram of $3$-SAT formulas. Finally, we show how to improve the ChainSAT solver by using our algorithm to provide a clause ordering.


\section{Bose-Einstein Distribution}\label{sec:bec-ksat}

The analysis of the state of matter, from a quantum point of view, states that all particles of the same type are equal and indistinguishable. Let us consider an isolated system of $N$ identical and indistinguishable bosons confined to a space of volume $V$ and sharing a given energy $E$.
These latter are particles that do not obey the Pauli exclusion principle, since two or more bosons may have exactly the same quantum numbers. We assume that these bosons can be distributed into a set of energy levels, where each level $E_i$ is characterized by an {\it energy} $\epsilon_i$, i.e., the energy of each particle settled on that energy level, and a {\it degeneration} $g_i$, representing the number of different physical states that can be found at that level. Accordingly, the $N$ identical and indistinguishable particles are distributed among the energy levels, and each level $E_i$ contains $n_i$ particles, to be accommodated among its $g_i$ quantum states. For instance, if $n_i=2$ and $g_i=3$, the particles $a$ and $b$ can settle on $E_i$ in one of these ways:
$  ab \| -  \| -,  \hspace{0.3cm} - \| ab \| -,  \hspace{0.3cm} - \| - \| ab, \hspace{0.3cm} a\| -  \|b, \hspace{0.3cm} a\|b\| -,  \hspace{0.3cm} - \|a\|b$. (Permutations of particles must not be included, since $a$ and $b$ are indistinguishable.)

It is straightforward to check that $n_i$ particles may be put on the level $E_i$ (consisting of $g_i$ states) in $[n_i+(g_i-1)]!$ different ways. Since bosons are indistinguishable and the physical states are equivalent, the number of possible assignments of $n_i$ bosons on $E_i$ is:
\begin{equation}
w_i = \frac{(n_i+g_i-1)!}{n_i!(g_i-1)!} = \binom{n_i+g_i-1}{n_i}.
\end{equation}
By iterating for all the energy levels $E_i$, one can observe that a distribution $\{n_i\}$ (i.e., a distribution with $n_i$ particles on the level $E_i$, $\forall i$) can be obtained in 
\begin{equation}
W=\prod_i{w_i}=\prod_i{\frac{(n_i+g_i-1)!}{n_i!(g_i-1)!}}
\end{equation}
different ways. In other words, $w_i$ is the number of distinct microstates associated with the $i$-th level of the spectrum, while $W$ is the number of distinct microstates associated with the whole distribution set $\{n_i\}$.
The particles distribution corresponding to the statistical equilibrium is the most probable one, thus it is the one that may be reached in the largest number of possible ways. Hence, in order to find it, we compute the maximum $W$ subject to the conservation of the number of particles $\sum_{i}n_i = N$, and to the preservation of the system energy $\sum_{i}\epsilon_i n_i = E$. We adopt the method of Lagrange's undetermined multipliers, but rather than maximizing $W$ directly, we maximize $\log{W}$, since $\log{}$ is a monotone transformation. This method results in the following condition:
\begin{equation}
\sum_i{\left[\log{\left(\frac{n_i+g_i}{n_i}\right)} - \alpha - \beta \epsilon_i \right] \delta n_i} = 0,
\end{equation}
where $\alpha$ and $\beta$ are the Lagrangian undetermined multipliers associated with the two restrictive conditions of conservation. Since the variations $\delta n_i$ are completely arbitrary, this condition can be satisfied if and only if all their coefficients vanish identically, namely:
$$ \log{\left(\frac{n_i+g_i}{n_i}\right)} - \alpha - \beta \epsilon_i = 0,\quad \forall i. $$
This equality leads to the following definition of {\it Bose-Einstein distribution}:
\begin{equation}
n_i=\frac{g_i}{e^{\alpha+\beta \epsilon_i}-1},
\end{equation}
where $\alpha=-\frac{\mu_C}{k_B T}$ and  $\beta=\frac{1}{k_B T}$ are inversely proportional (by means of Boltzmann's constant $k_B$) to the absolute temperature $T$ of the system at the equilibrium, and $\mu_C$ represents the chemical potential.

Given an ideal Bose-Einstein gas in equilibrium below its transition temperature, the {\it Bose-Einstein condensation} (BEC) is the property that a finite fraction of particles occupies the lowest energy level. According to Penrose and Onsager \cite{penrose1956bose}, we can provide a criterion of BEC for an ideal gas in equilibrium: BEC $\iff
 \frac{\left\langle n_0 \right\rangle}{N} = e^{O(1)}$, No BEC $\iff \frac{\left\langle n_0 \right\rangle}{N} = o(1)$,
where $\left\langle n_0 \right\rangle$ is the average number of particles that occupy the lowest energy level $E_0$. (The first equation is equivalent to $\frac{\left\langle n_0 \right\rangle}{N} = \mbox{constant}$, but it is weaker and easier to apply.) For low values of temperature, i.e., when \mbox{$T \rightarrow 0$ K}, the BEC takes place \cite{pathria}.
This phenomenon consists of a very unusual state of aggregation of particles, called {\it Bose-Einstein condensate}. Its characteristic is different from those of the solid state, liquid state, gas and plasma, thus it is known as ``{\it the fifth state of matter}''.
In particular, below a critical temperature $T_{BEC}$, all the particles settle on the same quantum state and occupy the same energy level. Hence, they are absolutely identical, inasmuch as there is no possible measurement that can tell them apart. In other words, they lose their individuality, and the single-particle perception is missing. 

Inspired by Bianconi and Barab{\'a}si's work \cite{bianconi2001}, we provide an algorithm to investigate the BEC phenomenon in the $k$--SAT problem. By translating a SAT formula into a graph, we define the condensation of the formula over its fittest clause as the emergence of a star-like topology in the graph. This phenomenon is associated with the condensation of bosons on the lowest energy level (see examples in Supporting Information).

\section{The SAT to Graph Algorithm}

An instance of the $k$--SAT problem consists of:
\begin{itemize}
\item [$\cdot$] a set $X$ of variables, with $\left| X \right| =n$;
\item [$\cdot$] a set $C$ of clauses over $X$, where $\left| C \right|=m$, such that each clause $C_i \in C,\ \forall i=1,...,m$, has $k$ literals and can be written as $C_i = L_1 \vee L_2 \vee ... \vee L_k $. Each literal $L_\mu \in L,\ \forall \mu=1,...,l$, where $L=X \cup \overline{X} \cup \{True, False\}$ is the set of literals, $\left| L \right|=l$.
\end{itemize}
\noindent The problem is to find a satisfying truth assignment for the following formula:
\begin{equation}
F=C_1 \wedge C_2 \wedge ... \wedge C_m.
\end{equation}

The {\it SAT to Graph Transformation Algorithm} (S2G) translates a $k$--SAT instance into a graph $G=(V,E)$, where $V$ are the vertices and $E$ are the edges. A vertex $v_i$ is a clause $C_i$ of the formula $F$, i.e., $v_i=v(C_i)$, whereas each edge $e_{jh}$ represents a relation between two clauses, i.e., $e_{jh} = (v(C_j), v(C_h))$, as we see later. Let us introduce two functions for literals and clauses. Firstly, we define the {\it global frequency of literals} as:
\begin{equation}
\varphi^{G}(L_\mu)= \mbox{occurrences of $L_\mu$ in $F$}, \qquad \mu=1,...,l,
\end{equation}
which reports the frequency of a literal into a $k$--SAT formula. Secondly, we define the {\it global fitness of clauses} as:
\begin{equation}
f^{G}(C_i)=\sum_{\mu=1}^k \varphi^{G}(L_\mu), \qquad L_\mu \in C_i,\qquad i=1,...,m,
\end{equation}
which is a fitness function to evaluate clauses and  grows with a monotonic behavior with respect to the $\varphi^{G}$ of its literals. The construction of the graph $G=(V,E)$ is an iterative process in which each clause $C_i$ is assigned to a vertex (node) $v_i,$ and edges $e_{jh}$ are links established according to an {\it affinity function}, as we see below. Since the construction is dynamical, we need to define the \emph{local frequency of literals} and the \emph{local fitness of clauses}.
While the global ones are determined on the complete formula $F$, the local ones concern only the   
clauses that have been added as vertices in $V$ using a subset $F'$ of the clauses of $F$. In particular, we define the {\it local frequency of literals} as follows:
\begin{equation}
\varphi^{L}(L_\mu)= \mbox{occurrences of $L_\mu$ in $F'$},\qquad \mu=1,...,l .
\end{equation}
Analogously, the {\it local fitness of clauses} is defined as:
\begin{equation}
f^{L}(C_i)=\sum_{\mu=1}^k \varphi^{L}(L_\mu), \qquad L_\mu \in C_i,\qquad i=1,...,m.
\end{equation}
It is obvious that, at iteration $i$, a literal $L_\mu$ has $\varphi^{L}(L_\mu)=0$ in case it belongs to a clause that has not been added to $V(G)$ yet; when the algorithm ends, $\varphi^{G}(L_\mu)=\varphi^{L}(L_\mu),\ \forall \mu=1,..,l$.

Hereinafter we need to suppose that the order of literals in a clause has no importance. However, since the OR operator is commutative, it is possible to define a distance metric that states how many literals are not in common between two clauses. Let $C_i, C_j$ be two clauses made up of literals $L_{\mu}^{i}$ and $L_{\mu}^{j}$ respectively; we define the following distance:
\begin{equation}
d(C_i,C_j) = \left| \, \left\{\mu \in \{1,...,k\} : L_{\mu}^{i} \neq L_{\mu}^{j} \right\} \, \right|,
\label{distance}
\end{equation}
which is a metric distance that can be related to the well-known Hamming distance \cite{roman1992coding}. In Supporting Information \ref{appendixthm}, we prove that $d$ is a metric.\\

Let $G=(V,E)$ be the graph obtained at the $(i-1)$-th iteration, and $F' \subset F$ be the temporary $k$--SAT subformula $F' = C_{t_1} \wedge C_{t_2} \wedge ... \wedge  C_{t_{i-1}}$.
In order to add a clause $C_{t_i}$ to $G$ as a node $v(C_{t_i})$, we estimate the probability of being connected to a node that already belongs to the graph; this probability must be computed for each node (clause) added to $G$, since it is the criterion to build edges between nodes. We define the probability that a new node $v(C_{t_i})$ is connected to the node $v(C_{t_j}) \in V(G)$ as:
\begin{equation}
\Pi_{t_j} = \frac{ k_{t_j} \cdot f^{L}(C_{t_j}) }{\displaystyle \sum_{\nu=1}^{|V|} k_{t_\nu} \cdot f^{L}(C_{t_\nu}) },
\label{eqPi}
\end{equation}
where $k_{t_j} = \mbox{degree}(v(C_{t_j}))$ is the connectivity of $C_{t_j}$ (i.e., the number of links shared by node $v(C_{t_j})$),  and $f^{L}(C_{t_j})$ is the fitness of the clause $C_{t_j}$. This probability distribution ensures that a new vertex is likely linked to an existing one with high fitness value or/and high connectivity \cite{bianconi2001}. We deduce that this process brings to a model in which the attractiveness and evolution of a node are determined by its fitness and by its number of links.

In order to assign the new node-clause $C_{t_i}$ an appropriate number representing an energy level \cite{bianconi2001}, it is necessary to normalize the local fitness values as $f_r^{L}(C_{t_i})= f^{L}(C_{t_i})/f^{L}(C_{t})$, where $C_{t}$ is the {\it fittest clause} in the temporary graph already built using $F'$. As a result, as soon as the node $v(C_{t_i})$ enters the system, it has the following energy (see \cite{bianconi2001}):
\begin{equation}
\epsilon_{t_i} = -T \cdot \log{ f_r^{L}(C_{t_i}) },
\end{equation}
where $T = \frac{1}{\beta}$,  and  $\beta$ is a parameter used to model the temperature of the system. (In this work, when comparing two or more energy levels, we omit the multiplicative factor $T$.) If {\it two different} nodes are assigned {\it the same} energy value in our model, it means (from a physical point of view) that they represent {\it two different} degeneration states of {\it the same} energy level, as shown in Table \ref{tab:dictionary}.

\begin{table}
\centering
\setlength{\tabcolsep}{5.8pt} 
\begin{small}
\begin{tabular}{c c c}
\hline
{$G=(V,E)$}  & {$k$--SAT} & {Statistical physics} \\
\hline
node & clause & degeneration state of the energy level of the node\\
edge & link between two clauses & one particle for each degeneration state involved\\
node weight & fitness of a clause & value of the energy level\\
edge weight & probability of being established & weight on particles\\
\hline
\end{tabular}
\vspace{0.1cm}
\end{small}
\caption{Dictionary translating the graph (left) into the $k$--SAT problem (centre) and statistical physics language (right).}
\label{tab:dictionary}
\end{table}

The definition of probabilities and the linking criterion are the building blocks of the S2G algorithm, which consists of three main steps. 

\paragraph{{\bf Step I.}}Let $\Lambda=\emptyset$, $V=\emptyset$, $E=\emptyset$, and $F'=\emptyset$. Let $i$ be the index representing the number of the iteration. Here, we set $i=1$. The first clause $C_{t_1}$ to add to $F'$ is chosen randomly among the $m$ clauses of the given formula $F$. After computing the local fitness of the clause, we assign to it the normalized local fitness $f_r^{L}(C_{t_1})$. Since $C_{t_1}$ is the only clause added to the graph so far, its $f_r^{L}$ is set to 1. After that, we compute the energy level $\epsilon_{t_1}$, which in this case is equal to $0$. The variable $t$ is used to store the index of the fittest clause. Obviously, at the first iteration, it must be set to $t_1$. The pseudo-code of the first step is presented in Algorithm \ref{alg:step1}.

\paragraph{{\bf Step II.}}Successively, we perform another step of the algorithm, in order to establish the first link between two clauses, as shown in Algorithm \ref{alg:step2}. This step and the following ones include two procedures, shown in Algorithm \ref{alg:proc1} and Algorithm \ref{alg:proc2}. The second clause is chosen such that it is the closest to $C_{t_1}$, in terms of the distance defined in (\ref{distance}). If two or more clauses have the same minimum distance from $C_{t_1}$, then a random clause is chosen among them. Notice that at every iteration $i$ all the local frequencies of literals are updated, therefore the local fitness and the energy level of clauses are updated as well. We perform the normalization of the fitness in order to obtain a non-negative energy level. Indeed, the logarithm function, when its base is greater than $1$ and its argument belongs to the interval $]0;1]$, returns a non-positive value; since the absolute temperature $T$ is non-negative, the energy level becomes a non-negative value, as expected.

\paragraph{{\bf Step III (general step).}}
The main loop of the S2G algorithm shown in Algorithm \ref{alg:step3} will be performed after a link is established. The purpose, as in the previous step, is to choose an index $t_i$ such that the $C_{t_i}$ is the clause closest to the clause with highest fitness among those that are in the network so far (after the ($i-1$)-th step). For each established link, we put a particle on each of the two degeneration states of the two clauses involved. Moreover, the probability of establishing a link becomes the weight on the edge representing that link. The general step differs from the second step because it needs at least one edge in the graph to work properly. This prerequisite allows us to have at least two non-zero vertex connectivities, permitting to compute the probabilities $\Pi_{t_j}$, since the denominator in (\ref{eqPi}) is surely nonzero. This is the reason why in Step II we forced $C_{t_1}$ and $C_{t_2}$ to link together.

\begin{algorithm}
\caption{S2G Algorithm}
\begin{algorithmic}[1]
\State Selecting\_the\_First\_Clause-Node()
\State Connecting\_First\_two\_Clauses-Nodes()
\Statex
\While {$i < m $}
	\State $ i \gets i+1 $
	\State Find\_Closest\_Clause()
	\For {$j \gets 1$ to $i-1$}
		\State $ \Pi_{t_j} \gets \displaystyle\frac{ k_{t_j} \cdot f^{L}(C_{t_j}) }{\displaystyle \sum_{\nu=1}^{i-1} k_{t_\nu} \cdot f^{L}(C_{t_\nu}) } $
		\State try to connect $v(C_{t_i})$ to $v(C_{t_j})$ with probability $\Pi_{t_j}$ \label{lis3}
		\State $k_{t_j} \gets$ degree($v(C_{t_j})$)  \hskip1.3cm /* update connectivity of node $v(C_{t_j})$ */
	\EndFor
	\State $k_{t_i} \gets$ degree($v(C_{t_i})$)  \hskip1.8cm /* update connectivity of node $v(C_{t_i})$ */
	\Statex
	\State Update\_Fitness()	
\EndWhile

\end{algorithmic}
\label{alg:step3}
\end{algorithm}

\paragraph{}
The S2G algorithm is based on a probabilistic approach, which could even lead to an unexpected network. According to this process, the graph is built in such a way as to involve dynamical energy levels, i.e., the numerical value of each energy level changes at each iteration, due to the dynamical changes of the local frequencies.

The first clause added to the graph, i.e. $C_{t_1}$, could be chosen differently. For instance, the $f^{G}(C_i)$ could be taken into account in the clause choice, and the global fittest clause would be then selected as first clause in $G$. In this way, the {\it first mover advantage} principle is emphasized, since the first clause is also the fittest one, therefore it is easier for it to acquire the majority of the links of the whole network. It follows that this technique would lead to more BEC networks 
but prejudicing the unpredictability of the overall process.

When the graph is completed, we consider the connectivity of the richest node (the node that has the maximum number of links) 
in order to decide whether a Bose-Einstein condensation has occurred. If the connectivity is large enough (the thresholds have been determined experimentally, see Working hypothesis \ref{hyp_fraction}), we say that a BEC has taken place in the graph, i.e., one node has a huge fraction of edges and the remaining fraction is shared among all the other nodes. If the graph does not show any condensation, we compute the degree distribution in order to understand what kind of network has developed. Moreover, we compute the mean and the standard deviation of all nodes except the winner (i.e., the richest), so as to obtain simple statistics involving the rest of the degree distribution.

The computational complexity of our algorithm is polynomial. The Step II procedure has a complexity of $O(N^3)$, where $N= \mbox{max} \{n,m,k\}$, due to the subprocedure that computes the distance $d$ between the clauses eligible to join the graph and the fittest clause already added to it. The main loop of the S2G algorithm has $O(N^4)$ time complexity, since it consists of the Step II procedure applied (with slight modifications) to all of the remaining clauses of the $k$--SAT formula.
\section{Fitness-Based Preferential Attachment}
In this section we extend the S2G algorithm by including the concept of preferential attachment, thus obtaining a new algorithm called S2G-PA. 
Even this model starts with two nodes connected by an edge. Exactly as in the previous model, at each iteration a new node is added to the graph. The preferential attachment implemented in the new algorithm is based on the same principle of the algorithm used so far: if we consider a single node of the network, the probability of acquiring new edges is positively correlated with its degree. According to the previous section, in the fitness-based model the connectivity is not the only parameter taken into account, but also the fitness plays an important role in computing the probability of acquiring new edges.

The main difference between this model and the model presented before consists of the {\it preferential out-degree} ($\rho$), a technique applicable to directed graphs. At each iteration $i$, the node that joins the graph {\it is forced to connect} at most to $\rho$ existing nodes and at least to one node.
Recall that in the previous model there were no restrictions to the number of outgoing links (od($v$)) that a node could have. It follows that a number of nodes, when they joined the existing network, did not link to any other node of the graph. This caused the probability $\Pi$ (probability of linking a new node to them) to remain always $0$, therefore their degree remained equal to $0$ during the whole process, i.e., they never linked to the main connected component of the graph.
On the contrary, the new algorithm ensures that all the nodes will be part of the network, i.e., {\it all the nodes will have at least one link} and $G$ has only one connected component. The output networks of S2G and S2G-PA can be compared in Supporting Information \ref{appendixS2G} and Supporting Information \ref{appendixS2G-PA}. When the most connected nodes have the highest number of particles, and the winner node is identified with the lowest energy level, we obtain a clear ``signature'' of BEC in a preferential attachment scheme with fitness, as proved by Borgs et al.\ \cite{borgs2007first}. These facts help us confirm that when the BEC occurs there is a clear mapping between the Bose gas and the graph derived by the S2G algorithm.

The preferential attachment ensures that the condition $1 \leq \mbox{od}(v(C_{t_i})) \leq \rho, \ \forall i=1,...,m$ holds at each iteration, where od is the node out-degree.
In practice, our new algorithm sets the out-degree to $\rho$, but two or more links may be directed to the same node, depending on the probabilities computed (nevertheless, multiple links are considered as simple ones). Hence, the resulting out-degree of the new node may be less than $\rho$, but it is always $\ge 1$. Conversely, the in-degree has no restrictions. Generally, the standard preferential attachment leads to random scale-free Barab\'asi and Albert networks \cite{barabasi:99}, in which the distribution of degree decreases with a power law, that is reduced to a line in logarithmic scale.
In our case, the preferential attachment is accompanied by a fitness function (that is why the algorithm has been called {\it fitness based} preferential attachment), so the resulting network is not exactly a scale-free network.
Furthermore, the new model causes {\it competition} among nodes \cite{barabasi:2001}. Indeed, a new node has a fixed number $\rho$ of links available, therefore the old nodes have to compete to acquire one link from the new node. This competition gets more and more challenging as the graph increases, since the number of nodes increases but the number of available links from a new node remains the same.
It is evident that the resulting network obeys the widely known {\it first mover advantage} principle \cite{bianconi2001}, according to which the first nodes of the graph have more time to gain links than the last ones. Finally, our fitness-based model ensures a lot of unpredictability to the system, since the fitness of each node changes at each iteration, as explained before. Thanks to this mechanism, a node with high fitness may get into the graph at a later time and become richer and richer till it overcomes the richest nodes. On the other hand, once that a node has entered the graph, its fitness may remain the same in the iterations after, thus other nodes may overcome it. These features lead to a dynamic and erratic evolution of the network.

\section{Non-Integer Out-Degree}

Let us consider the $i$-th iteration of the graph generation, when the node $v(C_{t_i})$ is added to the network. Suppose that, according to the probabilities $\Pi$, the new node $v(C_{t_i})$ must be linked to the existing node $v(C_{t_j})$. In this section, we make the following hypothesis.

\begin{hyp} \nonumber \label{hyp_google}
An outgoing link is less important than an incoming link, that is, the incoming links are rewarded more than the outgoing links.
\end{hyp}

\begin{figure}[H]
\begin{tabular}{c|l}%
\multirow{3}{*}{\hspace{1cm}\includegraphics[scale=0.35]{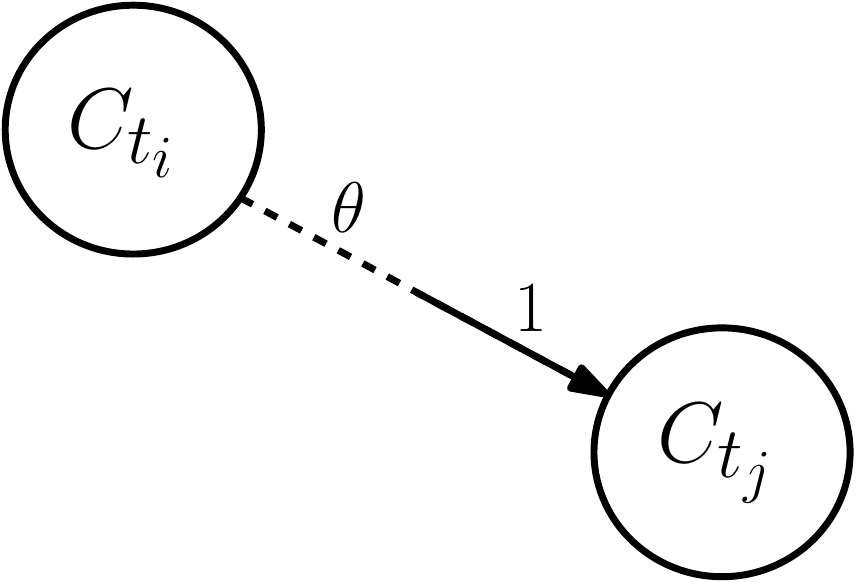}\hspace{1.2cm}
} &\hspace{0.8cm} $k_{t_i} \gets k_{t_i} + \theta, \ \ \ \mbox{where } 0<\theta<1 $ \\
& \\
& \\
& \\
& \hspace{0.8cm} $k_{t_j} \gets k_{t_j} + 1$
\end{tabular}
\caption{Link between $C_{t_i}$ and $C_{t_j}$. The dashed line represents the non-integer out-degree $\theta$ of $C_{t_i}$, while the continuous line represents the integer in-degree of $C_{t_j}$.}
\label{rational_out_degree}
\end{figure}

\noindent This hypothesis implies that our graph must be regarded as a directed graph in order to maintain the correspondence between a $k$--SAT instance and its graph, as well as to distinguish between outgoing and incoming links. According to the Google-like reference \cite{langville}, the same edge between the new node $v(C_{t_i})$ and the existing node $v(C_{t_j})$ does not increase their connectivity $k_{t_i}$ and $k_{t_j}$ (respectively) in the same way (see Figure \ref{rational_out_degree}).

\begin{figure}
\centering
\includegraphics[scale=0.9]{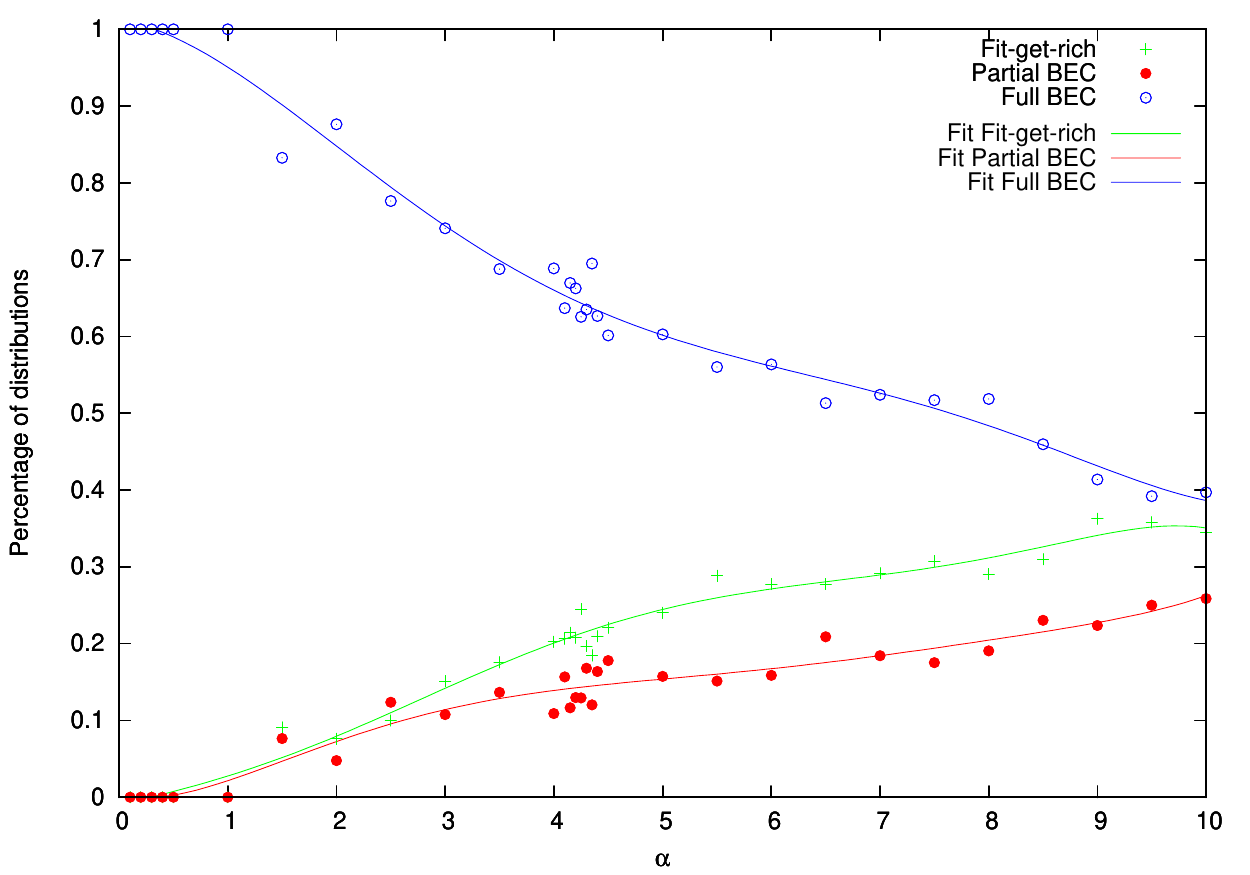} 
\caption{Percentage of each kind of network against ratio of clauses to variables, with fixed number of variables $n=100$ and non-integer out-degree $\theta=0.33$. The lines represent a sixth-order polynomial regression to fit the data. Full BEC occurs when the winner node gets greater or equal to 90\% of links of the network. Partial BEC occurs when the winner node gets greater or equal to 75\% and less than 90\% of links of the network. If the percentage is less than 75\%, the network has a Fit-get-rich topology.} 
\label{fig:frac_out_degree}
\end{figure}

\noindent Nevertheless, we continue to represent our graph as an undirected graph, making use of the relation $k_i = \theta \cdot \mbox{od}(v(C_i)) + \mbox{id}(v(C_i))$, where od and id are the node out-degree and in-degree respectively.
It is evident that a non-integer connectivity (i.e., a non-integer degree) leads to a new kind of evolution of the network. In this new model, nodes aim to connect to a particular node in the network, and when they manage to connect to it, that node gets richer and richer more rapidly than in the previous models. In fact, as incoming links are rewarded more than outgoing links, the connectivity of the node that acquires links raises much more than the connectivity of the nodes linking to it. We set $\theta=0.33$ so that an outgoing link is rewarded a third of an incoming link. The plot in Figure \ref{fig:frac_out_degree} has been obtained by fixing the number of variables $n=100$ and letting the number of clauses $m$ vary from $0$ to $1000$, so $\alpha=\frac{m}{n}$  (number of clauses over number of variables) varies from $0$ to $10$. The plot depicts the relationship between $\alpha$ and the percentage of each of the three classes of network returned by our algorithm, according to the following hypothesis.

\begin{hyp} \label{hyp_fraction}
Let us call {\it fraction-winner} $f$ the percentage of links acquired by the winner node over the whole set of links. We say that:
\begin{itemize}
\item [$\cdot$] a {\it Fit-get-rich topology} takes place when $f < 0.75$;
\item [$\cdot$] a {\it Partial BEC} takes place when $0.75 \leq f < 0.90$;
\item [$\cdot$] a {\it Full BEC} takes place when $f \geq 0.90$.
\end{itemize}
\end{hyp}
\noindent As shown in Figure \ref{fig:frac_out_degree}, we obtain a large number of networks that can undergo the full BEC (in which one node is an evident ``winner'' node). Figure \ref{fig:frac_out_degree} also shows that the number of BEC networks produced by our algorithm increases as $\alpha$ decreases.

In Algorithm \ref{alg:final}, we show the whole fitness-based preferential attachment algorithm with non-integer out-degree (S2G-PA). The preferential attachment technique replaces the $\Pi_{t_j}$ attachment scheme used in S2G. The S2G-PA algorithm uses the cumulative distribution function to ensure that, for each node, the probability of acquiring new links is directly proportional to its $\Pi$. Compared to the S2G model, the S2G-PA ensures that all the nodes have a nonzero connectivity, and also that an existing node-clause $v(C_j)$ with high fitness and connectivity (i.e., with high $\Pi_{t_j}$) has a higher probability of acquiring links from new nodes, inasmuch as the non-integer out-degree method emphasizes this behavior. Instructions \ref{istr1} and \ref{istr2} show how the non-integer out-degree has been implemented in our S2G-PA algorithm.

\begin{algorithm}[H]
\caption{S2G-PA Algorithm}
\begin{algorithmic}[1]

\State Selecting\_the\_First\_Clause-Node()
\State Connecting\_first\_two\_Clauses-Nodes()
\Statex
\While {$i < m $}
	\State $ i \gets i+1 $
	\State Find\_Closest\_Clause()
	\For {$j \gets 1$ to $i-1$}
		\State $ \Pi_{t_j} \gets \displaystyle\frac{ k_{t_j} \cdot f^{L}(C_{t_j}) }{\displaystyle \sum_{\nu=1}^{i-1} k_{t_\nu} \cdot f^{L}(C_{t_\nu}) } $

		\Statex
		\State $\Pi cum_{t_0} = 0 $
		\For {$j=1$ to $i-1$}
			\State $\Pi cum_{t_j} = \Pi cum_{t_{j-1}} + \Pi_{t_j} $ \hskip0.4cm /* compute cumulate probabilities */
		\EndFor
		\Statex
		\For {$z=1$ to $\rho$}
			\State $ x \gets random(]0;1]) $
			\State find $j \in \{1,...,i-1\}$ such that $ \Pi cum_{t_{j-1}} < x \leq \Pi 		cum_{t_j} $
			\State connect  $v(C_{t_i})$ to $v(C_{t_j})$
			\State\label{istr1} $k_{t_j} \gets k_{t_j} + 1$    \hskip1.8cm /* update connectivity of node $v(C_{t_j})$ */
			\State\label{istr2} $k_{t_i} \gets k_{t_i} +  \theta$    \hskip1.8cm /* update connectivity of node $v(C_{t_i})$ */
		\EndFor
	\EndFor
	\Statex
	\State Update\_Fitness()	
\EndWhile

\end{algorithmic}
\label{alg:final}
\end{algorithm}

\section{S2G-driven SAT Solvers}
Using the information provided by the S2G algorithm, in this section we show the improvement obtained in the performance of the ChainSAT algorithm \cite{alava2008circumspect}.
The S2G algorithm assigns an energy value to each clause of a $k$--SAT random instance. 
As seen before, the fitness  value of a clause is negatively correlated with its energy, 
and positively correlated both with the probability of having a high connectivity in the network and with the probability that its literals are frequently occurring in the instance.
Thus, the probability of satisfying all the linked clauses by assigning truth values only to one of them is larger if 
we assign truth values to one with the lowest energy value.
Consequently, in order to solve an instance we order the clauses by energy level.
If we find two or more clauses having the same energy, we  put first the one with the largest connectivity in the graph provided
by the S2G algorithm.
If they  have also the same connectivity, then we order them randomly.
As a result, an order is established among clauses of a random $k$--SAT instance. In the following  we refer to the order of the clauses 
as
their ``weight''. In particular, the heaviest clause will be the one on the lowest energy level.
%
%

%
ChainSAT \cite{alava2008circumspect} is a heuristic that never moves up in energy, since
the number of unsatisfied clauses is a non-increa\-sing function of the sequence of trial configurations traversed by the algorithm. 
For $k=4$, $k=5$, and $k=6$, ChainSAT is shown to solve random $k$--SAT problems almost surely in time linear in the number  of  variables. 
The ChainSAT algorithm, given  in pseudo-code in Algorithm \ref{ChainSAT}, $(i)$ never increases the energy of the current configuration $S$, and $(ii)$ exercises circumspection in decreasing the energy.
The ChainSAT algorithm has two adjustable parameters: $p_1$ for controlling the rate of descent (by accepting energy-lowering flips), and $p_2$ for limiting the length of the chains, in order to avoid looping. In our experiments, we set $p_1 = p_2$ \cite{alava2008circumspect}.
%

%
We present two new versions of the ChainSAT algorithm.
In these new versions we replace the random choice of clauses (see lines $5$ and $18$ of  Algorithm \ref{ChainSAT})  with an ordered one. Since ChainSAT is based on a non-increasing energy principle \cite{alava2008circumspect}, and given the energy levels provided by the S2G algorithm, our idea is to select clauses with minimum energy even when ChainSAT performs a random selection.

Let us introduce  a set $A = \{ a_1, a_2,...,  a_m \}$, where $m$ is the number of  clauses of the $k$--SAT formula.
The set $A$  is used  to record which clauses of the $k$--SAT instance have already been  chosen,
so that  loops (consisting of choosing  the same clause repeatedly) are avoided.
Let $H =\{C_1$, $C_2$, ..., $C_r\}$ be the set of  clauses  among which the algorithm
 chooses (line $5$ or $18$ of Algorithm \ref{ChainSAT}).
We suppose that these clauses are arranged in a decreasing order  of weight. 
In particular, $C_1$ is one of the heaviest clauses in $H$, i.e., 
$C_1$ is one of the clauses of $H$ with the lowest energy.
The steps for selecting a clause in the line $5$ or $18$ of Algorithm \ref{ChainSAT} are
the following. Initially we set $a_i  =  0$,\ $\forall$ $i=1,..., m$. At each step we require that the algorithm
chooses the heaviest clause $C_i$ among those in $H$ such that $a_i = 0$. Every time a clause $C_i$ is chosen, we set $a_i = 1$.
Step by step, the number of elements in $A$ equal to $1$ increases. When
$a_{i} = 1,\ \forall$ $C_{i} $ $\in$ $H$, then the clause is chosen randomly.
This random choice is necessary to prevent that our algorithm always analyzes the same chains clause-variable-clause.

Our modified version of ChainSAT presented so far, selects the new clause using the same set of elements $a_i  =  0$,\ $\forall$ $i=1,...,m$, both for 
the satisfied and for the unsatisfied clauses.
We call this version LC-ChainSAT, where LC stands for ``Linked Clauses'', since the choice of a clause in lines $5$ and $18$ of Algorithm \ref{ChainSAT} is based on the same set $A$.

We also present a second new version of the ChainSAT algorithm, called NLC-ChainSAT, where NLC stands for ``Non Linked Clauses''.
This new version differs from the first one because we replace $A$ with two sets $A_{sat}$ and $A_{unsat}$, with the same structure of $A$. We use $A_{unsat}$ to store the clauses chosen (as not satisfied) by line $5$ of Algorithm \ref{ChainSAT}, and $A_{sat}$ to store the ones chosen (as satisfied) by line $18$. The new algorithm runs exactly like the previous one but when
it must select a new clause, it examines the set $A_{unsat}$ or $A_{sat}$ depending on
whether the new clause is chosen by line $5$ or by line $18$ respectively.

\section{Experimental Results}
\label{sec:results}

In this section we investigate the outcomes of our algorithms. First, we give numerical evidence of the presence of Bose-Einstein condensation in the $k$--SAT problem, focusing on the phase transition region. We evaluate the phase diagram of the S2G algorithm to show the transition between a fit-get-rich phase and a winner-takes-all phase. Second, we analyze the SAT solvers proposed above by evaluating their performance on both random and real-life SAT instances.

\subsection{S2G Results}
For the $3$--SAT problem there is strong evidence \cite{nature:99} that the phase transition between solvable and unsatisfiable instances is located at $\alpha = 4.256$, where $\alpha=\frac{m}{n}$ is the ratio between the number of clauses $m$ and the number of variables $n$.
For our experiments we use the A.\ van Gelder's $k$--SAT instance generator {\sc mkcnf.c}\footnote{Available at\\ ftp://dimacs.rutgers.edu/pub/challenge/satisfiability/contributed/UCSC/instances. \\ The program {\sc mkcnf.c} takes four  inputs: $r$, the random seed;  $k$, the number of literals in each clause; $n$, the number of Boolean variables; $m$, the number of clauses.}. We asked the program to generate uniformly satisfiable and unsatisfiable formulas to obtain a purely uniform random $k$--SAT distribution.
For our experimental protocol, we consider $\alpha\in ]0,10]$ and $n\in\{10, 25, 50, 75, 100\}$. For each value of $\alpha$, we consider $100$ formulas and perform $30$ independent graph $G$ constructions per formula. We make use of the S2G-PA algorithm by imposing $\theta = 0.33$ and $\rho=1$.

\begin{figure}[h]
\centering
\includegraphics[scale=0.95,angle=0]{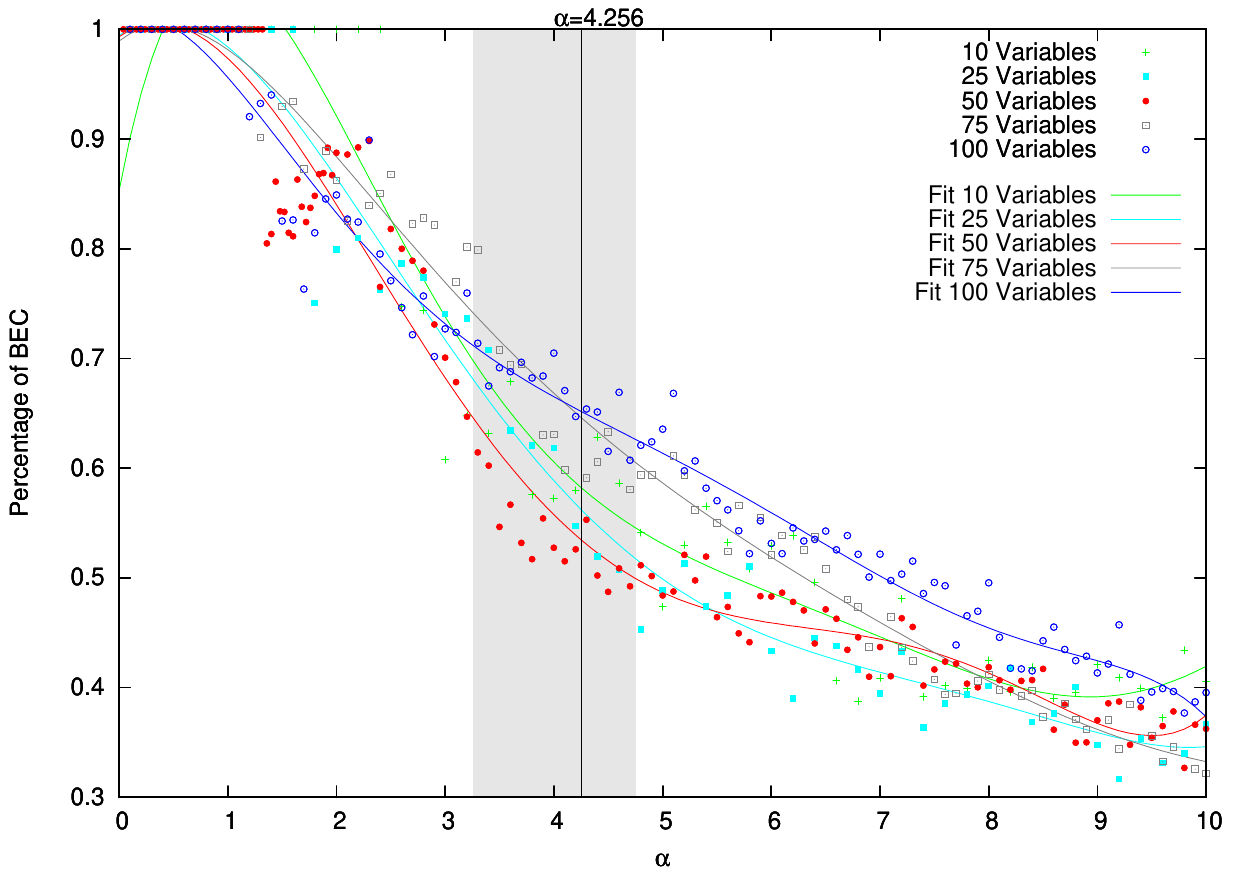}
\caption{Bose-Einstein condensation (BEC) in $3$--SAT. We report on the $x$-axis the ratio $\alpha$ of clauses to variables, and on the $y$-axis the percentage of BEC networks found. The points have been fitted through a sixth-order polynomial regression. The gray stripe shows the region where the critical temperature $T_{BEC}$ for Bose-Einstein condensation could be located.}
\label{fig:bec2final}
\end{figure}

In Figure \ref{fig:bec2final} we plot the percentage of BEC networks observed by varying $\alpha$. In general, when $\alpha < 3$ the resulting graph most likely undergoes a clear Bose-Einstein condensation; in this phase, the fittest clause maintains a large number of links even if the graph expands. Moreover, if $\alpha$ increases and enters the phase transition region, it is evident that the drop in the number of Bose-Einstein condensations becomes smoother. This different behavior seems to match with the increasing complexity of formulas with $\alpha \approx 4.256$ (the locus of the phase transition for $3$--SAT), and therefore we investigate it more thoroughly later. For $\alpha > 5$, the graph shows a fit-get-rich behavior, i.e., there is an increasing number of fittest nodes (clauses), but there is no more a unique winner node. This behavior remarks that, for increasing $\alpha$ values (close to the UNSAT phase of $3$--SAT), we have to find a truth assignment for many clauses to obtain a satisfiable formula. These empirical evidences are consistent with the transition between SAT and UNSAT instances \cite{science:02}.

\begin{figure}[h]
\centering
\includegraphics[scale=0.95]{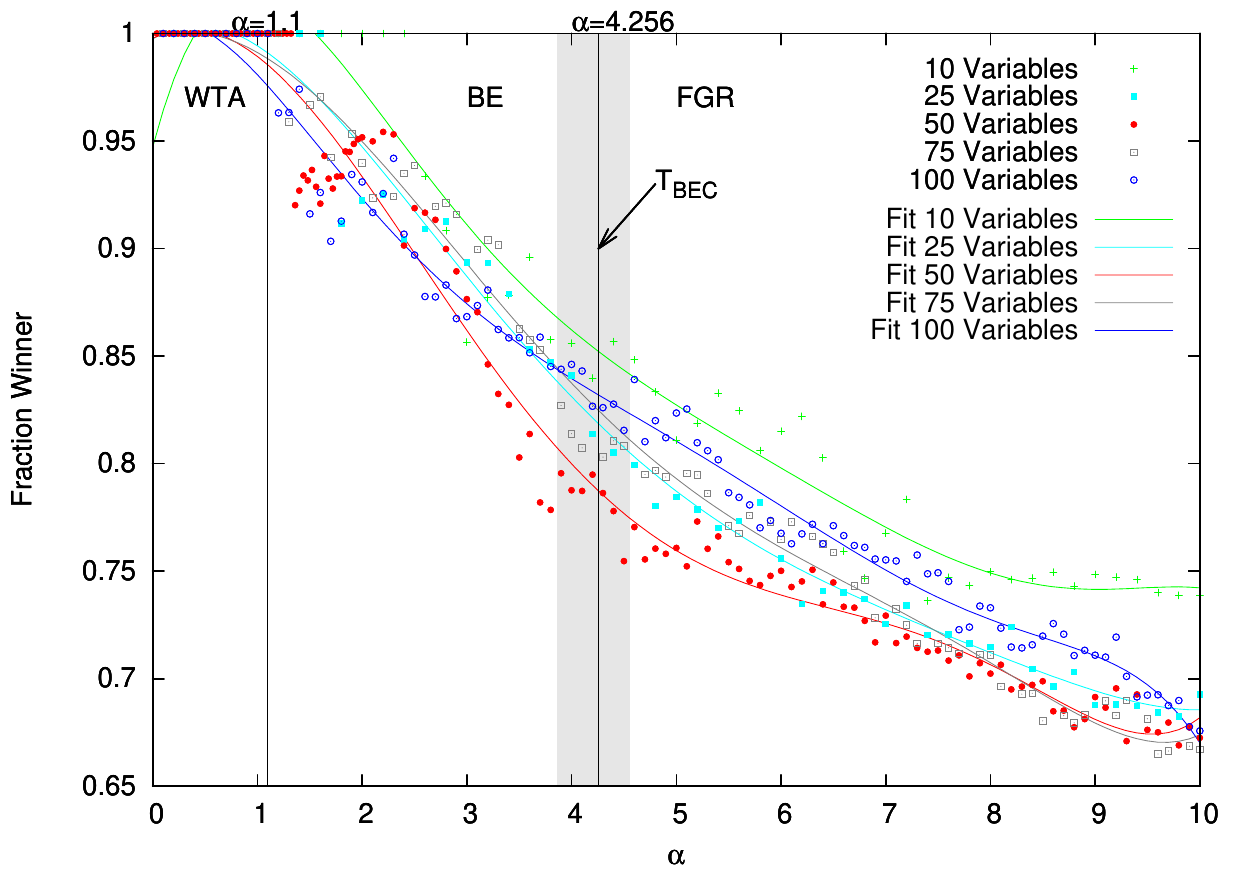}
\caption{Phase diagram of $3$--SAT. We report the fraction of links shared by the winner against $\alpha$ (the ratio of clauses to variables). Each point is an average over $1000$ $3$--SAT instances with $30$ graphs per instance. We have performed a sixth-order polynomial regression to fit the data. Satisfiable instances (with high probability) belong to the winner-takes-all phase. Unsatisfiable instances (with high probability) belong to the fit-get-rich phase. The critical temperature $T_{BEC}$ for Bose-Einstein condensation could be located in the grey area in the neighborhood of the SAT-UNSAT phase transition $\alpha=4.256$. Below the critical temperature, the fraction winner increases at a higher rate.}
\label{fig:final_graph}
\end{figure}

In order to evaluate the way in which $\alpha$ (i.e., the ratio clauses to variables) influences the evolution of the graph associated with $k$--SAT instances, we examine the {\it fraction winner} $f$ defined as the ratio of the number of links shared by the winner (i.e., the highest degree node) to the number of links of the whole graph. Figure \ref{fig:final_graph} shows how the fraction winner varies as function of the ratio of clauses to variables. 
We let the number of variables vary in the set $ \{10,25,50,75,100\}$. Each point of the plot has been computed by averaging over $1000$ different $3$--SAT instances, with $30$ graph generations per instance. 
The plot shows that the fraction-winner $f$ decreases with $\alpha$. When a $3$--SAT instance is satisfiable (with high probability), the S2G-PA algorithm produces a graph condensed over the fittest clause. Conversely, when a $3$--SAT instance is unsatisfiable (with high probability), its graph exhibits a winner node incapable of maintaining the whole connectivity of the network, and some hubs appear and grow following the fit-get-rich model.
By looking at the plots in Figure \ref{fig:final_graph} from right to left, one can observe that when $\alpha$ becomes smaller than the critical value $4.256$, the winner node holds the vast majority of links. In this case, the Bose-Einstein condensation takes place regardless of the number of variables. Moreover, the plot concerning the case of $50$ variables clearly shows a smooth drop for $4 < \alpha < 5$, indicating that the $50$-variables graphs undergo the slowest Bose-Einstein condensation (provided that $\alpha$ decreases).
It is possible to note that for $\alpha<1.1$ the fraction winner is equal to 1, since the winner node holds all the links in the network, i.e., all the edges have the winner node as a vertex. Our results suggest that this phase, called {\it winner-takes-all} (WTA), starts at $\alpha=1$ for large values of $n$.

\begin{figure}
\centering
\newcolumntype{S}{>{\centering\arraybackslash} m{0.45\linewidth} }
\begin{tabular}{S S} 
\includegraphics[scale=0.57]{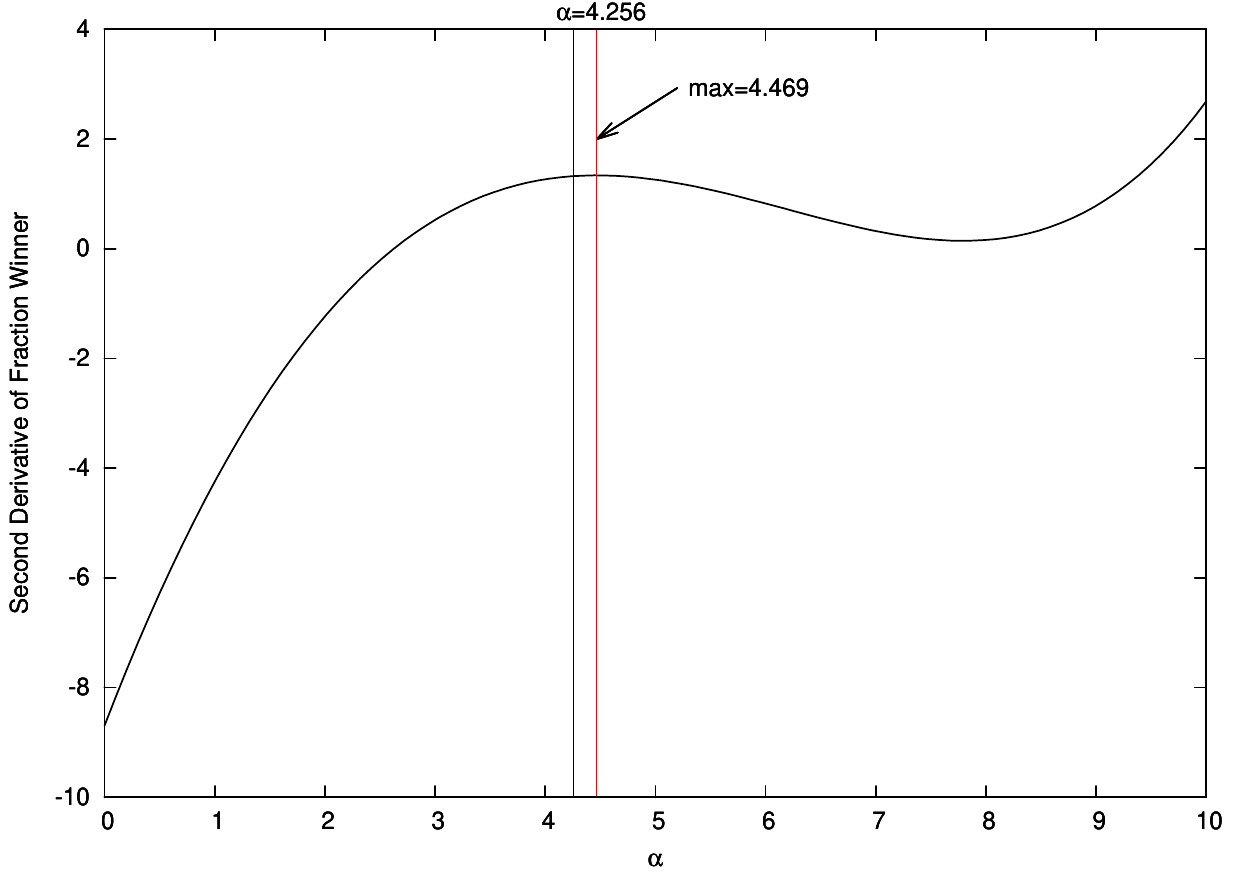} & \includegraphics[scale=0.57]{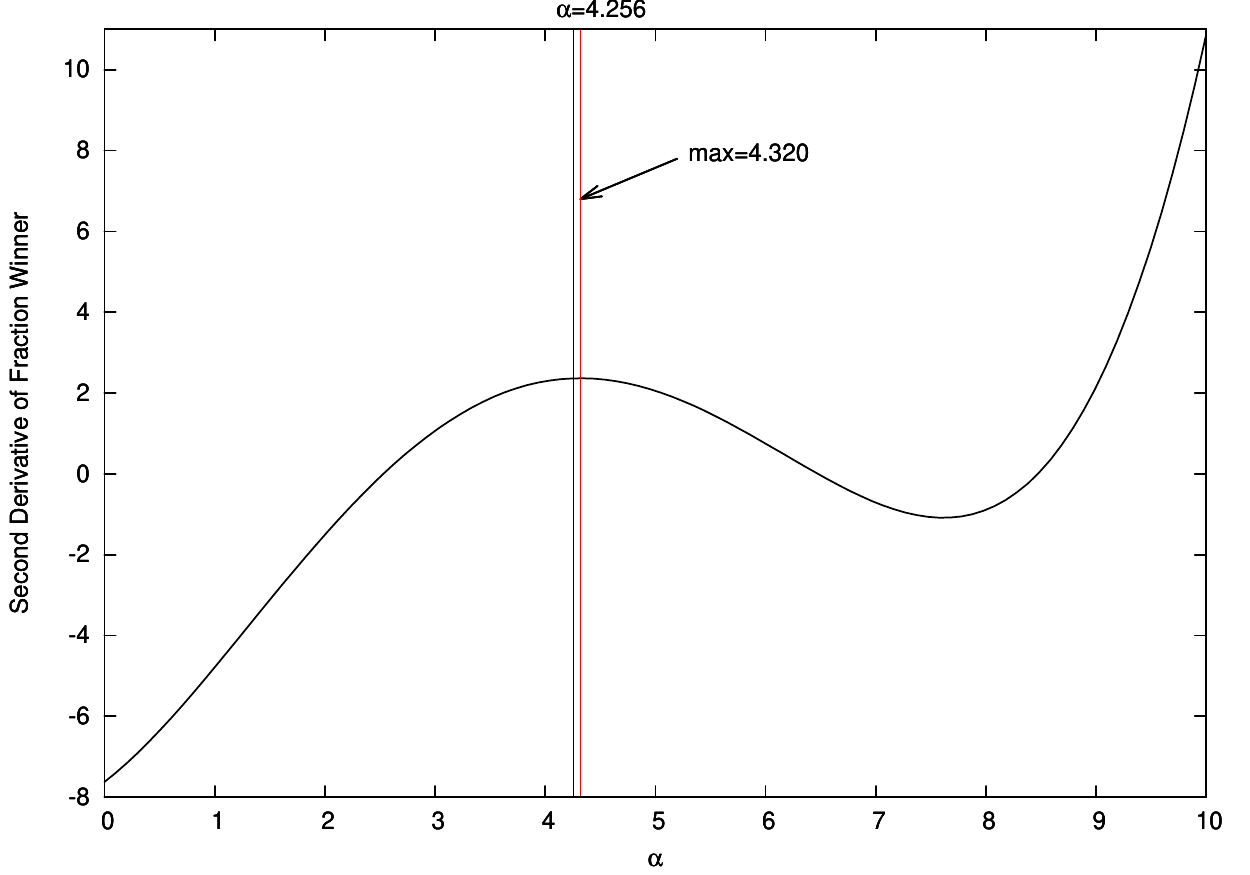} \\
($n=25$) & ($n=50$) \\
\includegraphics[scale=0.57]{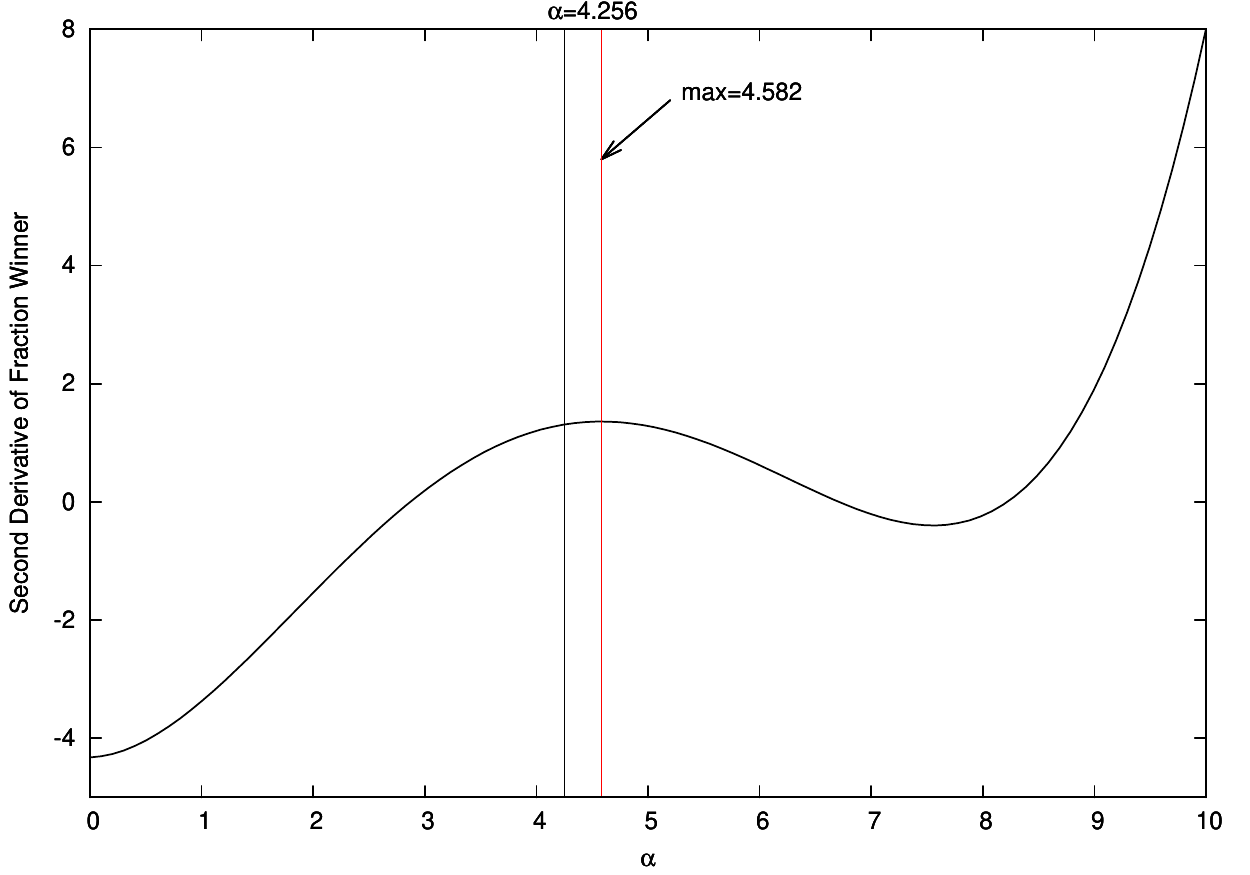} & \includegraphics[scale=0.57]{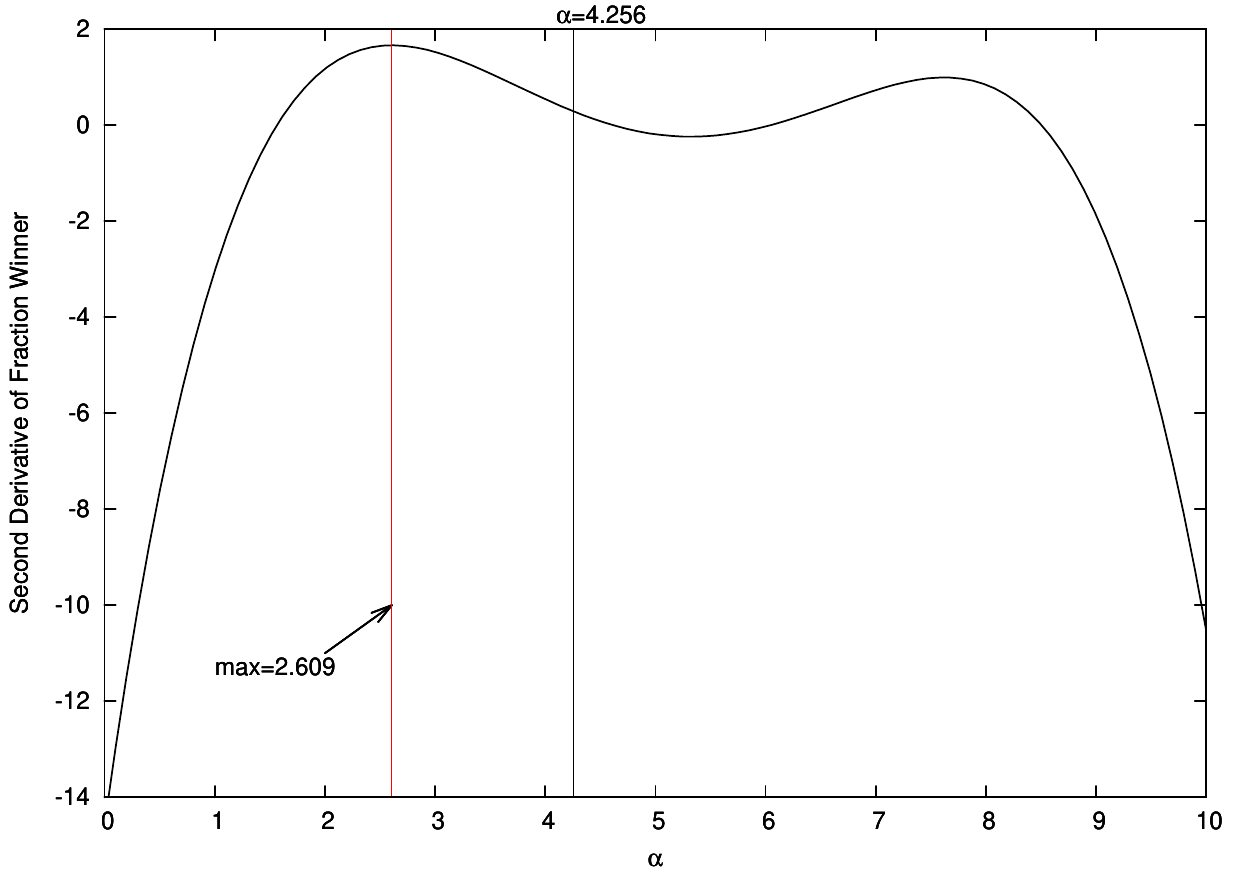} \\
($n=75$) & ($n=100$) \\
\end{tabular}
\caption{$3$--SAT Bose-Einstein condensation locus. We plot the second derivative of the fraction winner as function of the ratio $\alpha$ of clauses to variables. The SAT-UNSAT phase transition $\alpha=4.256$ is near the local maximum of the second derivative, and therefore corresponds to a quick change of the fraction-winner slope. In other words, the $3$--SAT phase transition acts as the BEC critical temperature.}
\label{fig:second_derivative}
\end{figure}

One can notice that below the phase transition region the slopes of the plots exhibit a different behavior than in the other regions. Specifically, below the phase transition of $3$--SAT, located at $\alpha=4.256$, the fraction winner increases at a higher rate. 
In order to evaluate the slopes, in Figure \ref{fig:second_derivative} we plot the second derivative of the polynomial curves of Figure \ref{fig:final_graph} as function of $\alpha$. A high value of the second derivative indicates a rapid change of the fraction-winner slope. For $25$, $50$, and $75$ variables, the $3$--SAT phase transition found by M{\'e}zard et al.\ \cite{science:02} approximates the local maximum of the fraction-winner second derivative. This local maximum represents the value of $\alpha$ corresponding to the most rapid change in the fraction-winner slope in the neighborhood of $4.256$. 
Therefore, the phase transition of $3$--SAT seems to be the critical temperature for Bose-Einstein condensation. These outcomes confirm the experimental findings above-mentioned, and are consistent with those referring to the plots in Figure \ref{fig:bec2final}.
The behavior of the plot of $100$ variables, slightly different from the others, is due to the unexpected values of the fraction winner obtained as $\alpha$ approaches $10$, which cause the sixth-order curve to exhibit a high curvature in the neighborhoods of $\alpha=2.6$ and $\alpha=8$.

\begin{figure}[h]
\centering
\includegraphics[scale=0.95]{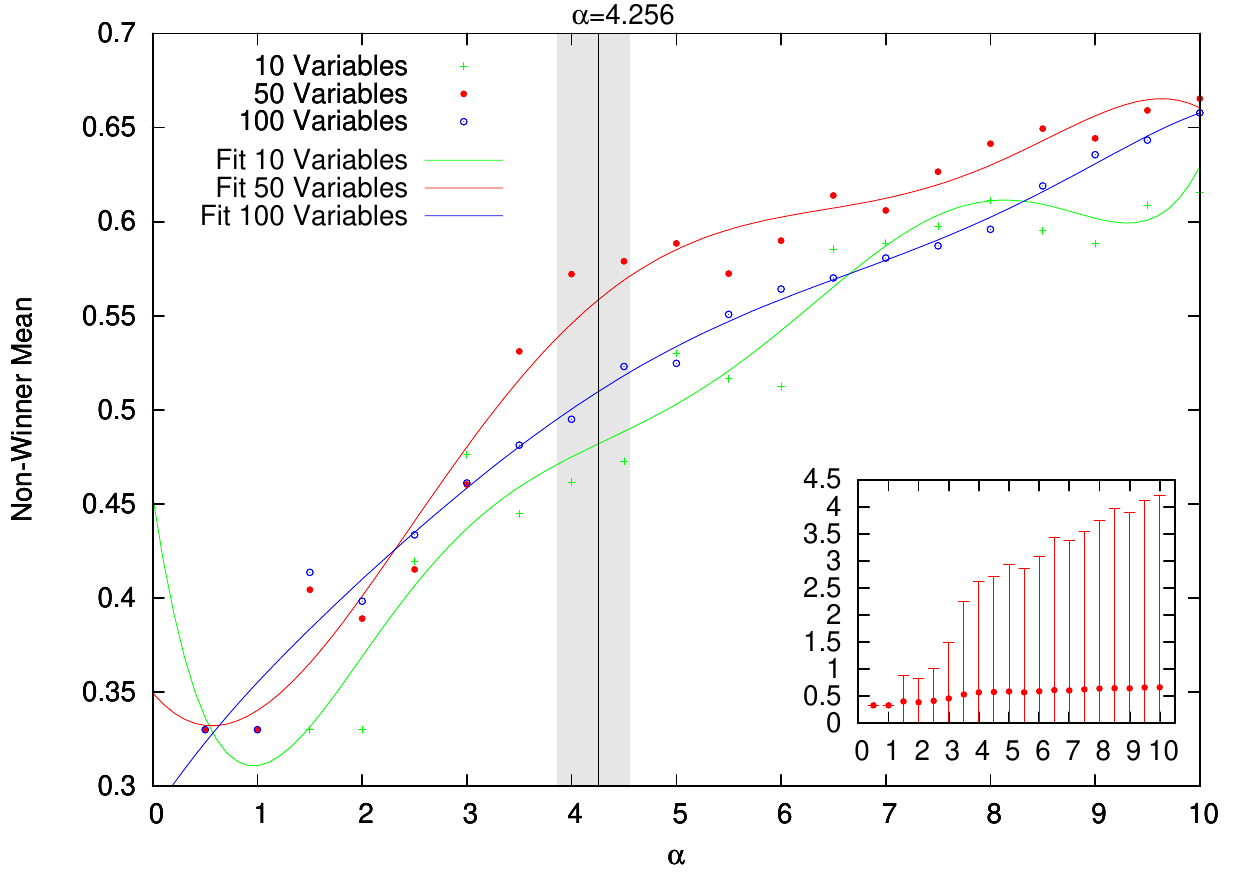}
\caption{Average non-winner connectivity in $3$--SAT networks. We report the mean of the connectivity of all the nodes except the winner, as function of the clauses-to-variables ratio. Each point is an average over $100$ $3$--SAT instances with $30$ graphs per instance. We have performed a sixth-order polynomial regression to fit the data. In the inset, we report the standard deviation of the mean connectivity for $50$-variables instances. Satisfiable instances (with high probability) are translated into condensed graphs, as all the connectivities are equal to $\theta$ and the standard deviation is zero. Conversely, unsatisfiable instances (with high probability) are translated into fit-get-rich networks with high standard deviation, thus with emerging hubs. In agreement with the fraction winner, to the left of the condensation area of Figure \ref{fig:final_graph} both the mean and the standard deviation exhibit a higher slope.}
\label{fig:mean}
\end{figure}

In Figure \ref{fig:mean} we plot the mean of the connectivity of the network nodes, computed on the degree distribution without considering the winner node. As previously discussed, for increasing $\alpha$ the winner node decreases its connectivity, therefore the other nodes acquire links. In the inset, we plot the standard deviation of the $50$-variables degree distribution (the plots concerning $10$ and $100$ variables are shown in Figure \ref{fig:typical} in Supporting Information \ref{appendixstd}). High standard deviation indicates that the connectivity values are scattered, hence the network exhibits highly connected hubs. More precisely, instances with high constraint density $\alpha$ not only have the winner node less rich than low constraint density instances, but also show higher spreading in the non-winner node connectivity. In other words, the number of hubs is positively correlated with $\alpha$, and this result is consistent with the plot in Figure \ref{fig:bec2final}, which shows that the number of condensed network decreases as $\alpha$ increases. Remarkably, the rate at which the standard deviation increases is higher to the left of the Bose-Einstein condensation region. The growing hubs of typical S2G-PA output networks are shown in Figure \ref{fig:typicalS2GPA} in Supporting Information \ref{appendixtyp}.

\subsection{LC-ChainSAT and NLC-ChainSAT Results}
We evaluate each algorithm on a collection of $6885$ $k$--SAT instances obtained from publicly available sources.
This benchmark consists of (i) $40$ Intel sequential circuits and $95$ l2s benchmarks used in the 2007 and 2008 hardware model checking
competition \cite{hwmcc},
(ii) $2250$ random instances for each  value of $k$ ($k=3$, $k=4$, and $k=5$), generated uniformly at random using the A.\ van Gelder's generator. We use {\sc aigtocnf} \cite{aig2cnf} to convert instances from AIG format to CNF.
Then, we convert them into $3$-CNF instances.
We set $n \in \{25, 50, 75, 100, 125\}$ and $m$ such that $\alpha=\frac{m}{n}$ $\in [\alpha_{sat}(k)-4 ; \alpha_{sat}(k) + 2]$, where
$\alpha_{sat}(k)$ has the estimated values  $\alpha_{sat}(3) = 4.256$, $\alpha_{sat}(4) = 9.931$, $\alpha_{sat}(5) = 21.117$ (see \cite{mertens2006threshold}).
For each value of $\alpha$, we generate $30$ different $k$--SAT instances.
We also introduce the following stop criterion \cite{nature:99}: we stop the algorithm either when a solution is found or when $10^6$ cycles of the main body of the  algorithm (i.e., $10^6$ formula evaluations) have been carried out.

The comparison between ChainSAT and our two modified versions is based on the following definition.
Let $Z_1$ and $Z_2$  be two algorithms tested on the same set of instances. We say that  \emph{$Z_1$ performs better than $Z_2$} if one of the following conditions
is met: $(i)$ $Z_1$ satisfies more instances than $Z_2$; $(ii)$ both $Z_1$ and $Z_2$ satisfy the same number of instances, but the average number of clauses satisfied by $Z_1$ is greater than the average number of clauses satisfied by $Z_2$; $(iii)$  both $Z_1$ and $Z_2$ satisfy the same number of instances with the same average number of clauses satisfied, but $Z_1$ performs less flips than $Z_2$.
The parameters of ChainSAT have been chosen to be small enough to work at least up to the ``dynamical transition'' \cite{krzakala2007gibbs}: we have set $p_1=p_2=0.005$ ($k=3$), $0.0001$ ($k=4$), and $0.0002$ ($k=5$) \cite{alava2008circumspect}.

\begin{figure}
\centering
\begin{tabular}{c} 
 \includegraphics[scale=0.7]{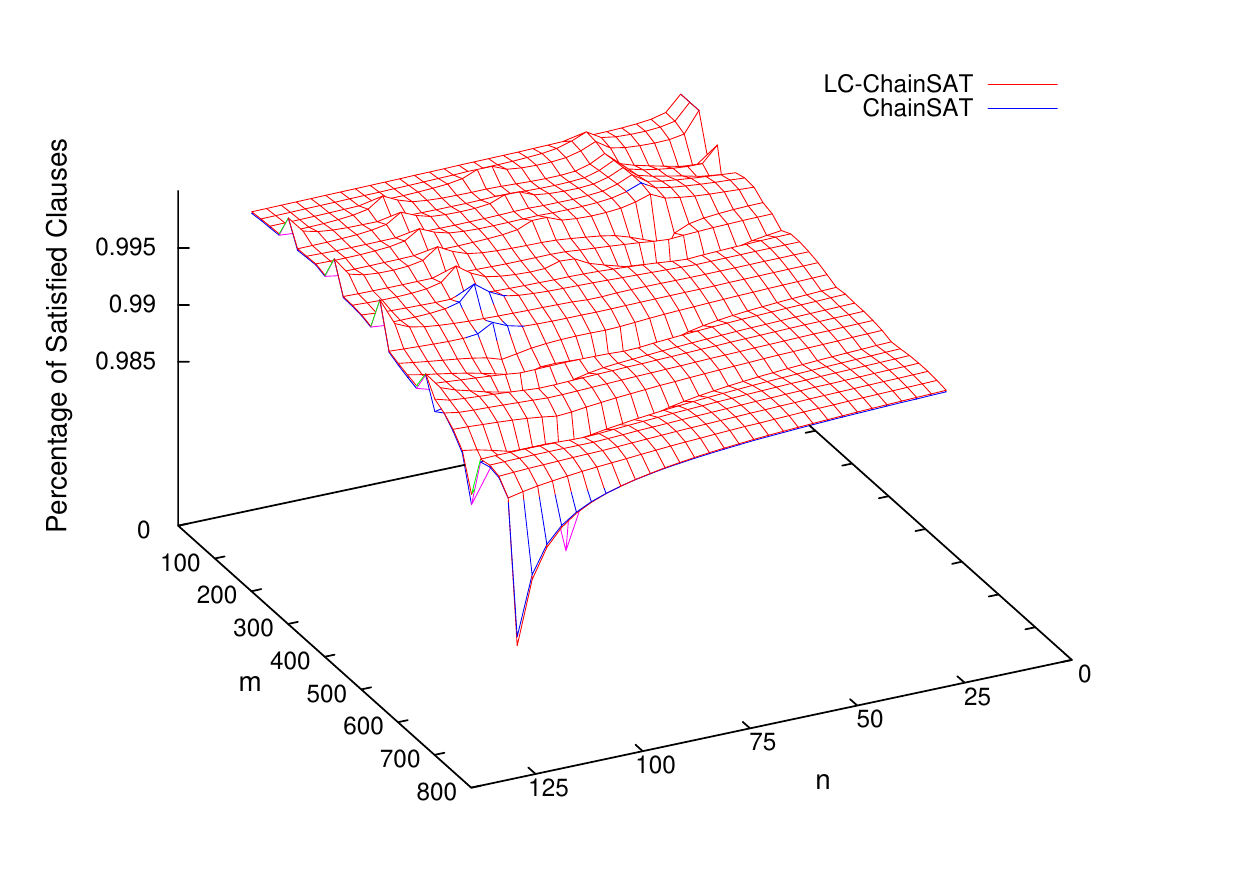}   \\
 (a) $3$--SAT  \\ 
 \includegraphics[scale=0.7]{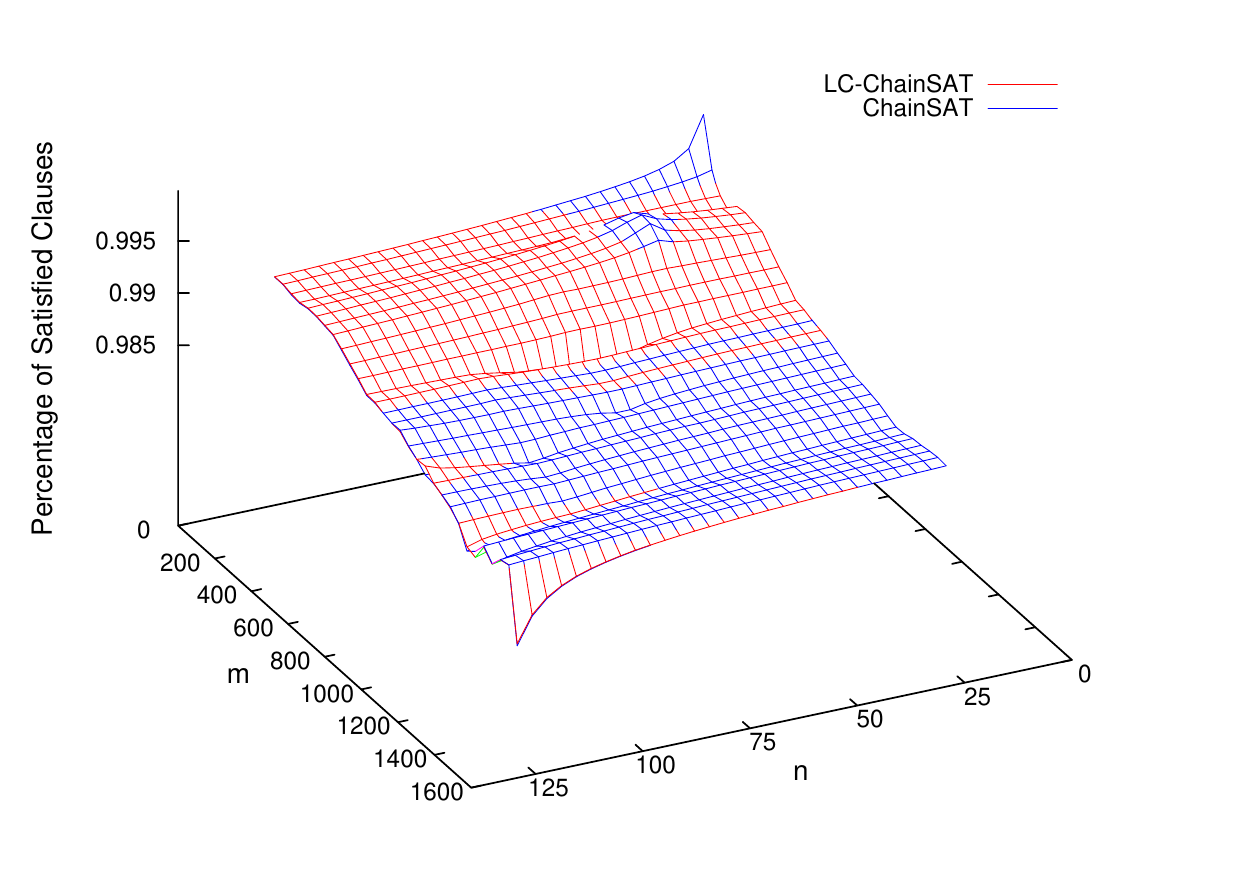} \\
  (b)  $4$--SAT\\
  \includegraphics[scale=0.7]{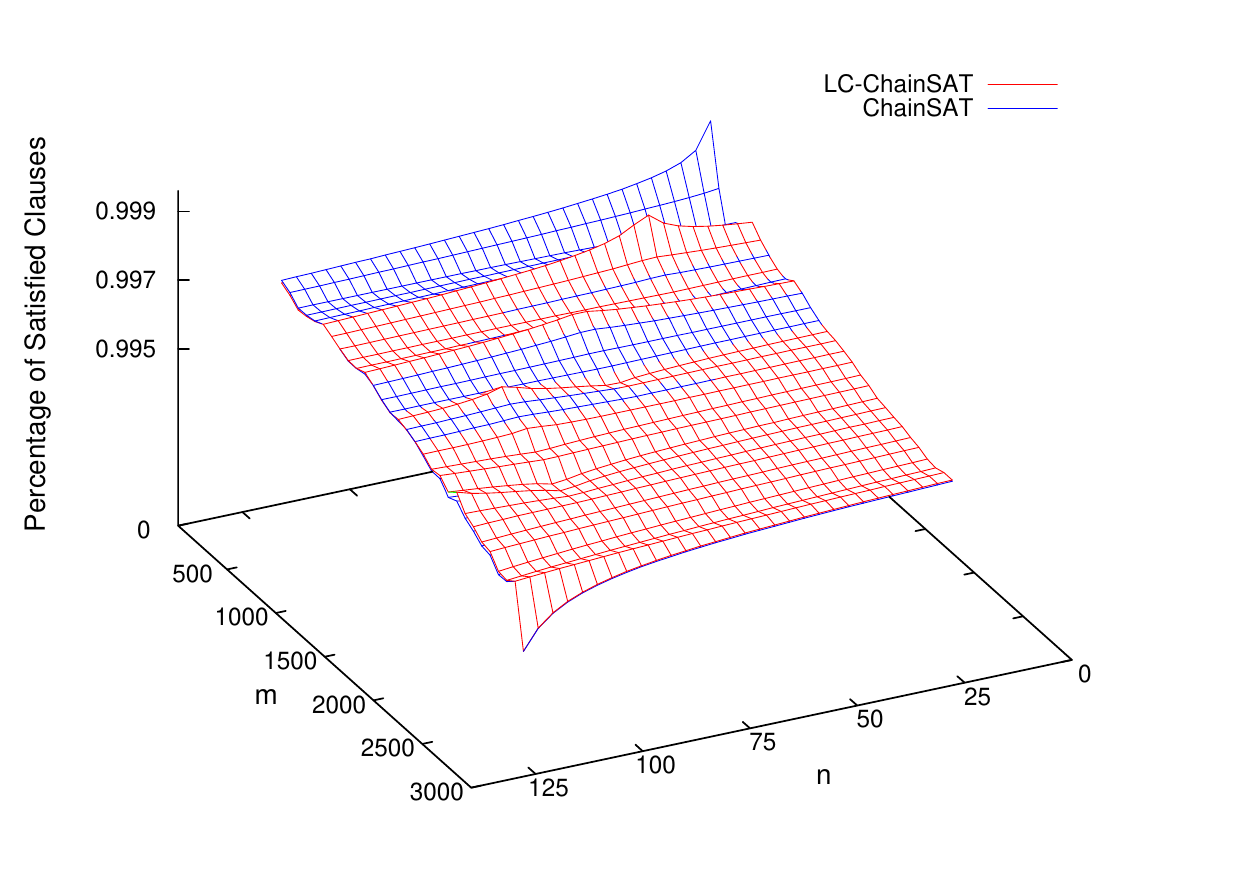} \\
  (c) $5$--SAT \\
\end{tabular}
\caption{MaxSAT  for $k=3$, $k=4$, and $k=5$. These plots show the percentage of clauses satisfied by LC-ChainSAT and ChainSAT as function of the number of clauses $m$ and variables $n$. Remarkably, when solving $3$--SAT instances, LC-ChainSAT clearly outperforms ChainSAT.}
\label{MaxSAT}
\end{figure}

\begin{figure}[!t]
\centering
\begin{tabular}{c}
\includegraphics[scale=0.68]{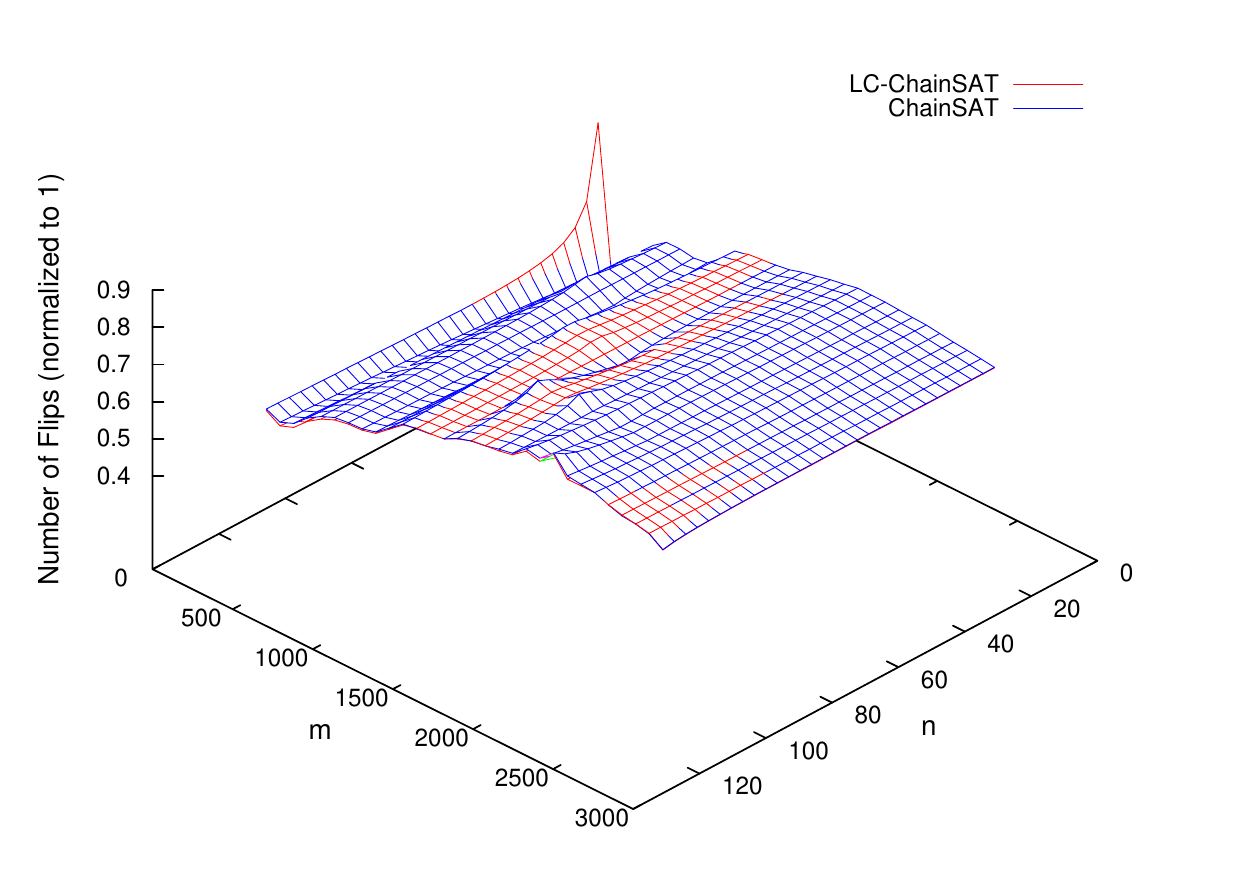} \\
\end{tabular}
\caption{Computational effort for $k = 5$.
We plot the number of flips (normalized to $1$) performed by the two algorithms. LC-ChainSAT improves ChainSAT employing almost the same numbers of flips, therefore requiring the same computational effort.  
}
\label{fig:MaxFlipsK5}
\end{figure}

The analysis of LC-ChainSAT and NLC-ChainSAT shows an improvement in the performance of $3$-SAT, $4$-SAT, and $5$--SAT solvers.
In particular,  LC-ChainSAT performs better than ChainSAT  in $56.8\%,\ 57.3\%$ and  $60\%$ of the benchmarks using $3$--SAT, $4$-SAT, and $5$--SAT instances respectively.
Likewise, NLC-ChainSAT performs better than ChainSAT in $58.3\%$, $58.7\%$, and  $54.1\%$ of the benchmarks respectively.
A more detailed analysis of the data is shown in Table \ref{tab:summary}. For each algorithm we report: (i) the number of instances satisfied; (ii) the average number of clauses satisfied in the whole set of instances (see the MaxSAT comparison in Figure \ref{MaxSAT}); (iii) the number of flips obtained  running the algorithm on the whole set of instances (see Figure \ref{fig:MaxFlipsK5}). The MaxSAT problem consists of determining a truth assignment that maximizes the number of clauses satisfied \cite{escoffier2007differential}. In order to confirm our results, in Table \ref{tab:summary2} we compare LC-ChainSAT and ChainSAT on further $171$ instances \cite{hwmcc}.
\begin{table}[!t]
	\centering
	\setlength{\tabcolsep}{10pt} 
		\begin{tabular}{clccc}
		\hline
			&Solver & Solved & MaxSAT & Flips \\
			\hline
		&	ChainSAT& 2117 & 38633.57 & 178793 \\
$k=3$		&	LC-ChainSAT& \textbf{2129} &  \textbf{38646.77} & 179421\\
		&	NLC-ChainSAT& \textbf{2132} &\textbf{38647.47} & \textbf{178543} \\
			\hline
				&	ChainSAT&  2089 & 84043.90 & 11379888 \\
$k=4$		&	LC-ChainSAT& \textbf{2103} &\textbf{84054.10} &11389576  \\
		&	NLC-ChainSAT& \textbf{2104} &  \textbf{84053.92} &  11383184\\
			\hline
					&	ChainSAT& 2047& 166720.88 & 16210343  \\
$k=5$		&	LC-ChainSAT& \textbf{2057} & \textbf{166726.64} & \textbf{16206307} \\
		&	NLC-ChainSAT& \textbf{2055} & \textbf{166725.50} & 16254044 \\
			\hline
		\end{tabular}
\vspace{0.1cm}
\caption{Summary of SAT solvers performance. Both LC-ChainSAT and NLC-ChainSAT outperform ChainSAT in terms of clauses satisfied by the algorithm. For $k=4$, although ChainSAT performs better than our modified versions in terms of number of flips carried out, it does not maximize the number of satisfied clauses.}
\label{tab:summary}
\end{table}
We obtain another confirmation of our results if we run the algorithms with stop criterion set as $10^4$ formula evaluations.
In this case, the number of satisfied clauses is almost equal to zero for all $\alpha$ values, due to the descent circumspect
that characterizes the ChainSAT algorithm.
Thus, by comparing the percentage of the clauses satisfied,  both of our modified algorithms
are able to satisfy more clauses  than the ChainSAT, though all algorithms evaluate each instance the same number
of times ($10^4$ times at most).

Even if we are not yet able to establish which of the two versions is the best,
our results point out that ordering clauses with the information provided by the S2G algorithm results in an improvement of the ChainSAT performance. 

\section{Discussion}
\label{sec:discussion}

Our work, using a mapping between the $k$--SAT problem and the Bose gas, shows numerical evidence for Bose-Einstein condensation in a network model for $k$--SAT.
Analogously to complex networks \cite{bianconi2001}, the graphs of $k$--SAT instances follow Bose statistics and can undergo Bose-Einstein condensation.
It is evident that the total number of links shared by the most connected node (also called {\it winner} node) varies as function of the ratio $\alpha$ of clauses to variables. In particular, the fraction winner, plotted as function of $\alpha$, indicates the difference between the Bose-Einstein phase and the fit-get-rich phase. When $\alpha<1.1$, the winner node shares all the edges of the graph (winner-takes-all phase). For low $\alpha$ values, the fittest node maintains a finite fraction of links even though the number of variables increases (Bose-Einstein phase); for high $\alpha$ values, the fraction of links connected to the winner decreases with increasing $\alpha$ (fit-get-rich phase). Moreover, the mean and the standard deviation of the non-winner degree distribution increase with increasing $\alpha$, as growing hubs appear in the graph.

It is well known that the phase transition of $3$--SAT occurs when the ratio $\alpha$ of clauses to variables belongs to a neighborhood of $4.256$. In our work we experimentally proved that the critical temperature for Bose-Einstein condensation in a $k$--SAT graph also belongs to the same neighborhood. This fact allows us to draw an important conclusion: there is a strict correspondence between the phase transition of $k$--SAT and the critical temperature for Bose-Einstein condensation. 
To our knowledge, this is the first time that complex networks and Bose-Einstein condensation are related to the $k$--SAT problem without a priori examination of its truth assignments. 

We have also presented a hybrid SAT solver that combines  the ChainSAT algorithm and the information provided by the S2G algorithm.
Our approach is based on the analysis of the energy level related to each clause.
We demonstrate that by ordering clauses according to their energy we outperform one of the best SAT solvers (ChainSAT, see results \cite{alava2008circumspect}) on  the majority of the benchmarks.
This means that we enhance an algorithm that is able to solve $k$--SAT problems almost surely in time linear in the number of variables. Hence, our algorithms could also be a good tool from an application point of view, e.g., checking satisfiability of formulas in hardware and software verification.
 

\bibliographystyle{unsrt}
\bibliography{ksat-biblio}


\newpage

\renewcommand{\thetable}{S\arabic{table}}
\renewcommand{\thefigure}{S\arabic{figure}}
\renewcommand{\thealgorithm}{S\arabic{algorithm}}
\setcounter{table}{0}
\setcounter{algorithm}{0}
\setcounter{figure}{0} 

\appendix 
\label{sec:appendix}
\noindent {\large \bf Supporting Information}

\section{}
\label{appendixthm}

\begin{theo} \nonumber$d$ is a metric.\end{theo}
\noindent {\bf Proof.} \
Let $X=(X_1,...,X_k)$, $Y=(Y_1,...,Y_k)$, $Z=(Z_1,...,Z_k)$ be three clauses, each made up of $k$ literals.
In this proof, we assume that literal permutations are always possible in every clause, so we do not make any assumption on the order of the literals in a clause. Indeed, this hypothesis comes from the fact that, in $k$--SAT instances, the OR operator is commutative. Let us verify the three conditions required for a metric.
\begin{enumerate}
\item {\it Definite positiveness}.\\
			$d(X,Y) \geq 0, \: \forall \, X,Y;\\ $
			$d(X,Y)=0 \Longleftrightarrow 
			\left| \, \left\{\mu \in \{1,...,k\} : (X_\mu) \neq (Y_\mu) \right\} \, \right| = 0 \Longleftrightarrow
			X_\mu=Y_\mu, \; \forall \mu=1,...,k \Longleftrightarrow
			X=Y$.\\
\item {\it Symmetry}.\\
			$d(X,Y) = \left| \, \left\{\mu \in \{1,...,k\} : X_\mu \neq Y_\mu \right\} \, \right| =
			\left| \, \left\{\mu \in \{1,...,k\} : Y_\mu \neq X_\mu \right\} \, \right| = d(Y,X)$.\\
\item {\it Triangle inequality}.\\
			Let $u(X,Y)$ be the function that gives back the number of literals that are in common between two clauses. We need to prove one of the following inequalities (which are equivalent to one another):\\
			$d(X,Y) \leq d(X,Z)+d(Z,Y) \Longleftrightarrow 
			k-u(X,Y) \leq k-u(X,Z) + k-u(Z,Y) \Longleftrightarrow 
			u(X,Y) \geq u(X,Z)+u(Z,Y)-k$. \\
			Let us define $U_{XY}$ as the number of places (ranging from $1$ to $k$) in which $X=Y$. It is clear that $ \left| \, U_{XY} \, \right| = u(X,Y)$ and $U_{XY} \supseteq U_{XZ} \cap U_{ZY}$ due to the transitivity of equality. Hence, if we use the dimension theorem for vector spaces, 
			it follows that:\\
			dim($U_{XY}$) $\geq$ dim($U_{XZ} \cap U_{ZY}$) $=$ dim($U_{XZ}$) $+$ dim($U_{ZY}$) $-$ dim($U_{XZ}+U_{ZY}$).\\
			But we have surely dim$(U_{XZ}+U_{ZY}) \leq k$. So after replacing the dimensions with the cardinalities of sets, we finally obtain:\\
			$u(X,Y) \geq u(X,Z)+u(Z,Y)-k$.
\hfill $\square$ 
\end{enumerate}

\newpage
\section{Supporting Algorithms and Tables}
\label{appendixalgo}

\begin{algorithm}
\caption{(Step I) -- Selecting\_the\_First\_Clause-Node}
\begin{algorithmic}[1]
\State $ \Lambda \gets V \gets E \gets F' \gets \emptyset $
\State $ i \gets 1 $
\State $ t_i \gets \mbox{random}(\{1,...,m\}) $
\State $ F' \gets F' \cup \{C_{t_i}\} $
\State $ \Lambda \gets \Lambda \cup \{L_\mu : L_\mu \in C_{t_i}\} $
\State $ V  \gets V \cup \{v(C_{t_i})\} $
\For {$L_\mu \in  \Lambda$}
	\State $ \varphi^{L}(L_\mu) \gets \mbox{occurrences of $L_\mu$ in $F'$}$  /* compute local frequency of $L_\mu$  */
\EndFor
\State $ f^{L}(C_{t_i})  \gets \displaystyle \sum_{\mu=1}^k \varphi^{L}(L_\mu), \quad L_\mu \in C_{t_i} $ \hskip0.4cm  /* compute local fitness of  $C_{t_i}$*/
\State $ f_r^{L}(C_{t_i}) \gets 1  $  \hskip3.5cm/* set initial normalized local fitness */
\State $ \epsilon_{t_i} \gets -T \cdot \log{ f_r^{L}(C_{t_i})}$ \hskip1.8cm /* set initial energy level */
\State $ t \gets t_i $  \hskip4.6cm /* index of the fittest clause */
\end{algorithmic}
\label{alg:step1}
\end{algorithm}

\begin{algorithm}
\caption{(Procedure I) -- Find\_Closest\_Clause}
\begin{algorithmic}[1]
	\State $t_i \gets t_i \in \{1,...,m\} \setminus \{t_1,t_2,...,t_{i-1} \}$ such that
	\Statex \hspace{0.5cm}  $ d(C_{t_i},C_{t}) = \mbox{min} \{ d( C_{t_i},C_{t} ) \; \mid \; t_i \in \{1,...,m\} \setminus \{t_1,t_2,...,t_{i-1}\} \} $
          \hskip2.8cm/* $t$ has been computed in the previous step */
	\If {$ \exists $ two or more clauses with minimum distance }
		\State $ t_i \gets $ random(one of the two or more clauses with minimum distance)
	\EndIf
	\Statex
	\State $ F' \gets F' \cup \{C_{t_i}\} $
	\State $ \Lambda \gets \Lambda \cup \{ L_\mu : L_\mu \in C_{t_i} \} $
	\State $  V \gets V \cup \{v(C_{t_i}) \} $
\end{algorithmic}
\label{alg:proc1}
\end{algorithm}

\begin{algorithm}
\caption{(Procedure II) -- Update\_Fitness}
\begin{algorithmic}[1]
\For {$L_\mu \in \Lambda$}
		\State $ \varphi^{L}(L_\mu) \gets \mbox{occurrences of $L_\mu$ in $F'$}  $ \hskip0.7cm /* update local frequency of $L_\mu$ */
	\EndFor
	
	\Statex
	
	\For {$j \gets 1$ to $i$}
	\State  $ f^{L}(C_{t_j}) \gets\displaystyle \sum_{\mu=1}^k \varphi^{L}(L_\mu), \qquad L_\mu \in C_{t_j} $ \hskip0.5cm /* update local fitness of $C_{t_j}$ */
	\EndFor
	\Statex
	
	\State $ t \gets t \in \{t_1,t_2,...,t_i \}$ such that	\hskip2.2cm/* index of the fittest clause */ \label{lineat}
\Statex \hspace{0.5cm} $ f^{L}(C_{t}) = \mbox{max} \{ f^{L}(C_{t_1}),f^{L}(C_{t_2}),...,f^{L}(C_{t_i}) \} $
	\Statex

	\For {$j \gets 1$ to $i$}
		\State $ f_r^{L}(C_{t_j}) \gets \displaystyle\frac {f^{L}(C_{t_j})} {f^{L}(C_{t})} $  \hskip0.6cm /*  normalize fitness */
		\State $ \epsilon_{t_j} \gets -T \cdot \log{ f_r^{L}(C_{t_j}) } $ \hskip0.4cm /* update energy level */
	\EndFor
\end{algorithmic}
\label{alg:proc2}
\end{algorithm}

\begin{algorithm}
\caption{(Step II) -- Connecting\_First\_two\_Clauses-Nodes}
\begin{algorithmic}[1]
\State $ i \gets i+1 $
\State Find\_Closest\_Clause()
\Statex
\State $ \Pi_{t_1} \gets 1 $ \hskip0.8cm/* Probability of linking the node $v(C_{t_2})$ to the node $v(C_{t_1})$  */
\State $ E \gets E \cup  \{( v(C_{t_1}), v(C_{t_2}) )\} $ \hskip0.9cm /* connect $v(C_{t_2})$ to $v(C_{t_1})$  */
\State $ k_{t_1} \gets$ degree($v(C_{t_1}$))  \hskip1.4cm/* update connectivity of node $v(C_{t_1})$ */
\State $ k_{t_2} \gets$ degree($v(C_{t_2}$))  \hskip1.4cm/* update connectivity of node $v(C_{t_2})$ */
\Statex
\State Update\_Fitness()	

\end{algorithmic}
\label{alg:step2}
\end{algorithm}

\begin{algorithm}
\caption{ChainSAT}
\begin{algorithmic}[1]
\algrenewcommand\algorithmicfor{\textbf{with}}

\State $S$ = random assignment of values to the variables
\State chaining = $False$
\While {$S$ is not a solution}
	\If {chaining = $False$}
		\State $C$ = a clause not satisfied by $S$ selected uniformly at random
		\State $V$ = a variable in $C$ selected u.a.r.
	\EndIf
	\State $\Delta E$ = change in the number of unsatisfied clauses if $V$ is flipped in $S$
	\State chaining = $False$
	\If {$\Delta E = 0$}
		\State flip $V$ in $S$
	\ElsIf{$\Delta E < 0 $}
			\For{probability $p_1$}
				\State flip $V$ in $S$
			\EndFor
	\Else
		\For{probability $1-p_2$}
			\State $C$ = a clause satisfied only by $V$ selected u.a.r. 
			\State $V'$ = a variable in $C$ other than $V$ selected u.a.r.
			\State $V = V'$
			\State chaining = $True$
		\EndFor
	\EndIf
\EndWhile
\end{algorithmic}
\label{ChainSAT}
\end{algorithm}

\begin{table}
	\centering
	\setlength{\tabcolsep}{10pt} 
		\begin{tabular}{lccc}
		\hline
			Solver & Solved & MaxSAT & Flips \\
			\hline
			ChainSAT&  163   &   3213.27  &  177820 \\
	 	LC-ChainSAT&  163  &    \textbf{3220.32}  &  179233\\
      \hline
		\end{tabular}
\vspace{0.1cm}
\caption{Further comparison between LC-ChainSAT and ChainSAT on $171$ $3$-SAT instances. Both algorithms solve the same number of instances, but once again LC-ChainSAT satisfies more clauses than ChainSAT.}
\label{tab:summary2}
\end{table}

\newpage
\section{S2G Examples}
\label{appendixS2G}

\begin{figure}
\centering
\newcolumntype{S}{>{\centering\arraybackslash} m{0.45\linewidth} }
\begin{tabular}{S S} 
\includegraphics[scale=0.29]{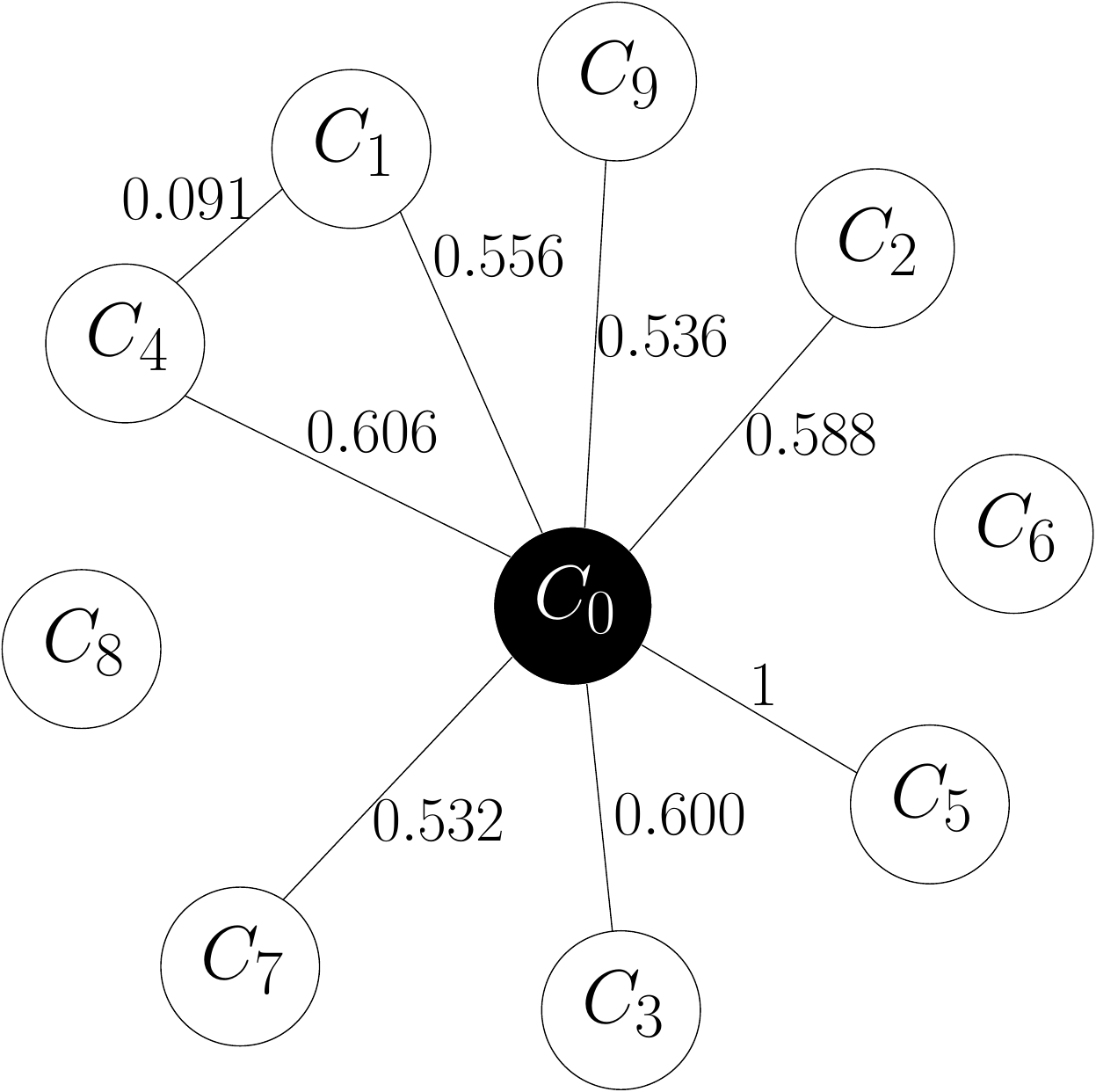} & \includegraphics[scale=0.40]{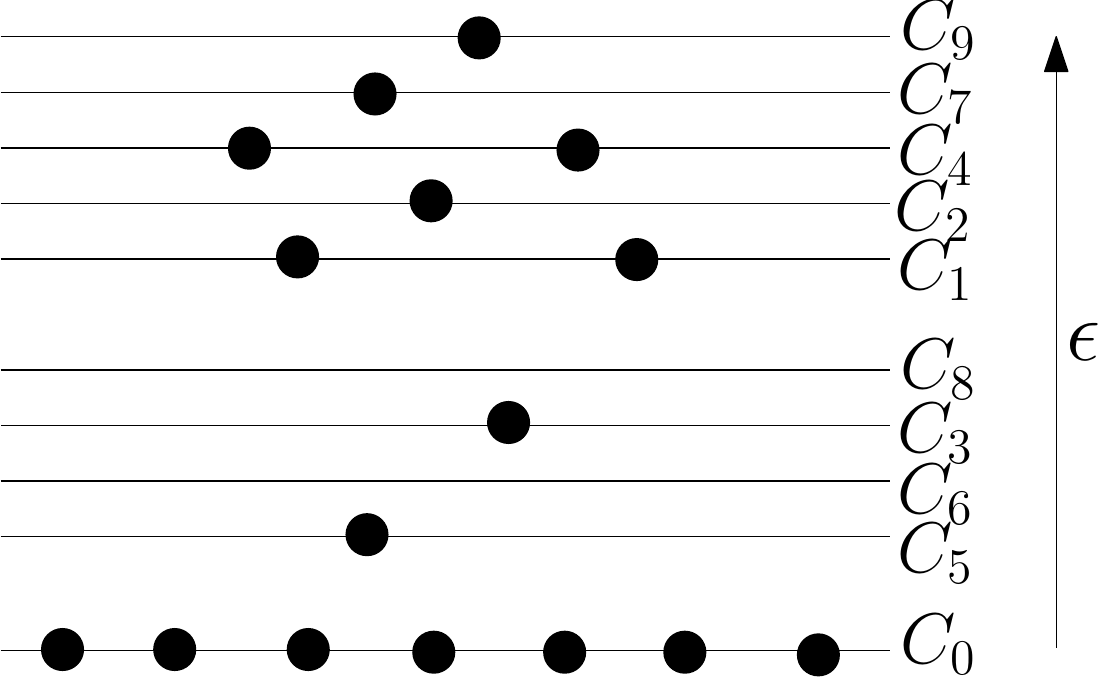}\\
(a) & (b) \\ \\
\includegraphics[scale=0.29]{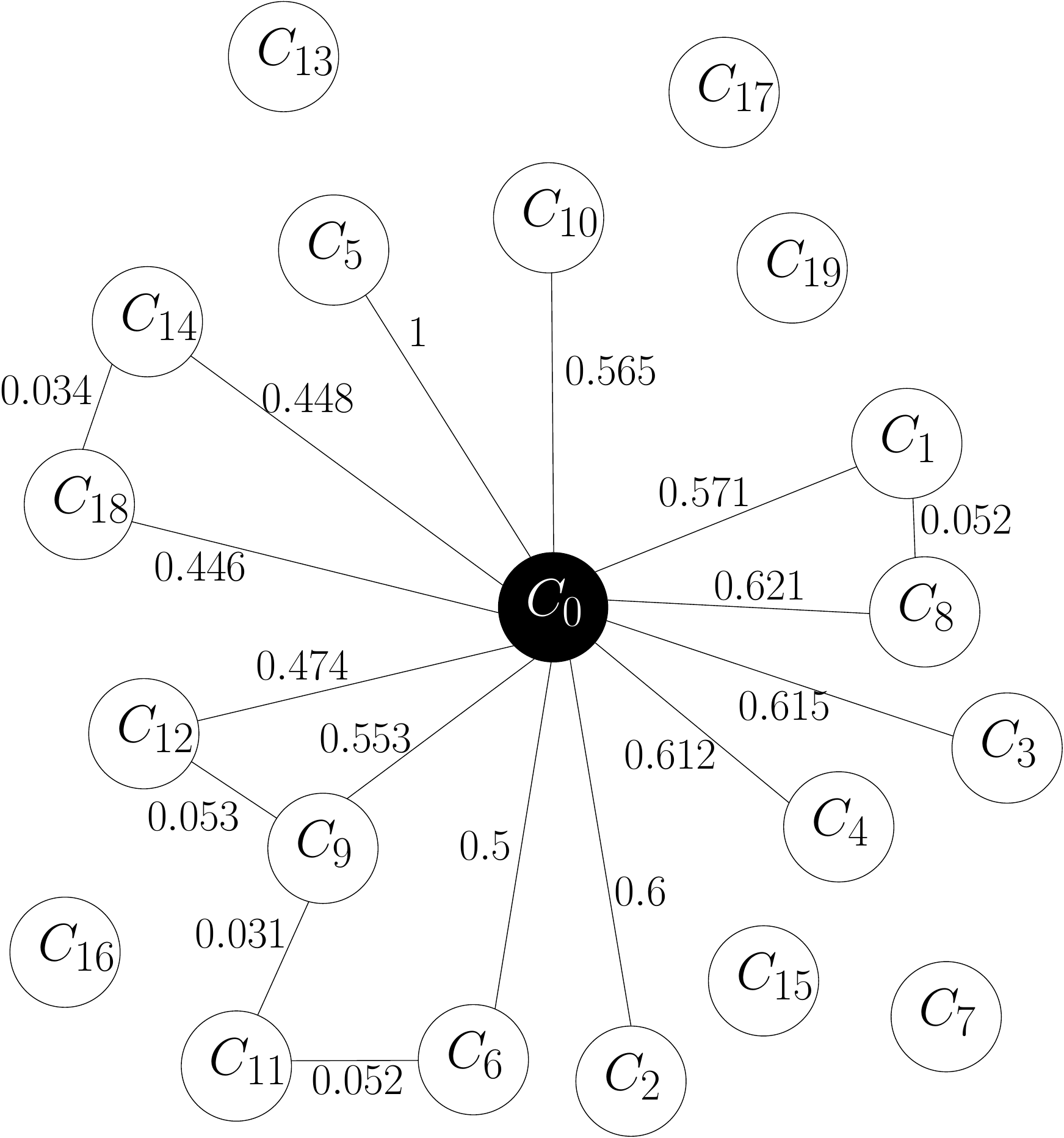} & \includegraphics[scale=0.40]{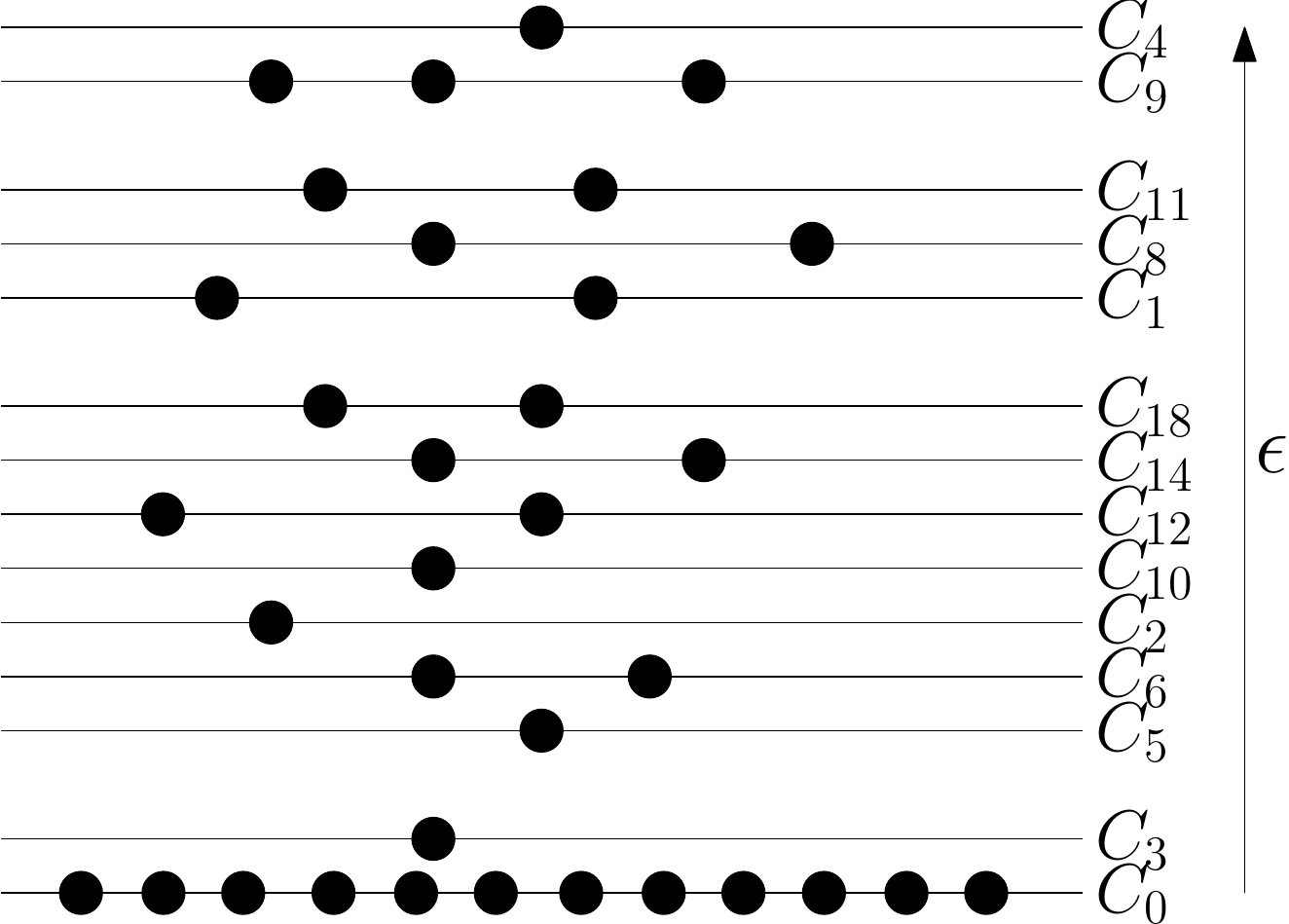} \\
(c) & (d) \\ \\
\includegraphics[scale=0.28]{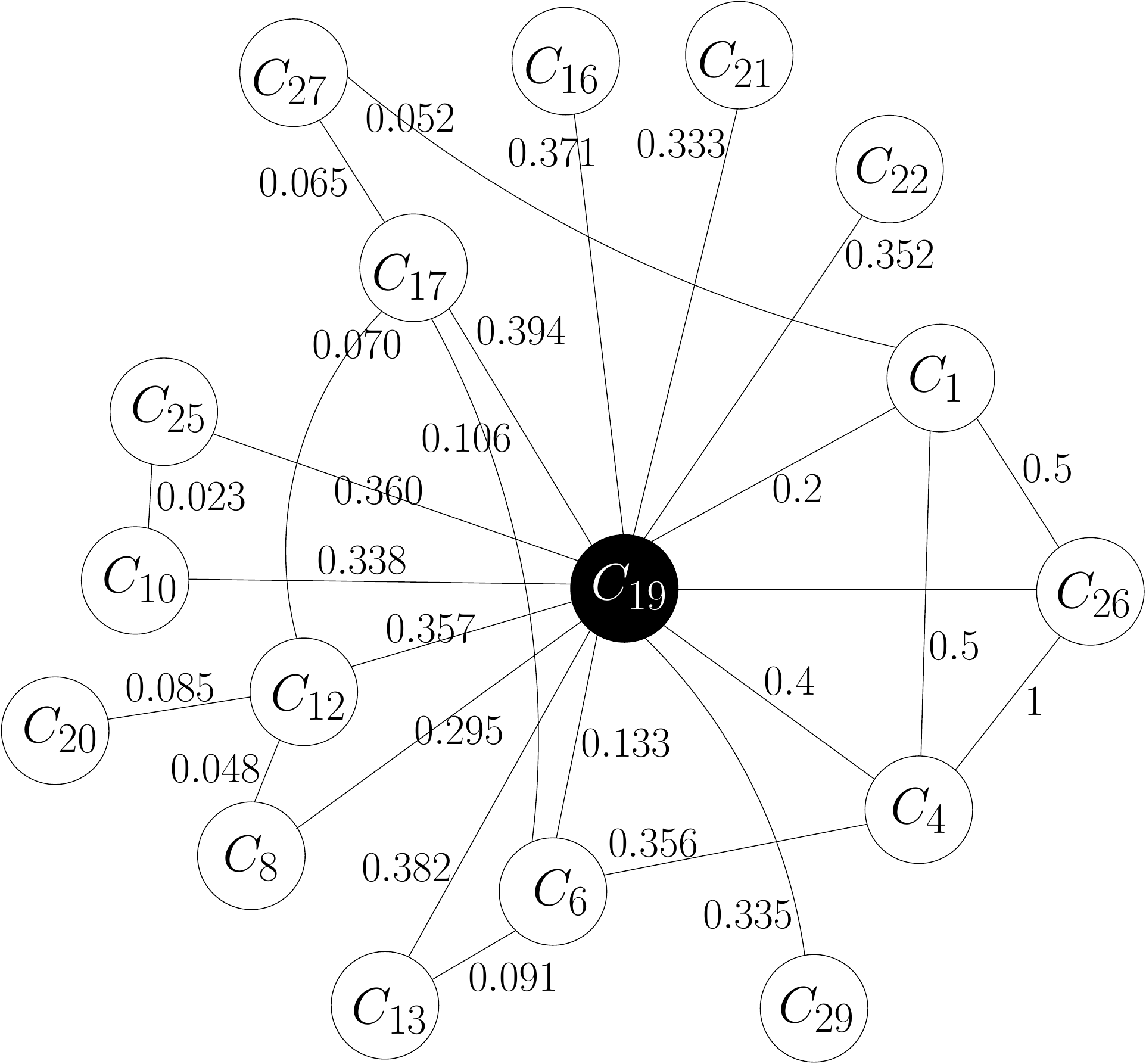} & \includegraphics[scale=0.38]{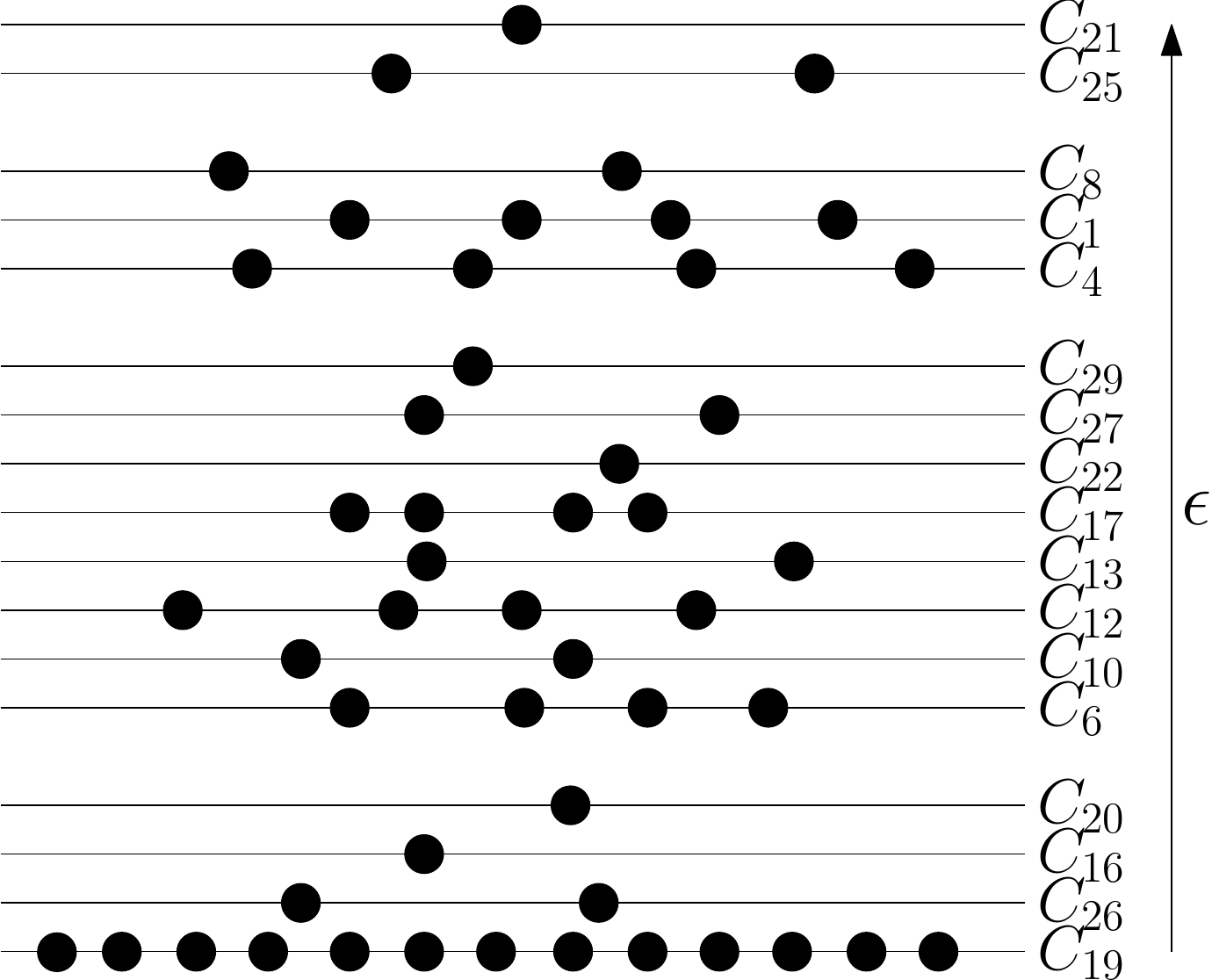}\\
(e) & (f) \\
\end{tabular}
\caption{$3$--SAT S2G graph and Bose gas. We show the graphs derived by a sample instance with $10$ (a), $20$ (c), $30$ (e) clauses, and the energy levels (b), (d), (f) identified by the S2G algorithm. In these figures, when two lines are close to each other, it means they are two distinct degeneration states of the same energy level. For instance, in (b) the clauses $C_8$, $C_3$, $C_6$ and $C_5$ represent four degeneration states of the same energy level. In (d) and (f), degeneration states without particles are omitted. In (e), isolated vertices are omitted. The weight on each edge represents the probability that the corresponding link is established. The node with the lowest energy gets the highest number of particles (BEC phase).}

\label{fig:satgraph}
\end{figure}

In this section we present four graphs obtained as a result of the S2G algorithm.
The plots in Figure \ref{fig:satgraph} highlight that the most connected nodes have the highest number of particles, and the winner node is identified with the lowest energy level. These facts help us confirm that when a Bose-Einstein condensation occurs there is a clear mapping between the graph derived by the S2G algorithm and the Bose gas at low temperatures.
\vspace{1cm}

\noindent $3$--SAT BEC-instance with $10$ clauses and $60$ variables (Figure \ref{fig:satgraph} (a) and (b)):
\begin{tiny}
\begin{eqnarray*}
F &=&
\overbrace{(V_{59} \vee \overline{V_{55}} \vee V_{52})}^{C_0} \wedge
\overbrace{(\overline{V_{46}} \vee V_{31} \vee V_{41})}^{C_1} \wedge
\overbrace{(\overline{V_{56}} \vee \overline{V_{44}} \vee V_{18})}^{C_2} \wedge
\overbrace{(\overline{V_{42}} \vee \overline{V_{10}} \vee V_{27})}^{C_3} \wedge
\overbrace{(\overline{V_{14}} \vee \overline{V_{54}} \vee \overline{V_{22}})}^{C_4} \wedge  \\
&&  \wedge \overbrace{(\overline{V_{40}} \vee V_{52} \vee \overline{V_{27}})}^{C_5} \wedge
\overbrace{(V_{42} \vee \overline{V_{55}} \vee \overline{V_{29}})}^{C_6} \wedge
\overbrace{(V_{9} \vee \overline{V_{53}} \vee V_{39})}^{C_7} \wedge
\overbrace{(V_{48} \vee V_{19} \vee V_{27})}^{C_8} \wedge
\overbrace{(\overline{V_{34}} \vee V_{25} \vee V_{11})}^{C_{9}} 
\end{eqnarray*}
\end{tiny}

\noindent $3$--SAT BEC-instance with $20$ clauses and $60$ variables (Figure \ref{fig:satgraph} (c) and (d)):

\begin{tiny}
\begin{eqnarray*}
F  &=& \overbrace{(V_{59} \vee \overline{V_{55}} \vee V_{52})}^{C_0} \wedge
\overbrace{(\overline{V_{46}} \vee V_{31} \vee V_{41})}^{C_1}  \wedge 
\overbrace{(\overline{V_{56}} \vee \overline{V_{44}} \vee V_{18})}^{C_2}  \wedge 
\overbrace{(\overline{V_{42}} \vee \overline{V_{10}} \vee V_{27})}^{C_3}  \wedge 
\overbrace{(\overline{V_{14}} \vee \overline{V_{54}} \vee \overline{V_{22}})}^{C_4} \wedge  \\
&& \wedge  \overbrace{(\overline{V_{40}} \vee V_{52} \vee \overline{V_{27}})}^{C_5}  \wedge 
\overbrace{(V_{42} \vee \overline{V_{55}} \vee \overline{V_{29}})}^{C_6}  \wedge 
\overbrace{(V_{9} \vee \overline{V_{53}} \vee V_{39})}^{C_7}  \wedge 
\overbrace{(V_{48} \vee V_{19} \vee V_{27})}^{C_8}  \wedge 
\overbrace{(\overline{V_{34}} \vee V_{25} \vee V_{11})}^{C_{9}}  \wedge  \\
&& \wedge 
\overbrace{(\overline{V_{30}} \vee \overline{V_{46}} \vee {V_{60}})}^{C_{10}} \wedge
\overbrace{({V_{45}} \vee \overline{V_{23}} \vee {V_{18}})}^{C_{11}} \wedge
\overbrace{(\overline{V_{10}} \vee {V_{44}} \vee \overline{V_{8}})}^{C_{12}} \wedge
\overbrace{(\overline{V_{5}} \vee {V_{58}} \vee {V_{4}})}^{C_{13}} \wedge
\overbrace{(\overline{V_{48}} \vee \overline{V_{44}} \vee \overline{V_{40}})}^{C_{14}} \wedge \\
&& \wedge 
\overbrace{({V_{6}} \vee {V_{9}} \vee {V_{32}})}^{C_{15}} \wedge
\overbrace{(\overline{V_{28}} \vee {V_{50}} \vee {V_{35}})}^{C_{16}} \wedge
\overbrace{({V_{60}} \vee \overline{V_{13}} \vee {V_{54}})}^{C_{17}} \wedge
\overbrace{(\overline{V_{53}} \vee \overline{V_{5}} \vee {V_{1}})}^{C_{18}} \wedge
\overbrace{(\overline{V_{10}} \vee {V_{24}} \vee \overline{V_{55}})}^{C_{19}} 
\end{eqnarray*}
\end{tiny}

\noindent $3$--SAT BEC-instance with $30$ clauses and $60$ variables (Figure \ref{fig:satgraph} (e) and (f)):
\begin{tiny}
\begin{eqnarray*}
F &=&
\overbrace{(V_{59} \vee \overline{V_{55}} \vee V_{52})}^{C_0} \wedge
\overbrace{(\overline{V_{46}} \vee V_{31} \vee V_{41})}^{C_1} \wedge
\overbrace{(\overline{V_{56}} \vee \overline{V_{44}} \vee V_{18})}^{C_2} \wedge
\overbrace{(\overline{V_{42}} \vee \overline{V_{10}} \vee V_{27})}^{C_3} \wedge
\overbrace{(\overline{V_{14}} \vee \overline{V_{54}} \vee \overline{V_{22}})}^{C_4} \wedge  \\
&& \wedge
\overbrace{(\overline{V_{40}} \vee V_{52} \vee \overline{V_{27}})}^{C_5} \wedge
\overbrace{(V_{42} \vee \overline{V_{55}} \vee \overline{V_{29}})}^{C_6} \wedge
\overbrace{(V_{9} \vee \overline{V_{53}} \vee V_{39})}^{C_7} \wedge
\overbrace{(V_{48} \vee V_{19} \vee V_{27})}^{C_8} \wedge
\overbrace{(\overline{V_{34}} \vee V_{25} \vee V_{11})}^{C_{9}} \wedge  \\
&& \wedge
\overbrace{(\overline{V_{30}} \vee \overline{V_{46}} \vee {V_{60}})}^{C_{10}} \wedge
\overbrace{({V_{45}} \vee \overline{V_{23}} \vee {V_{18}})}^{C_{11}} \wedge
\overbrace{(\overline{V_{10}} \vee {V_{44}} \vee \overline{V_{8}})}^{C_{12}} \wedge
\overbrace{(\overline{V_{5}} \vee {V_{58}} \vee {V_{4}})}^{C_{13}} \wedge
\overbrace{(\overline{V_{48}} \vee \overline{V_{44}} \vee \overline{V_{40}})}^{C_{14}} \wedge  \\
&& \wedge
\overbrace{({V_{6}} \vee {V_{9}} \vee {V_{32}})}^{C_{15}} \wedge
\overbrace{(\overline{V_{28}} \vee {V_{50}} \vee {V_{35}})}^{C_{16}} \wedge
\overbrace{({V_{60}} \vee \overline{V_{13}} \vee {V_{54}})}^{C_{17}} \wedge
\overbrace{(\overline{V_{53}} \vee \overline{V_{5}} \vee {V_{1}})}^{C_{18}} \wedge
\overbrace{(\overline{V_{10}} \vee {V_{24}} \vee \overline{V_{55}})}^{C_{19}} \wedge \\
&& \wedge
\overbrace{(V_{35} \vee \overline{V_{28}} \vee \overline{V_{42}})}^{C_{20}} \wedge
\overbrace{(\overline{V_{19}} \vee \overline{V_{60}} \vee \overline{V_{35}})}^{C_{21}} \wedge
\overbrace{(\overline{V_{3}} \vee \overline{V_{45}} \vee \overline{V_{50}})}^{C_{22}} \wedge
\overbrace{(\overline{V_{59}} \vee \overline{V_{45}} \vee V_{55})}^{C_{23}} \wedge
\overbrace{(V_{10} \vee \overline{V_{50}} \vee V_{54})}^{C_{24}} \wedge \\
&& \wedge
\overbrace{(V_{38} \vee {V_{14}} \vee V_{43})}^{C_{25}} \wedge
\overbrace{(\overline{V_{5}} \vee \overline{V_{38}} \vee \overline{V_{14}})}^{C_{26}} \wedge
\overbrace{(V_{17} \vee {V_{33}} \vee \overline{V_{44}})}^{C_{27}} \wedge
\overbrace{(\overline{V_{57}} \vee {V_{28}} \vee V_{21})}^{C_{28}} \wedge
\overbrace{(V_{21} \vee \overline{V_{36}} \vee V_{50})}^{C_{29}} 
\end{eqnarray*}
\end{tiny}

\newpage

\begin{figure}[h]
\centering
\newcolumntype{S}{>{\centering\arraybackslash} m{0.45\linewidth} }
\begin{tabular}{S S} 
\includegraphics[scale=0.29]{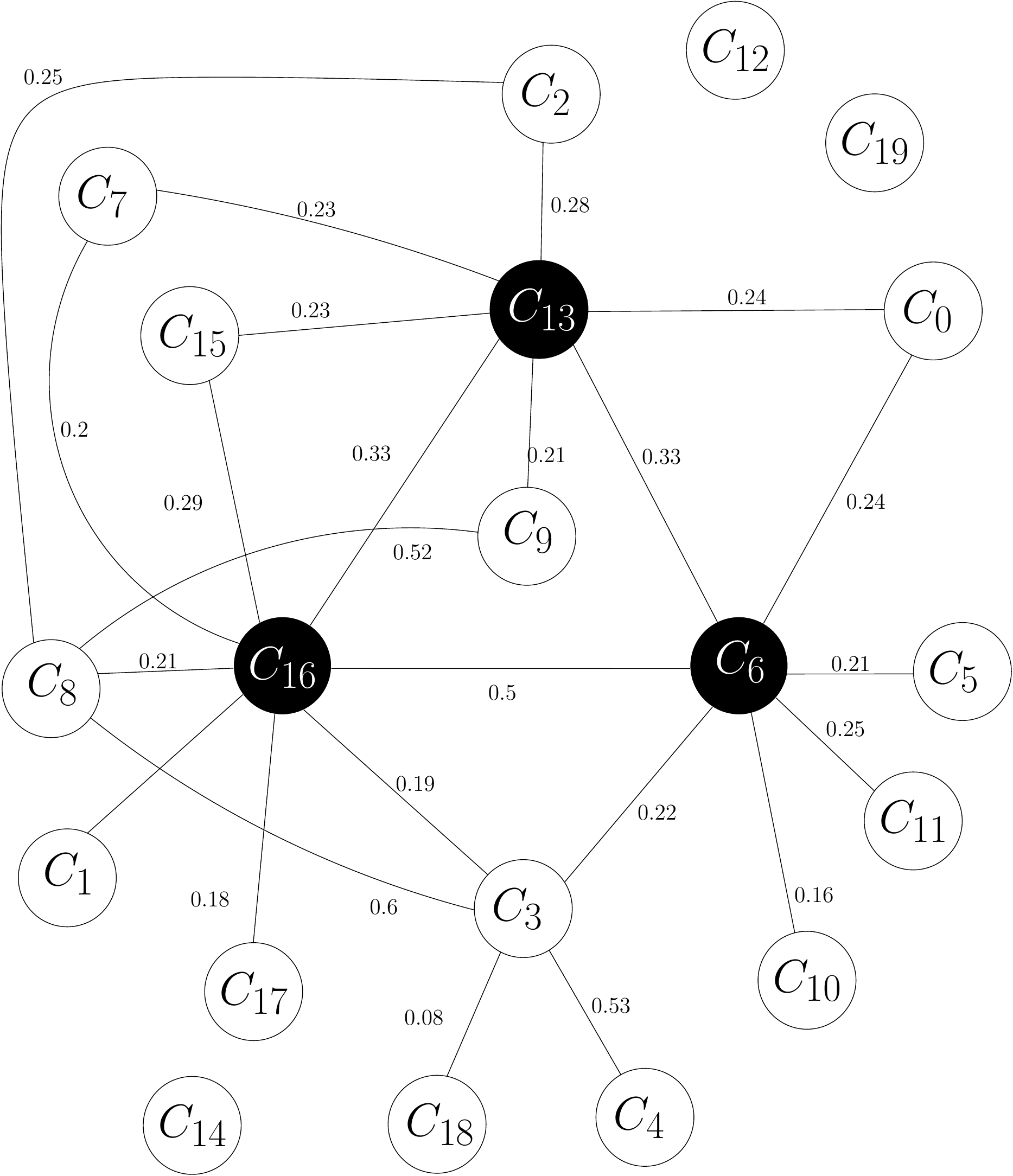} & \includegraphics[scale=0.35]{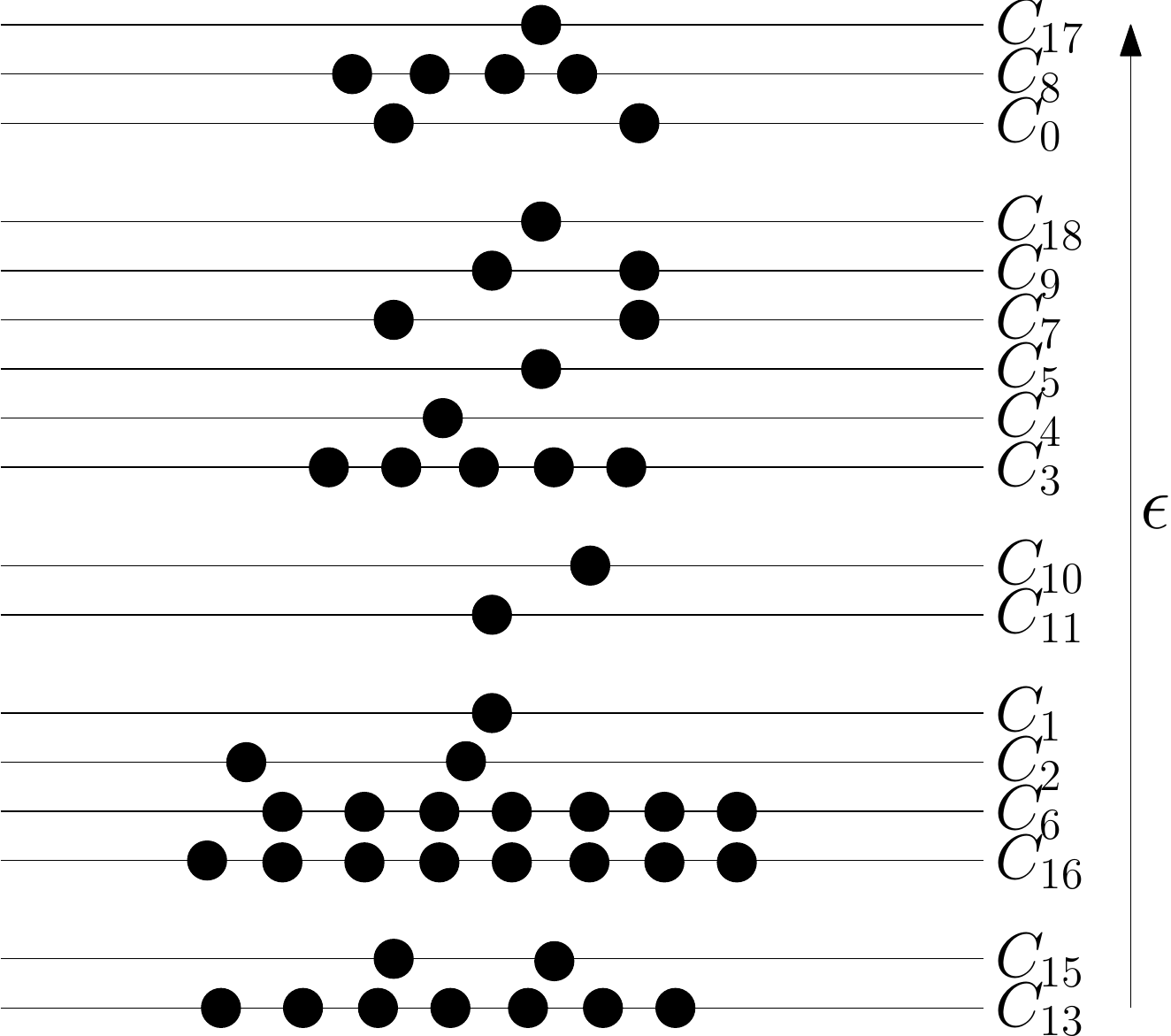}\\
(a) & (b) 
\end{tabular}
\caption{$3$--SAT S2G graph and fit-get-rich phase. In (a) we show the graph derived by a sample instance with $20$ clauses, and in (b) the energy levels identified by the S2G algorithm. When two lines are close to each other, it means they are two distinct degeneration states of the same energy level. For instance, the clauses $C_{13}$ and $C_{15}$ represent two degeneration states of the same energy level. Degeneration states without particles are omitted. The weight on each edge represents the probability that the corresponding link is established. In this example a fit-get-rich (FGR) phase happens, consisting of three hubs $C_6$, $C_{13}$ and $C_{16}$.}
\label{fig:satgraph2}
\end{figure}

\noindent $3$--SAT FGR-instance with $20$ clauses and $20$ variables (Figure \ref{fig:satgraph2}):
\begin{tiny}
\begin{eqnarray*}
F &=&
\overbrace{({V_{5}} \vee {V_{3}} \vee {V_{17}})}^{C_0} \wedge
\overbrace{(\overline{V_{3}} \vee \overline{V_{20}} \vee \overline{V_{5}})}^{C_1} \wedge
\overbrace{({V_{6}} \vee \overline{V_{13}} \vee {V_{11}})}^{C_2} \wedge
\overbrace{(\overline{V_{16}} \vee {V_{11}} \vee {V_{9}})}^{C_3} \wedge
\overbrace{(\overline{V_{17}} \vee \overline{V_{19}} \vee {V_{2}})}^{C_4} \wedge  \\
&& \wedge
\overbrace{({V_{4}} \vee {V_{14}} \vee {V_{18}})}^{C_5} \wedge
\overbrace{(\overline{V_{10}} \vee \overline{V_{2}} \vee \overline{V_{5}})}^{C_6} \wedge
\overbrace{(\overline{V_{6}} \vee \overline{V_{11}} \vee \overline{V_{8}})}^{C_7} \wedge
\overbrace{(\overline{V_{11}} \vee \overline{V_{1}} \vee \overline{V_{9}})}^{C_8} \wedge
\overbrace{({V_{6}} \vee \overline{V_{15}} \vee {V_{13}})}^{C_{9}} \wedge  \\
&& \wedge
\overbrace{({V_{9}} \vee {V_{18}} \vee \overline{V_{17}})}^{C_{10}} \wedge
\overbrace{(\overline{V_{8}} \vee \overline{V_{14}} \vee \overline{V_{20}})}^{C_{11}} \wedge
\overbrace{(\overline{V_{9}} \vee \overline{V_{19}} \vee \overline{V_{8}})}^{C_{12}} \wedge
\overbrace{(\overline{V_{10}} \vee {V_{5}} \vee \overline{V_{20}})}^{C_{13}} \wedge
\overbrace{(\overline{V_{13}} \vee {V_{9}} \vee {V_{6}})}^{C_{14}} \wedge  \\
&& \wedge
\overbrace{(\overline{V_{5}} \vee {V_{4}} \vee {V_{6}})}^{C_{15}} \wedge
\overbrace{(\overline{V_{19}} \vee \overline{V_{3}} \vee \overline{V_{10}})}^{C_{16}} \wedge
\overbrace{({V_{14}} \vee {V_{8}} \vee {V_{15}})}^{C_{17}} \wedge
\overbrace{(\overline{V_{12}} \vee {V_{5}} \vee \overline{V_{4}})}^{C_{18}} \wedge
\overbrace{(\overline{V_{4}} \vee {V_{15}} \vee \overline{V_{2}})}^{C_{19}}  
\end{eqnarray*}
\end{tiny}

\newpage
\section{S2G-PA Examples}
\label{appendixS2G-PA}
\begin{figure}
\centering
\newcolumntype{S}{>{\centering\arraybackslash} m{0.45\linewidth} }
\begin{tabular}{S S} 
\includegraphics[scale=0.32]{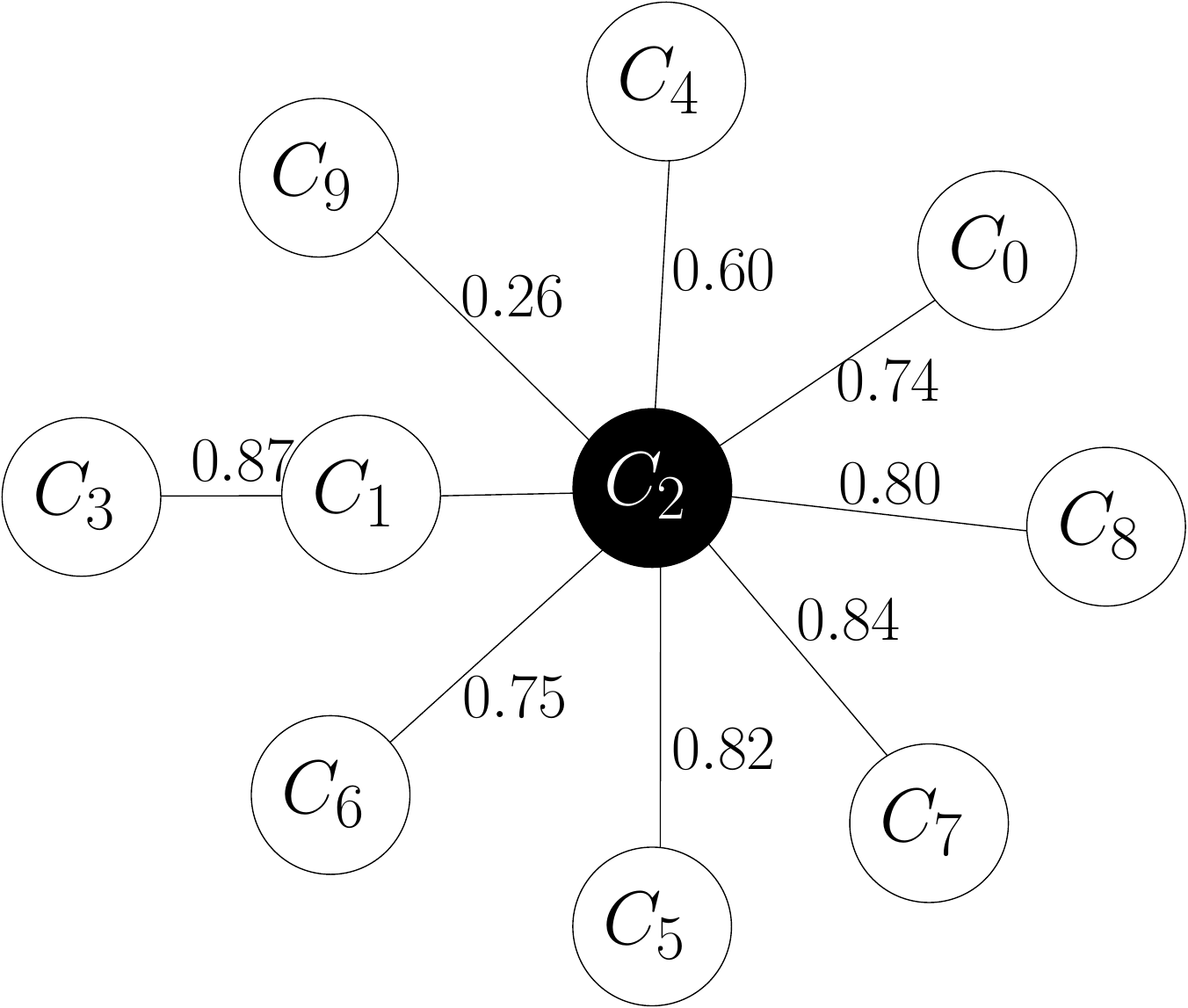} & \includegraphics[scale=0.32]{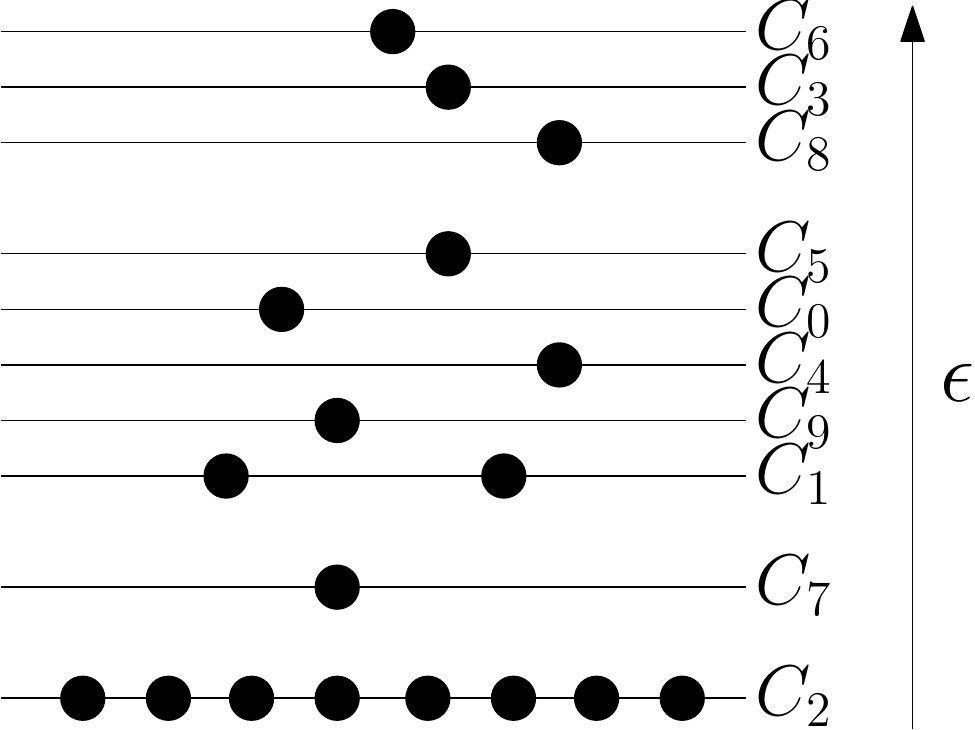}\\
(a) & (b) \\ \\
\includegraphics[scale=0.32]{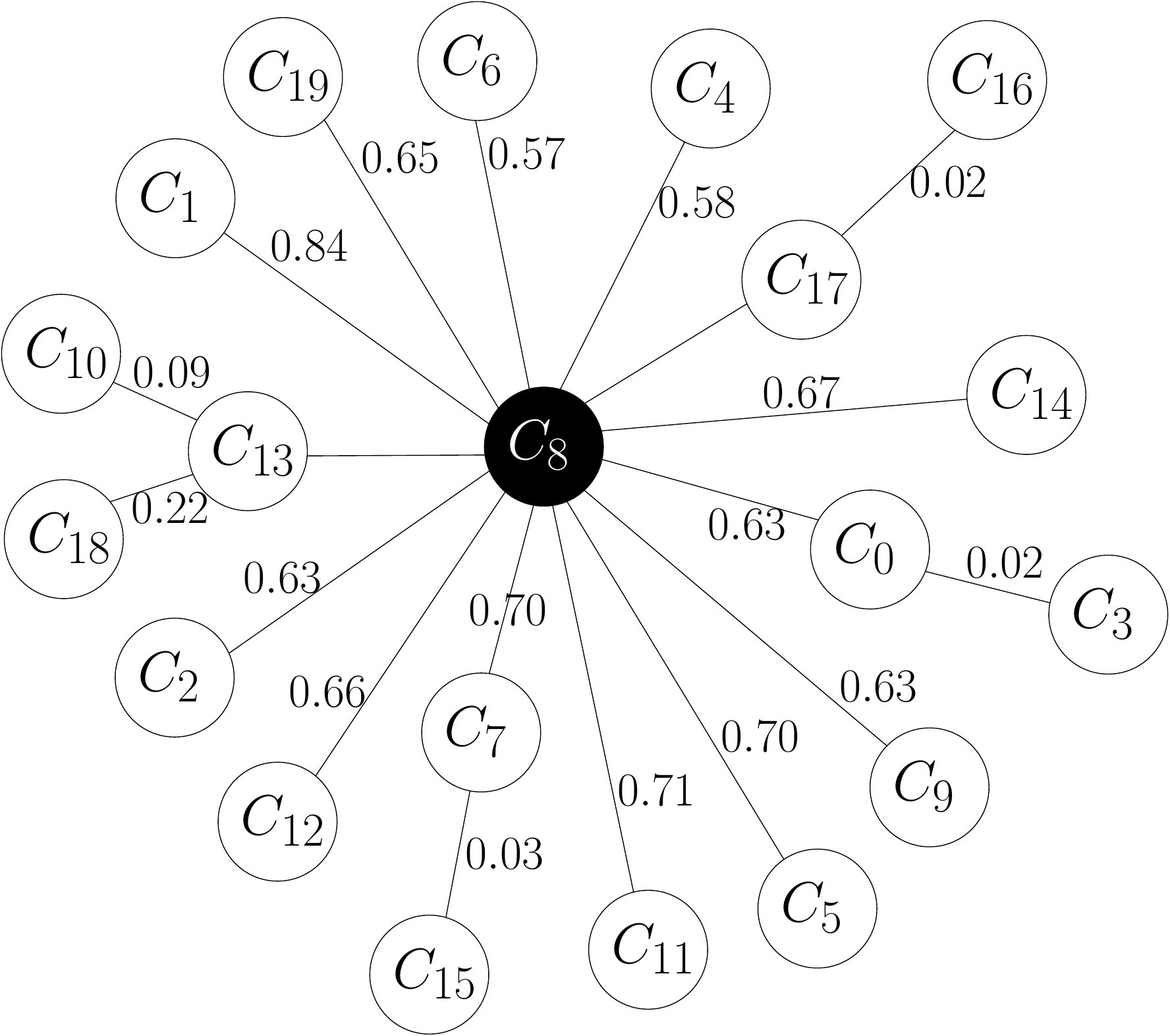} & \includegraphics[scale=0.32]{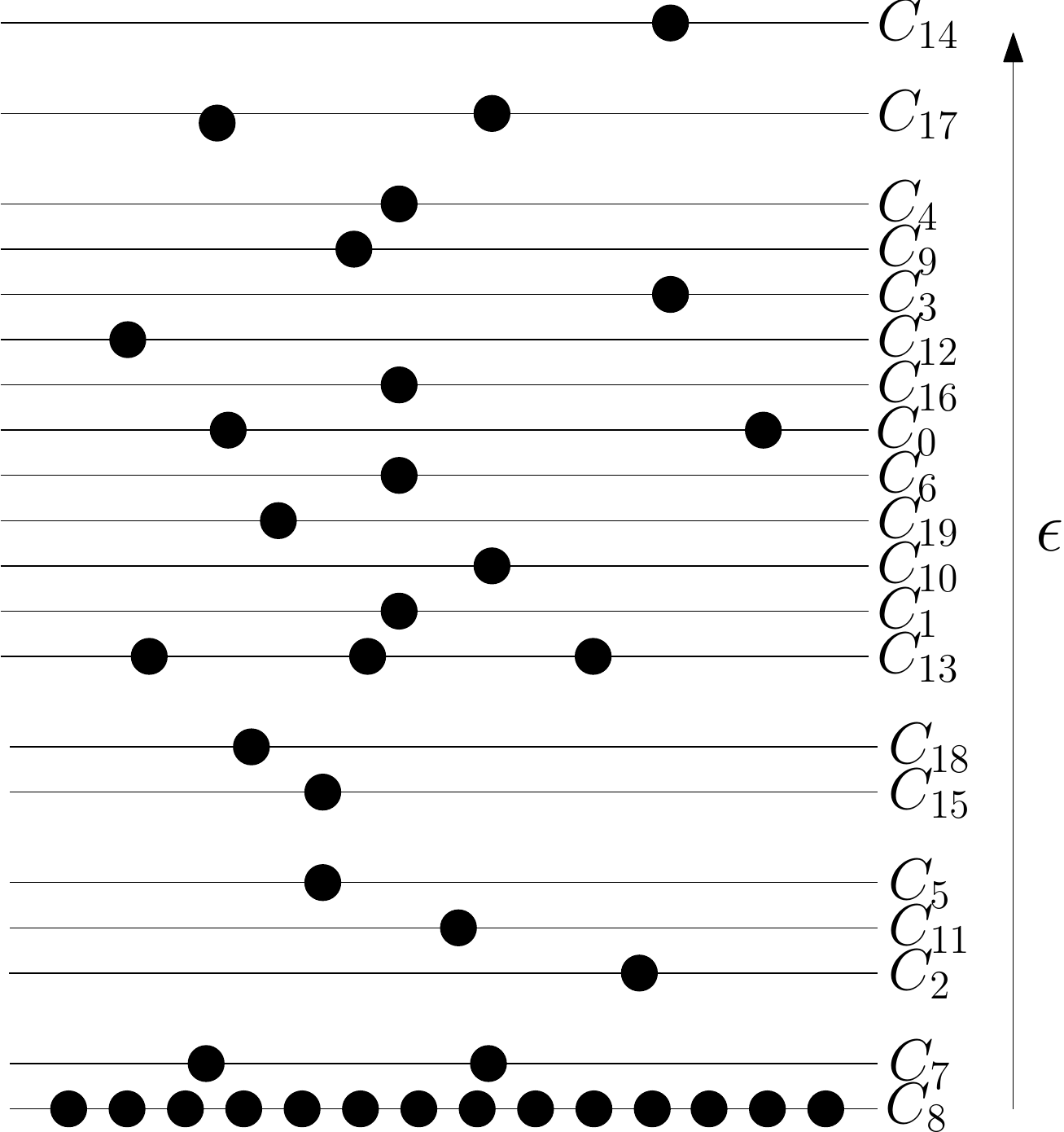} \\
(c) & (d) \\ \\
\includegraphics[scale=0.28]{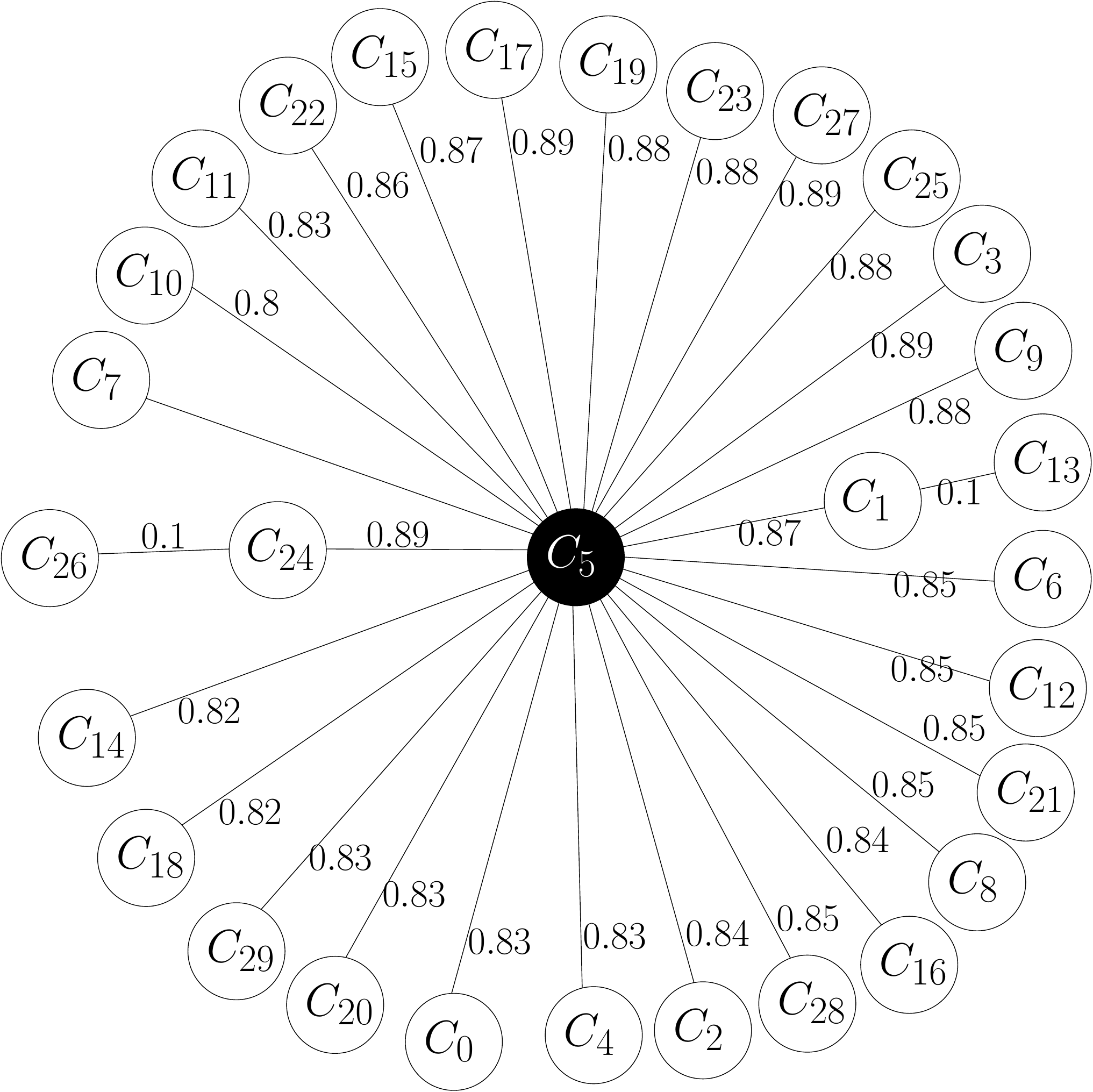} & \includegraphics[scale=0.25]{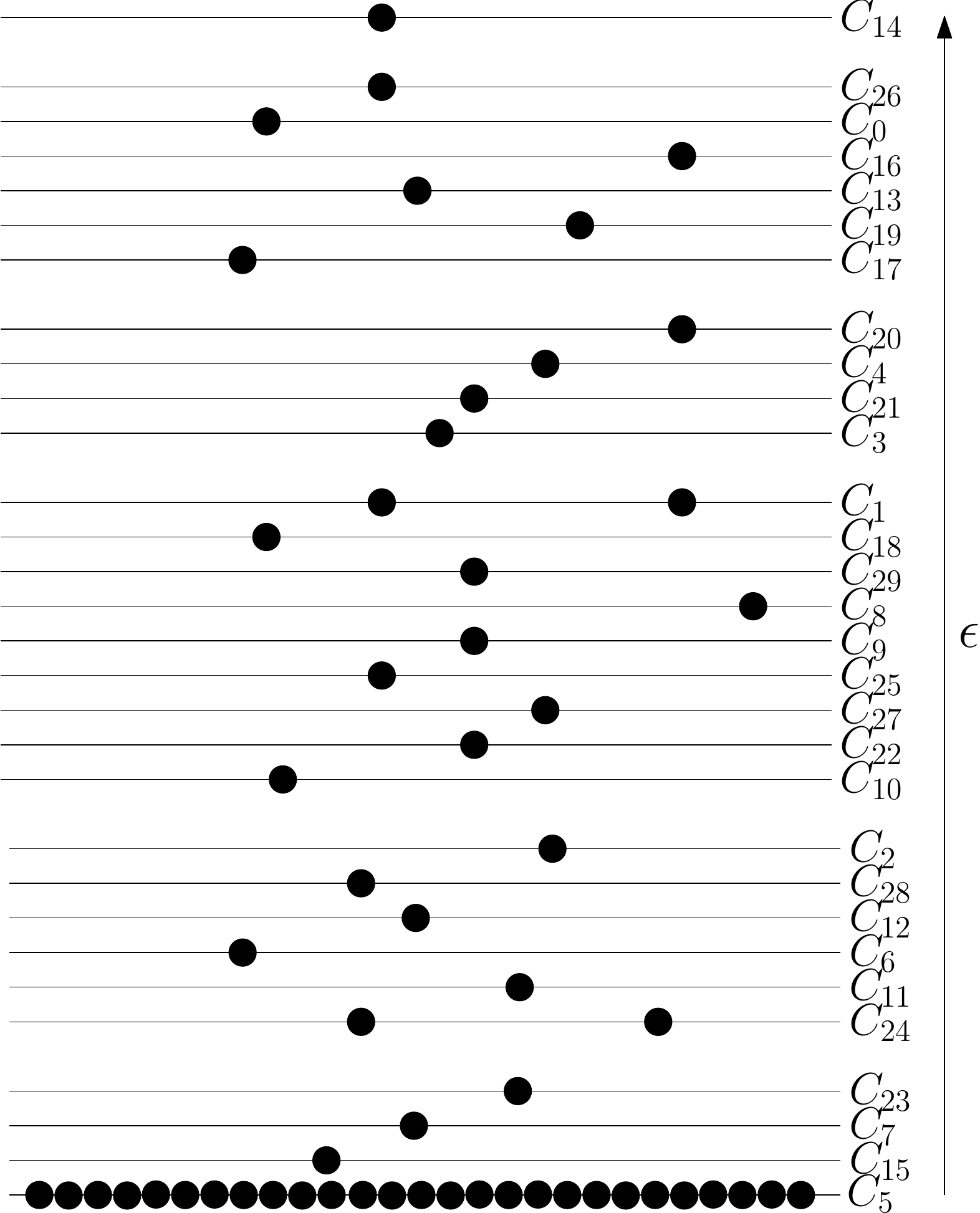} \\
(e) & (f)
\end{tabular}
\caption{$3$--SAT S2G-PA graph and Bose gas. We show the graphs derived by a sample instance with $10$ (a), $20$ (c), $30$ (e) clauses, and the energy levels (b), (d), (f) identified by the S2G-PA algorithm. Thanks to the fitness-based preferential attachment, there are no isolated vertices, thus every degeneration state is populated by at least one particle. These examples show an almost complete BEC.}
\label{fig:satgraphpref}
\end{figure}

$3$--SAT BEC-instance with $10$ clauses and $30$ variables (Figure \ref{fig:satgraphpref} (a) and (b)):
\begin{tiny}
\begin{eqnarray*}
F &=&
\overbrace{(\overline{V_{26}} \vee \overline{V_{19}} \vee {V_{9}})}^{C_0} \wedge
\overbrace{(\overline{V_{30}} \vee \overline{V_{9}} \vee {V_{24}})}^{C_1} \wedge
\overbrace{(\overline{V_{27}} \vee \overline{V_{9}} \vee {V_{11}})}^{C_2} \wedge
\overbrace{({V_{29}} \vee \overline{V_{16}} \vee \overline{V_{3}})}^{C_3} \wedge
\overbrace{({V_{11}} \vee {V_{2}} \vee {V_{1}})}^{C_4} \wedge  \\
&& \wedge
\overbrace{({V_{30}} \vee \overline{V_{28}} \vee {V_{26}})}^{C_{5}} \wedge
\overbrace{(\overline{V_{23}} \vee {V_{16}} \vee {V_{21}})}^{C_{6}} \wedge
\overbrace{(\overline{V_{28}} \vee \overline{V_{22}} \vee {V_{9}})}^{C_{7}} \wedge
\overbrace{(\overline{V_{21}} \vee \overline{V_{5}} \vee {V_{14}})}^{C_{8}} \wedge
\overbrace{(\overline{V_{7}} \vee \overline{V_{27}} \vee \overline{V_{11}})}^{C_{9}} 
\end{eqnarray*}
\end{tiny}

\noindent $3$--SAT BEC-instance with $20$ clauses and $30$ variables (Figure \ref{fig:satgraphpref} (c) and (d)):
\begin{tiny}
\begin{eqnarray*}
F &=&
\overbrace{(\overline{V_{26}} \vee \overline{V_{19}} \vee {V_{9}})}^{C_0} \wedge
\overbrace{(\overline{V_{30}} \vee \overline{V_{9}} \vee {V_{24}})}^{C_1} \wedge
\overbrace{(\overline{V_{27}} \vee \overline{V_{9}} \vee {V_{11}})}^{C_2} \wedge
\overbrace{({V_{29}} \vee \overline{V_{16}} \vee \overline{V_{3}})}^{C_3} \wedge
\overbrace{({V_{11}} \vee {V_{2}} \vee {V_{1}})}^{C_4} \wedge  \\
&& \wedge
\overbrace{({V_{30}} \vee \overline{V_{28}} \vee {V_{26}})}^{C_{5}} \wedge
\overbrace{(\overline{V_{23}} \vee {V_{16}} \vee {V_{21}})}^{C_{6}} \wedge
\overbrace{(\overline{V_{28}} \vee \overline{V_{22}} \vee {V_{9}})}^{C_{7}} \wedge
\overbrace{(\overline{V_{21}} \vee \overline{V_{5}} \vee {V_{14}})}^{C_{8}} \wedge
\overbrace{(\overline{V_{7}} \vee \overline{V_{27}} \vee \overline{V_{11}})}^{C_{9}} \wedge  \\
&& \wedge
\overbrace{(\overline{V_{20}} \vee {V_{26}} \vee \overline{V_{14}})}^{C_{10}} \wedge
\overbrace{({V_{21}} \vee \overline{V_{28}} \vee \overline{V_{15}})}^{C_{11}} \wedge
\overbrace{({V_{5}} \vee \overline{V_{27}} \vee {V_{20}})}^{C_{12}} \wedge
\overbrace{({V_{24}} \vee {V_{10}} \vee {V_{14}})}^{C_{13}} \wedge
\overbrace{(\overline{V_{17}} \vee {V_{13}} \vee {V_{6}})}^{C_{14}} \wedge  \\
&& \wedge
\overbrace{(\overline{V_{15}} \vee \overline{V_{23}} \vee {V_{30}})}^{C_{15}} \wedge
\overbrace{({V_{23}} \vee \overline{V_{12}} \vee {V_{9}})}^{C_{16}} \wedge
\overbrace{(\overline{V_{5}} \vee {V_{22}} \vee \overline{V_{4}})}^{C_{17}} \wedge
\overbrace{(\overline{V_{3}} \vee {V_{29}} \vee {V_{2}})}^{C_{18}} \wedge
\overbrace{(\overline{V_{24}} \vee \overline{V_{22}} \vee \overline{V_{20}})}^{C_{19}}  
\end{eqnarray*}
\end{tiny}

\noindent $3$--SAT BEC-instance with $30$ clauses and $30$ variables (Figure \ref{fig:satgraphpref} (e) and (f)):
\begin{tiny}
\begin{eqnarray*}
F &=&
\overbrace{(\overline{V_{26}} \vee \overline{V_{19}} \vee {V_{9}})}^{C_0} \wedge
\overbrace{(\overline{V_{30}} \vee \overline{V_{9}} \vee {V_{24}})}^{C_1} \wedge
\overbrace{(\overline{V_{27}} \vee \overline{V_{9}} \vee {V_{11}})}^{C_2} \wedge
\overbrace{({V_{29}} \vee \overline{V_{16}} \vee \overline{V_{3}})}^{C_3} \wedge
\overbrace{({V_{11}} \vee {V_{2}} \vee {V_{1}})}^{C_4} \wedge  \\
&& \wedge
\overbrace{({V_{30}} \vee \overline{V_{28}} \vee {V_{26}})}^{C_{5}} \wedge
\overbrace{(\overline{V_{23}} \vee {V_{16}} \vee {V_{21}})}^{C_{6}} \wedge
\overbrace{(\overline{V_{28}} \vee \overline{V_{22}} \vee {V_{9}})}^{C_{7}} \wedge
\overbrace{(\overline{V_{21}} \vee \overline{V_{5}} \vee {V_{14}})}^{C_{8}} \wedge
\overbrace{(\overline{V_{7}} \vee \overline{V_{27}} \vee \overline{V_{11}})}^{C_{9}} \wedge  \\
&& \wedge
\overbrace{(\overline{V_{20}} \vee {V_{26}} \vee \overline{V_{14}})}^{C_{10}} \wedge
\overbrace{({V_{21}} \vee \overline{V_{28}} \vee \overline{V_{15}})}^{C_{11}} \wedge
\overbrace{({V_{5}} \vee \overline{V_{27}} \vee {V_{20}})}^{C_{12}} \wedge
\overbrace{({V_{24}} \vee {V_{10}} \vee {V_{14}})}^{C_{13}} \wedge
\overbrace{(\overline{V_{17}} \vee {V_{13}} \vee {V_{6}})}^{C_{14}} \wedge  \\
&& \wedge
\overbrace{(\overline{V_{15}} \vee \overline{V_{23}} \vee {V_{30}})}^{C_{15}} \wedge
\overbrace{({V_{23}} \vee \overline{V_{12}} \vee {V_{9}})}^{C_{16}} \wedge
\overbrace{(\overline{V_{5}} \vee {V_{22}} \vee \overline{V_{4}})}^{C_{17}} \wedge
\overbrace{(\overline{V_{3}} \vee {V_{29}} \vee {V_{2}})}^{C_{18}} \wedge
\overbrace{(\overline{V_{24}} \vee \overline{V_{22}} \vee \overline{V_{20}})}^{C_{19}} \wedge  \\
&& \wedge
\overbrace{({V_{3}} \vee {V_{5}} \vee {V_{16}})}^{C_{20}} \wedge
\overbrace{(\overline{V_{14}} \vee {V_{25}} \vee {V_{18}})}^{C_{21}} \wedge
\overbrace{({V_{30}} \vee \overline{V_{7}} \vee {V_{27}})}^{C_{22}} \wedge
\overbrace{(\overline{V_{27}} \vee \overline{V_{3}} \vee {V_{1}})}^{C_{23}} \wedge
\overbrace{(\overline{V_{5}} \vee {V_{12}} \vee \overline{V_{28}})}^{C_{24}} \wedge  \\
&& \wedge
\overbrace{({V_{18}} \vee \overline{V_{14}} \vee \overline{V_{21}})}^{C_{25}} \wedge
\overbrace{(\overline{V_{10}} \vee \overline{V_{30}} \vee \overline{V_{18}})}^{C_{26}} \wedge
\overbrace{(\overline{V_{2}} \vee \overline{V_{23}} \vee \overline{V_{25}})}^{C_{27}} \wedge
\overbrace{(\overline{V_{30}} \vee \overline{V_{23}} \vee {V_{28}})}^{C_{28}} \wedge
\overbrace{({V_{5}} \vee \overline{V_{25}} \vee {V_{27}})}^{C_{29}} 
\end{eqnarray*}
\end{tiny}

\newpage
\section{Standard Deviation of Non-Winner Degree Distribution}
\label{appendixstd}

\begin{figure}[H]
\centering
\newcolumntype{S}{>{\centering \arraybackslash} m{0.65\linewidth} }
\centering
\begin{tabular}{S } 
\includegraphics[scale=0.6]{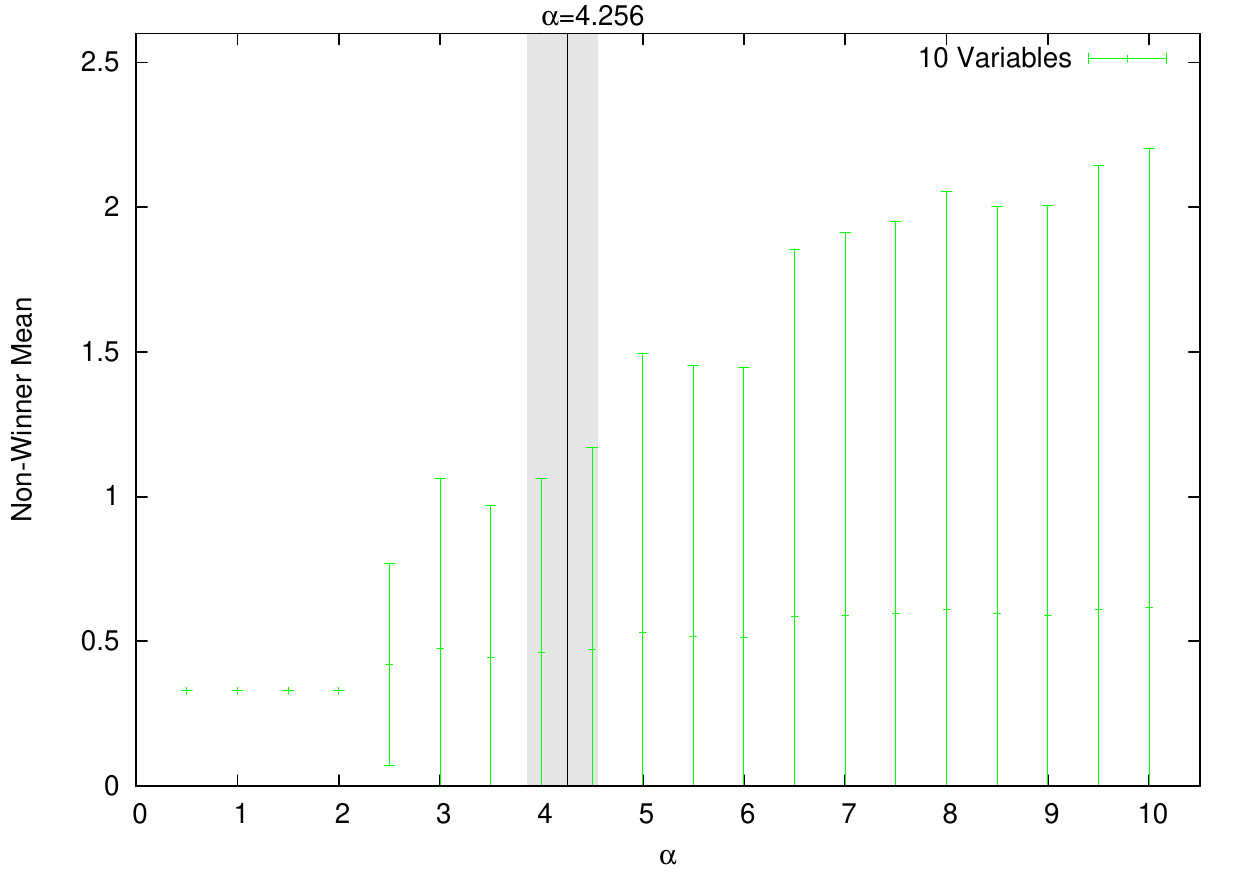} \\
\includegraphics[scale=0.6]{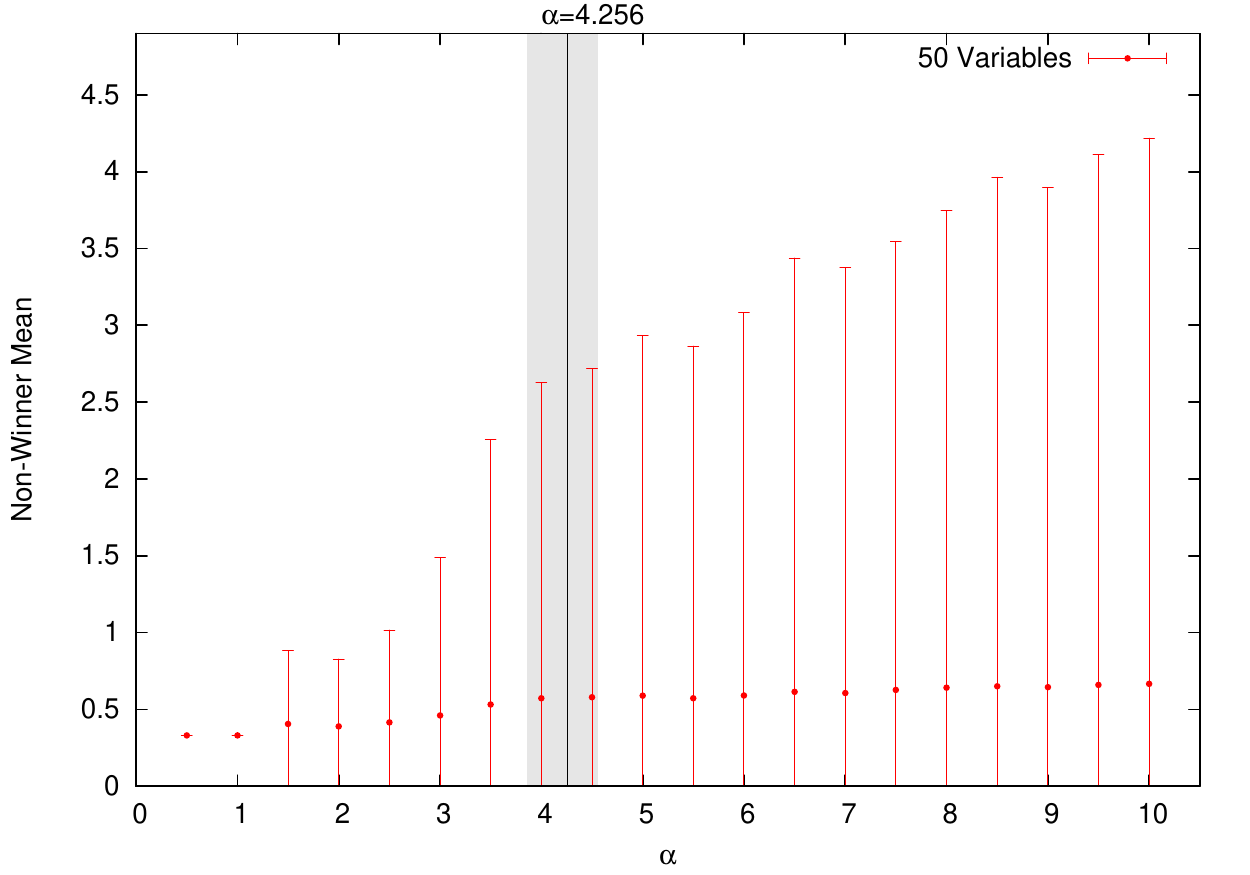} \\
\includegraphics[scale=0.6]{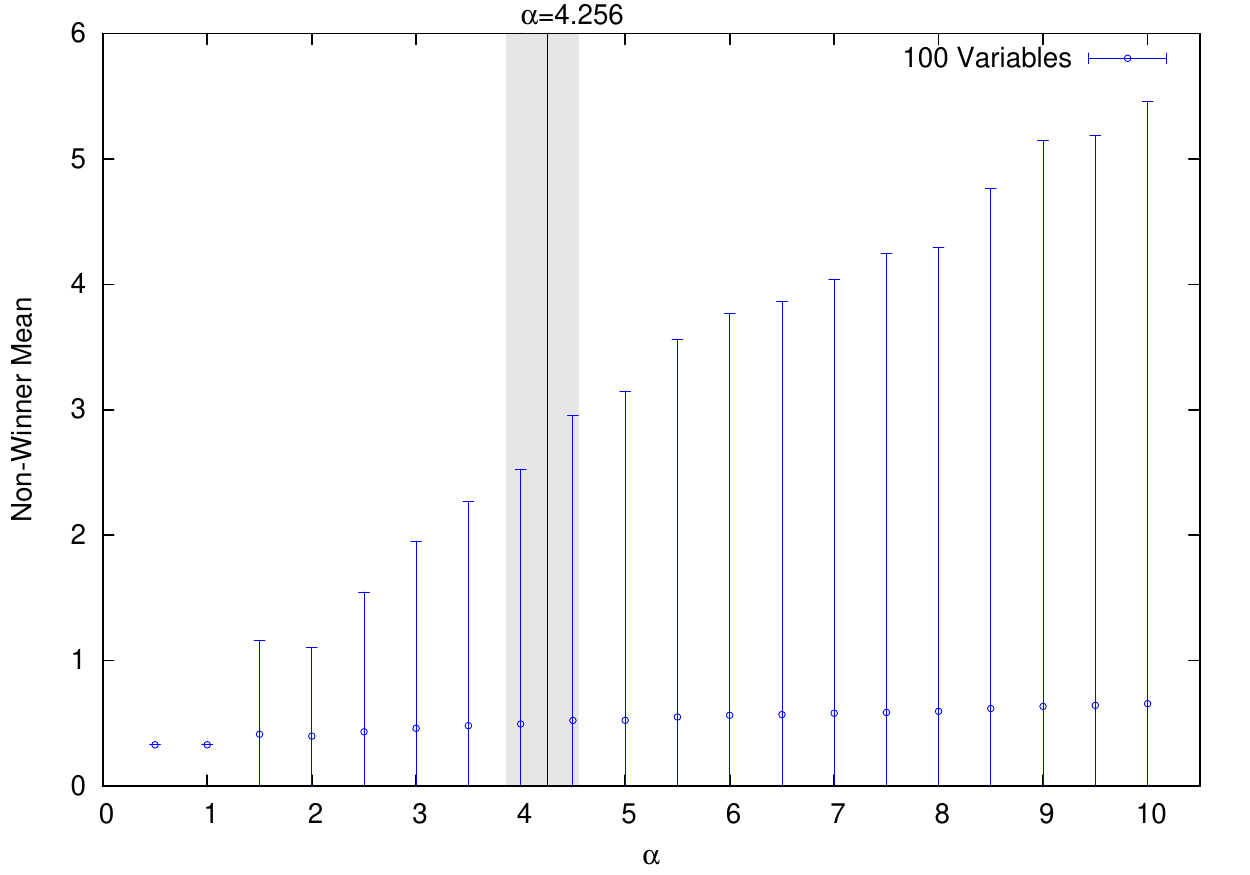} 
\end{tabular}
\caption{Standard deviation of the non-winner degree distribution. Each point is an average over $100$ $3$--SAT instances with $30$ graphs per instance. The distribution taken into account is the standard degree distribution except for the winner node, not considered in this analysis. Both the mean and the standard deviation increase as $\alpha$ increases, numerically showing the emergence of new hubs.}
\label{fig:typical}
\end{figure}

\newpage
\section{Typical S2G-PA Networks}
\label{appendixtyp}

\begin{figure}[H]
\centering
\newcolumntype{S}{>{\centering\arraybackslash} m{0.45\linewidth} }
\begin{tabular}{S S} 
\includegraphics[scale=0.2]{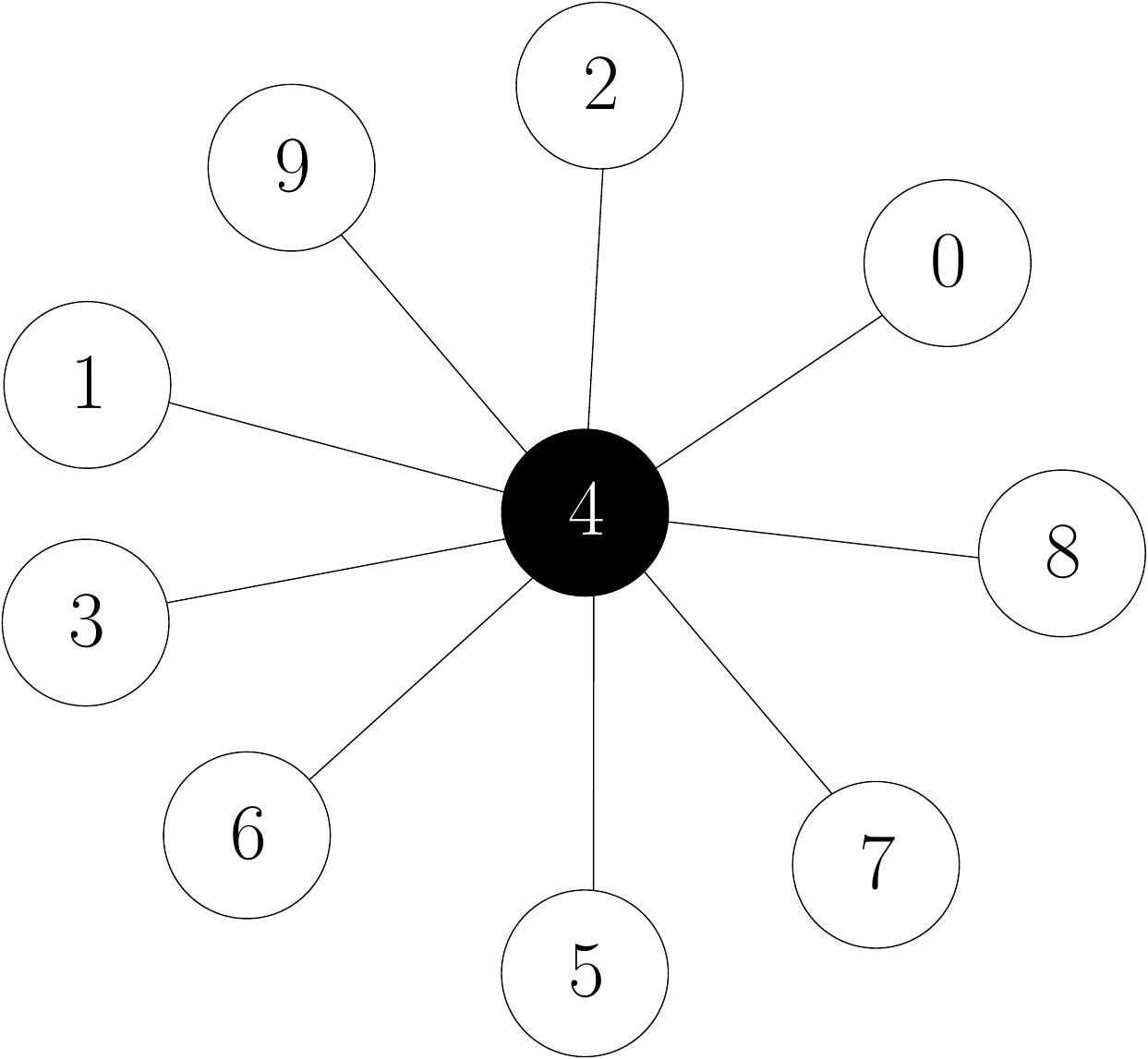} & \includegraphics[scale=0.17]{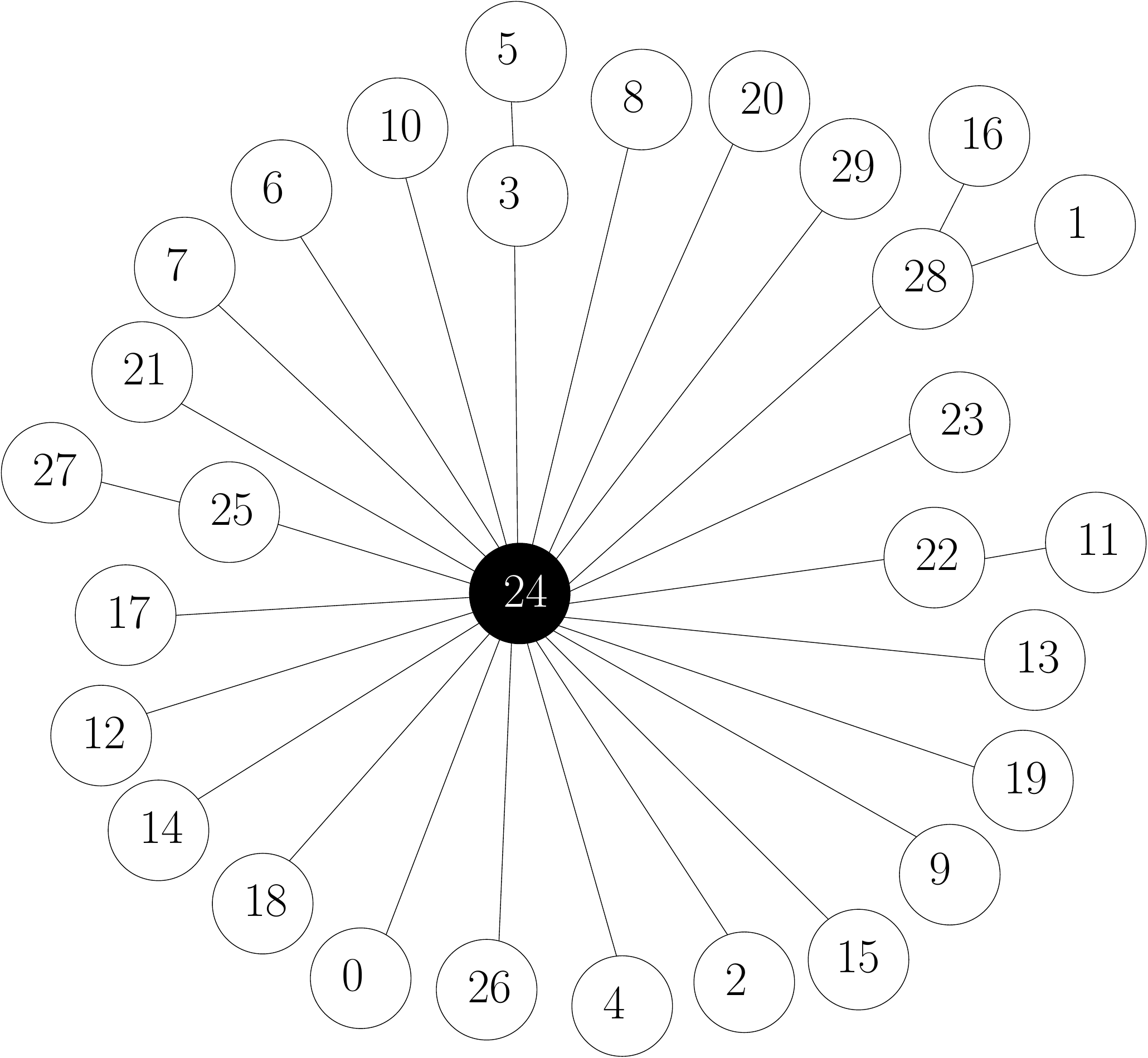}\\
$\alpha=1$ & $\alpha=3$ \\ \\
\includegraphics[scale=0.24]{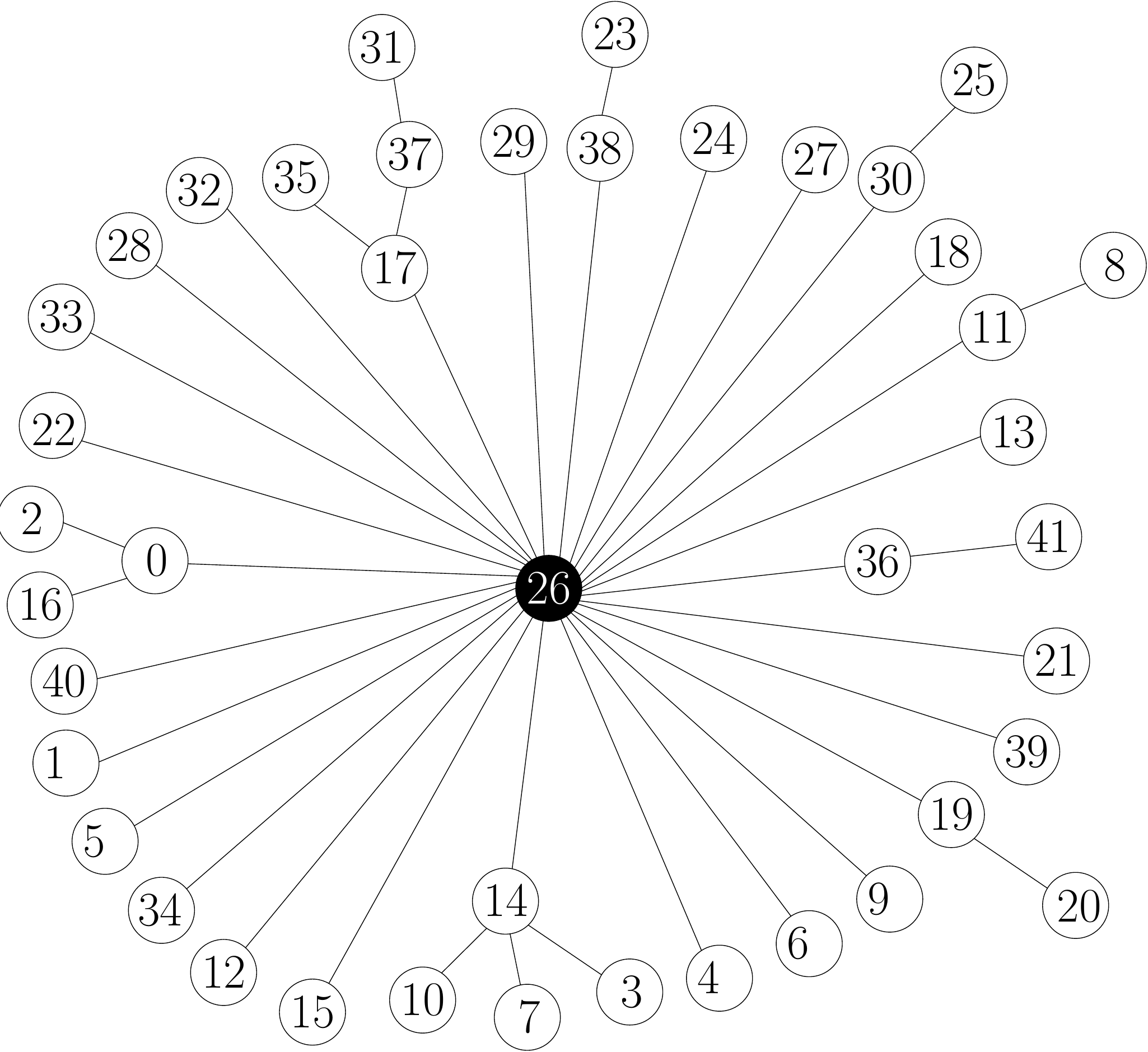} & \includegraphics[scale=0.21]{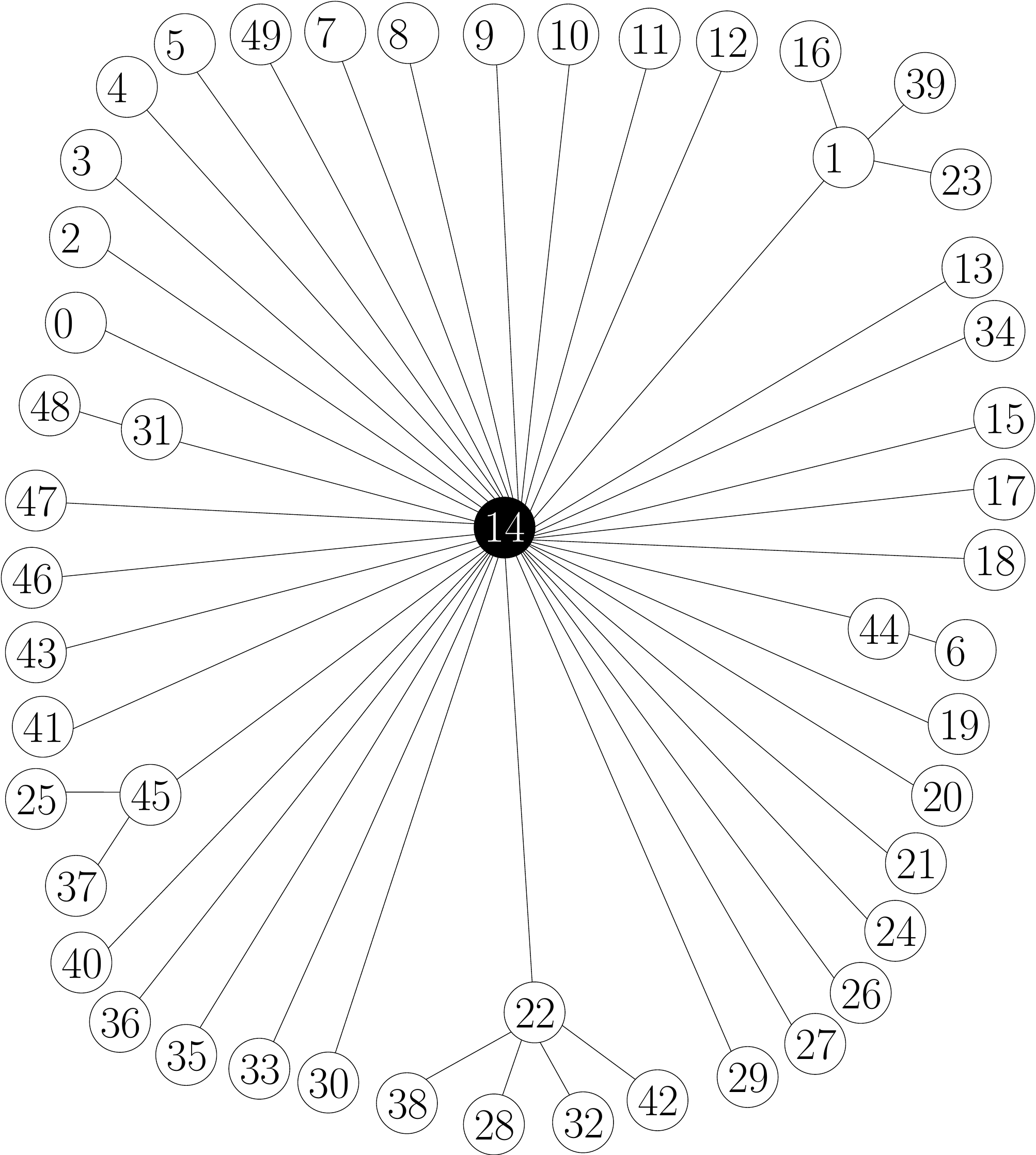} \\
$\alpha=4.2$ & $\alpha=5$ \\ \\
\includegraphics[scale=0.25]{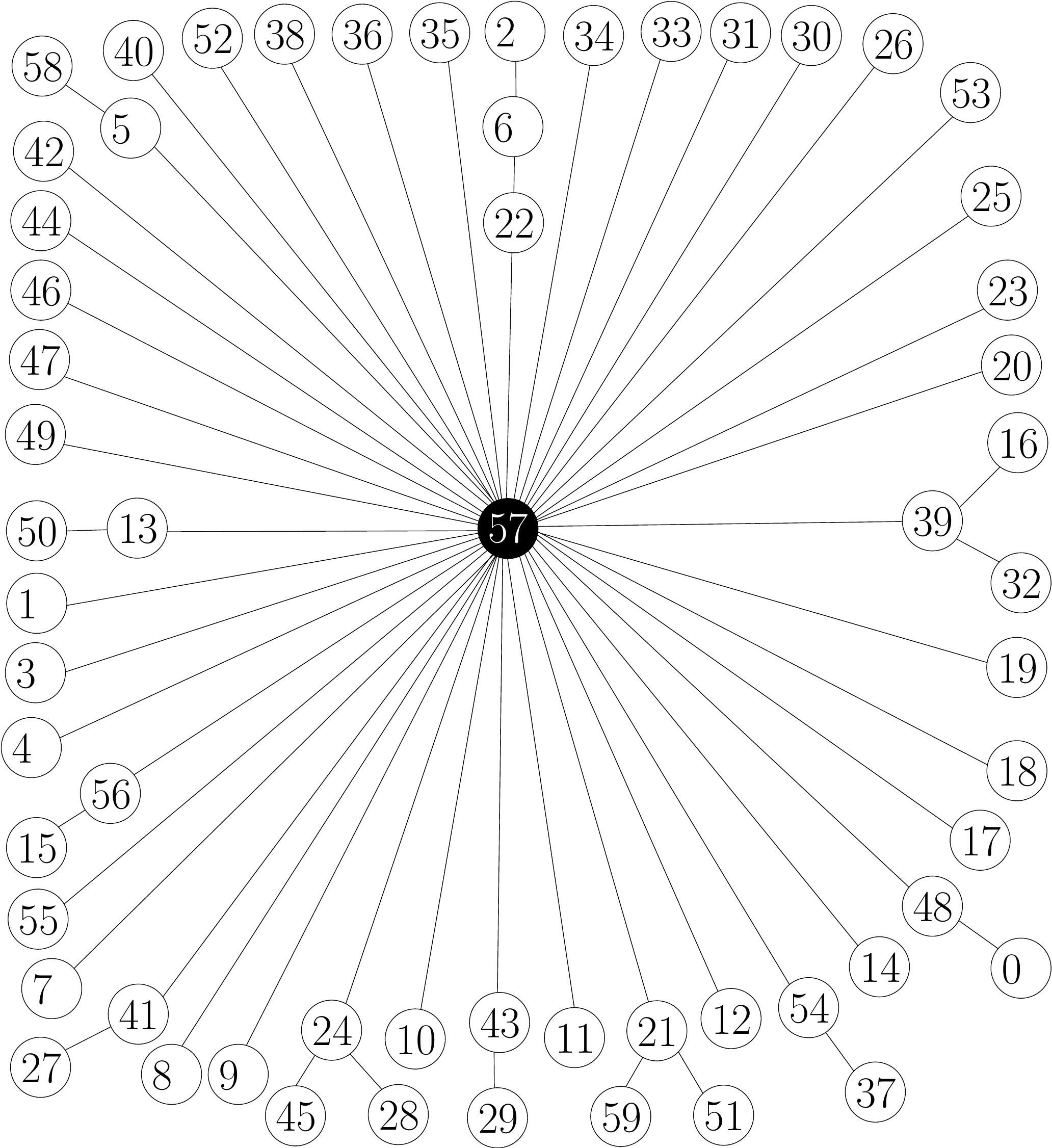} & \includegraphics[scale=0.28]{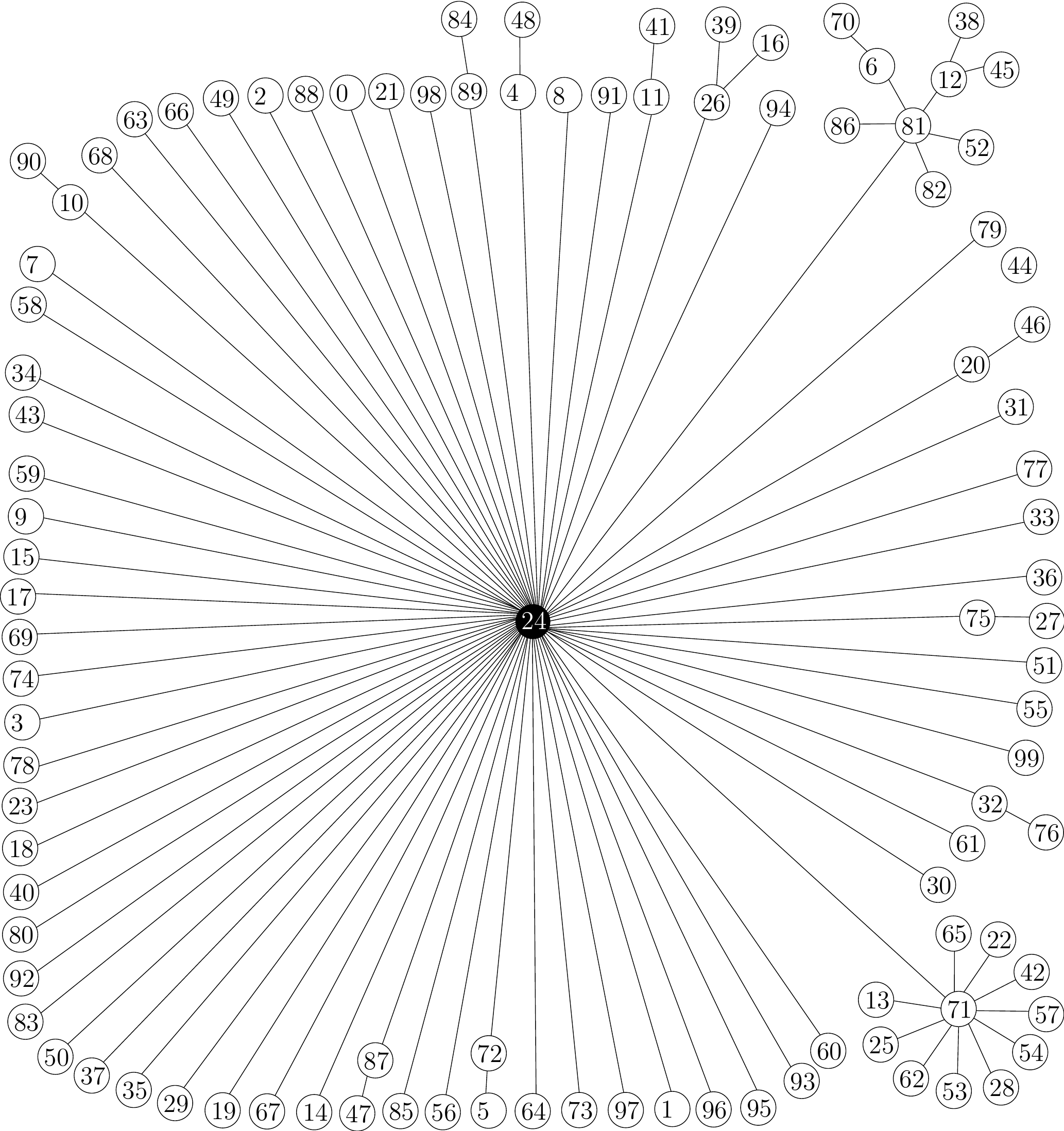} \\
$\alpha=6$ & $\alpha=10$
\end{tabular}
\caption{S2G-PA typical output networks with $10$ variables. The probability of undergoing a full BEC increases with decreasing $\alpha$. When $\alpha=1$, the network obtained by the S2G-PA algorithm features a BEC regardless of the SAT instance given as input. When $\alpha=10$, the typical network shows a partial condensation on the winner node with other growing hubs connected to it.}
\label{fig:typicalS2GPA}
\end{figure}

\end{document}